\documentclass[
  12pt,
  DIV=12,
  headings=standardclasses,
  headings=big
]{scrartcl}

\usepackage[colorlinks=true,citecolor=gray]{hyperref}

\usepackage{newtxtext,newtxmath}  

\usepackage{amsmath,amsfonts}
\usepackage[nofillcomment, linesnumbered, ruled]{algorithm2e}

\SetAlCapSkip{1em}
\SetKwInput{KwParam}{Parameters}
\SetKwInput{KwHyper}{Hyperparameters}
\SetCommentSty{mycommfont}

\title{Learning to Decode the Surface Code\\with a Recurrent, Transformer-Based Neural Network}

\author{\large
\hspace{-1.4cm}%
Johannes Bausch$^{1\ast\dagger}$,
Andrew W Senior$^{1}$\footnote{Corresponding authors: \texttt{andrewsenior@google.com}, \texttt{jbausch@google.com}}\ \,$^{\dagger}$,
Francisco J H Heras$^{1\dagger}$,
Thomas Edlich$^{1\dagger}$,\\\large
\hspace{-1.4cm}%
Alex Davies$^{1\dagger}$,
Michael Newman$^{2}$\footnote{Equal contribution}\ \,,
Cody Jones$^{2}$,
Kevin Satzinger$^{2}$,
Murphy Yuezhen Niu$^{2}$,\\\large
\hspace{-1.4cm}%
Sam Blackwell$^{1}$,
George Holland$^{1}$,
Dvir Kafri$^{2}$,
Juan Atalaya$^{2}$,
Craig Gidney$^{2}$,\\\large
\hspace{-1.4cm}%
Demis Hassabis$^{1}$,
Sergio Boixo$^{2}$,
Hartmut Neven$^{2}$,
Pushmeet Kohli$^{1}$\\
\\
\hspace{-1.4cm}%
\normalsize{$^{1}$Google DeepMind \& $^{2}$Google Quantum AI}
}

\date{}

\usepackage{graphicx}
\graphicspath{{figures/}}

\usepackage{placeins}

\usepackage{physics}
\usepackage{amsfonts}
\usepackage[capitalise]{cleveref}
\usepackage{booktabs}
\usepackage{tikz}
\usetikzlibrary{quantikz}
\usepackage{upgreek}
\usepackage{siunitx}
\usepackage{caption}
\usepackage{subcaption}
\newcommand\prob{\text{Prob}}

\newcommand\mmref[1]{see materials and methods, \cref{#1}}

\begin{document}
\baselineskip18pt
\maketitle 
\begin{abstract}
Quantum error-correction is a prerequisite for reliable quantum computation. Towards this goal, we present a recurrent, transformer-based neural network which learns to decode the surface code, the leading quantum error-correction code. Our decoder outperforms state-of-the-art algorithmic decoders on real-world data from Google’s Sycamore quantum processor for distance 3 and 5 surface codes. On distances up to 11, the decoder maintains its advantage on simulated data with realistic noise including cross-talk, leakage, and analog readout signals, and sustains its accuracy far beyond the 25 cycles it was trained on. Our work illustrates the ability of machine learning to go beyond human-designed algorithms by learning from data directly, highlighting machine learning as a strong contender for decoding in quantum computers.
\end{abstract}

\clearpage

\section{Quantum error correction}

\subsection{Context and background}\label{sec:intro}

The idea that quantum computation has the potential for computational advantages over classical computation, both in terms of speed and resource consumption, dates all the way back to Feynman \cite{Feynman1982}.
Beyond Shor's well-known prime factoring algorithm \cite{Shor1999} and Grover's quadratic speedup for unstructured search \cite{Grover1996}, many potential applications in fields such as material science \cite{Lloyd1996,Aspuru-Guzik2005, babbush2018low, rubin2023fault}, machine learning \cite{Bausch2020,Huang2021-jx,huang2022quantum}, and optimization \cite{Farhi2001,somma2008quantum}, have been proposed.

Yet, for practical quantum computation to become a reality, errors on the physical level of the device need to be corrected so that deep circuits can be run with high confidence in their result. Such fault-tolerant quantum computation can be achieved through redundancy introduced by grouping multiple physical qubits into one logical qubit \cite{Shor1995, Kitaev1997}.

One of the most promising strategies for fault-tolerant computation is based on the surface code, which has the highest-known tolerance for errors of any codes with a 2D nearest-neighbor connectivity \cite{bravyi1998quantum, Fowler2012, Kitaev2003}.
In the surface code, a logical qubit is formed by a $d\times d$ grid of physical qubits, called \emph{data qubits}, such that errors can be detected by periodically measuring $X$ and $Z$ stabilizer checks on groups of adjacent data qubits, using $d^2 - 1$ \emph{stabilizer qubits} located among the data qubits (\cref{fig:architecture}A).  A \emph{detection event} occurs when two consecutive measurements of the same stabilizer give different parity outcomes.
A pair of observables $X_L$ and $Z_L$, which commute with the stabilizers but anti-commute with each other, define the logical state of the surface code qubit.
The minimum length of these observables is called the \emph{code distance}, which represents the number of errors required to change the logical qubit without flipping a stabilizer check. In a square surface code, this is the side length $d$ of the data qubit grid.
The task of an error correction \emph{decoder} is to use the history of stabilizer measurements, the \emph{error syndrome}, to apply a correction to the noisy logical measurement outcome in order to obtain the correct one.

However, decoding quantum codes is a hard problem. For instance, they exhibit \emph{degeneracy}, whereby exponentially many configurations of errors may produce the same history of stabilizer measurements.
They must also contend with rich noise models induced by quantum circuits that include \emph{leakage}, qubit excitations beyond the computational states $\ket 0$ and $\ket 1$ that are long-lived and mobile \cite{Ghosh2013-oq}; and \emph{crosstalk}, unwanted interactions between qubits inducing long-range and complicated patterns of detection events \cite{Tripathi2022-ab}. 
Degeneracy, circuit-level correlations, leakage, and the difficulty in modeling these errors produce decoding problems that resist many tried-and-true methods commonly utilized for classical codes \cite{pearl2022reverend, panteleev2021degenerate, liu2019neural, higgott2022fragile}.

Despite significant progress on quantum error correction \cite{Waldherr2014, Luo2021, milestone2, Sivak2023, sundaresan2022matching, krinner2022realizing, egan2021fault, ryan2021realization, zhao2022realization, gupta2023encoding}, challenges remain. Ultimately, to perform fault-tolerant quantum computation such as the factorization of a 2\,000 bit number, the logical error rate needs to be reduced to less than $10^{-10}$ per logical operation~\cite{Fowler2012, gidney2021factor}.
Logical errors for the surface code are suppressed exponentially, $\sim \Lambda^{-d/2}$, when increasing the code distance $d$,  where $\Lambda$ is a `quality factor' determined by the accuracy of the device and the performance of the decoder. This means that improving the inference accuracy of the decoder will reduce the required size or required gate fidelity of a quantum processor to run a quantum algorithm.
Consequently, an accurate decoder is vital to realizing a fault-tolerant quantum computer using realistic noisy hardware and minimal resources.
In addition, the decoder must be fast enough to keep up with the rate of syndrome information produced by the quantum computer, lest it creates an exponentially increasing backlog of syndrome information to process \cite{terhal2015quantum}.

\subsection{Quantum error correction with machine learning}\label{sec:mldecoders}

In recent years, there has been an explosion of work applying machine-learning techniques to quantum computation, including decoding.
Initial decoders used restricted Boltzmann machines \cite{torlai2017neural}, and many subsequent approaches use reinforcement learning \cite{sweke2020reinforcement, fitzek2020deep, andreasson2019quantum, matekole2022decoding} or supervised learning~\cite{varsamopoulos2019comparing, maskara2019advantages, cao2023qecgpt, choukroun2023deep}.
Several previous works have focused on the surface code, observing that machine-learning techniques could utilize correlations introduced by $Y$-type errors to outperform popular minimum-weight perfect matching (MWPM) decoding \cite{krastanov2017deep}, and could even be used as a preprocessing step for graph-based decoders \cite{Meinerz_2022}.
Other works have focused more on speed and scalability \cite{overwater2022neural, Zhang2023, ni2020neural, gicev2021scalable}, including the use of symmetries to improve performance \cite{egorov2023end, Wagner2020}.
There have also been examples of machine-learning decoders applied to the fully fault-tolerant setting with more realistic noise models \cite{chamberland2018deep, Baireuther_2018, lange2023datadriven}.
In particular, on distance-7 color codes, a recurrent neural network architecture has been used to demonstrate good performance over many error correction cycles \cite{Baireuther_2019}.
However---unlike us---none of these works considered crosstalk or leakage.

More recently, \cite{varbanov2023neural} built on the architecture of \cite{Baireuther_2019}, and assessed its performance on the Sycamore surface code experiment \cite{milestone2}.
They trained their recurrent neural network-based decoder on a circuit-level depolarizing noise model (with noise parameters fitted to the experimental data) and evaluated on experimental data, demonstrating parity with the best previously-published decoder at code distance 3.
Furthermore, the authors directly quantified the benefits of modelling correlations.
They also explored the use of analog
inputs (modelled by a symmetric Gaussian I/Q readout noise model), which allowed a slight increase in accuracy.

In this work, we push the boundary of both scale and accuracy of machine-learning decoding in the fault-tolerant setting of circuit-level noise. We present a novel recurrent architecture using transformers and convolutions, which learns to decode the surface code.
We evaluate our machine-learning decoder on experimental data from a surface code logical qubit implemented on Google's Sycamore quantum computer \cite{milestone2}, and observe markedly better error suppression than state-of-the-art correlated matching and tensor network decoders \cite{milestone2, higgott2022fragile,bravyi2014efficient,arute2019quantum} by training on data directly.

Furthermore, for larger code distances up to 11, we benchmark our ML decoder against a noise model which realistically simulates the effects of leakage, crosstalk, and analog readout as present in a real-world quantum device.
The ML decoder outperforms state-of-the-art correlated matching decoders, and sustains its accuracy over at least 100\,000 error correction cycles (when only trained on up to 25 cycles).
Our recurrent transformer model achieves a further accuracy boost by being able to seamlessly incorporate raw measurement signals from the device (in the form of ``in-phase and quadrature'' readout signals \cite{jeffrey2014fast}), and leakage information from the syndrome.

These capabilities, emerging from the ML decoder's ability to learn from raw data---as well as the decoder's provision of calibrated error probabilities---come without computational overhead.
In short, our decoder expands the scope of fault-tolerant machine-learning decoders while offering better accuracy in realistic settings than the most competitive alternatives.

\section{A recurrent syndrome transformer}

\subsection{Model architecture}

\begin{figure}[p]
    \vspace{-1.3cm}
    \includegraphics[scale=0.5,trim={0cm 0.2cm 0.2cm 0.2cm},clip]{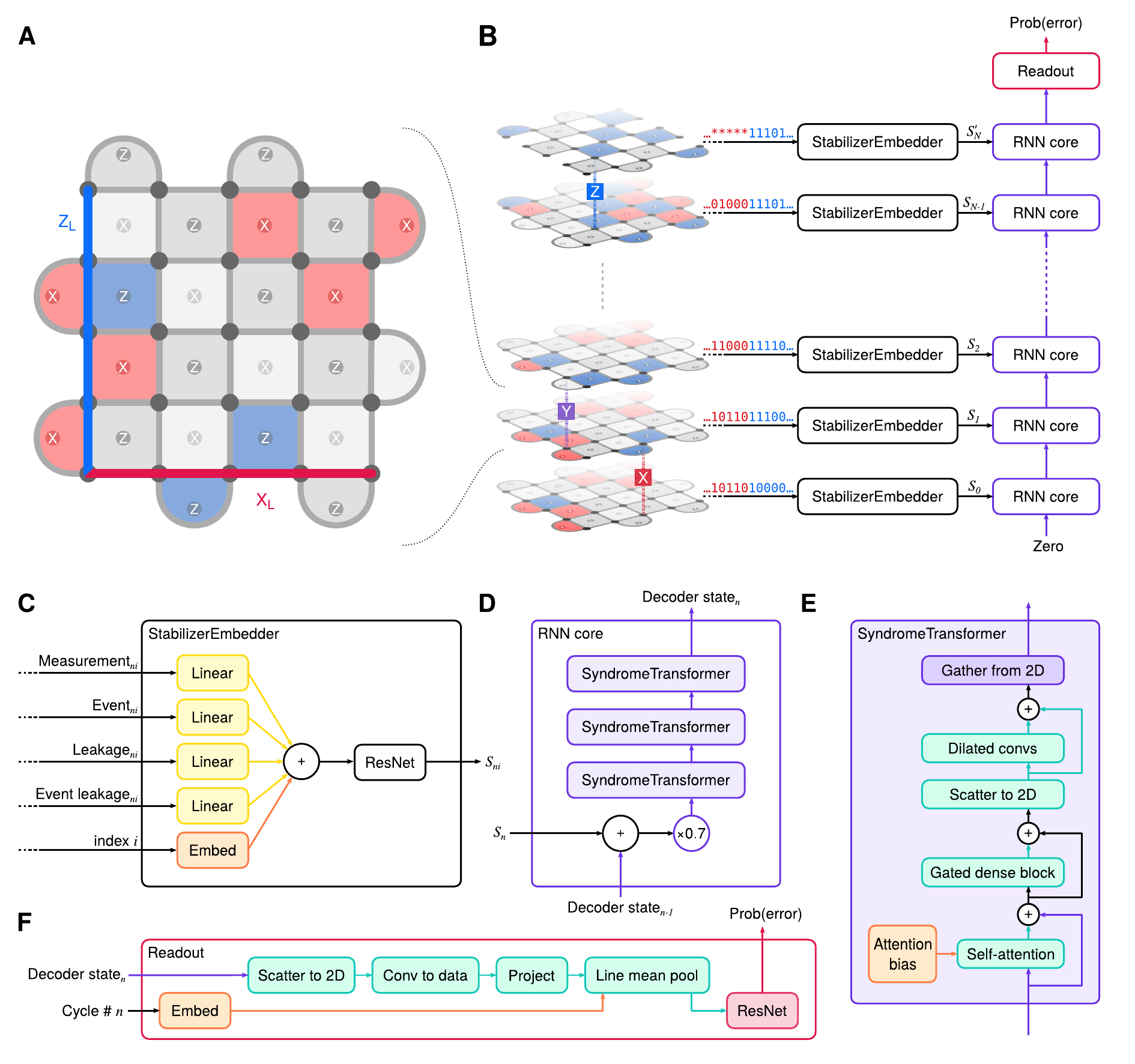}
    \caption{\textbf{The neural network architecture designed for surface code decoding.} (\textbf{A}) $5\times 5$ rotated surface code layout, with data qubits (dark grey dots), $X$ and $Z$ stabilizer qubits (labelled light grey dots, or highlighted in blue/red when they detect a parity violation) interspersed in a checkerboard pattern. Logical observables $Z_\mathrm{L}$ and\ $X_\mathrm{L}$ are shown as bold lines on the left and bottom grid edges respectively. (\textbf{B}) The recurrent network iterates over time updating a representation of the decoder state and incorporating the new stabilizers at each cycle. (\textbf{C}) Creation of an embedding vector $S_{ni}$ for each new stabilizer. (\textbf{D}) Each block of the recurrent network combines the decoder state and the stabilizers $S_n$ for one cycle (scaled down by a factor of $0.7$). The decoder state is updated through three Syndrome Transformer layers. (\textbf{E}) Each Syndrome Transformer layer updates the stabilizer representations through multi-headed attention modulated by a learned attention bias followed by a dense block and dilated 2D convolutions. (\textbf{F}) Logical errors are predicted from the final decoder state.}
    \label{fig:architecture}
    
\end{figure}

Our machine-learning decoder is a neural network with a structure designed to match the structure of the error correction problem, which learns to decode the surface code by training on experimental and simulated data.
Its design mirrors the time-invariance of the stabilizer readout by repeated iteration of a fixed computational block (\cref{fig:architecture}B). This {\em recurrent} architecture constructs a fixed-size \emph{decoder state} representation which stores information about the stabilizers observed up to the current cycle.

Since the data received at each cycle correspond to individual stabilizers, our model maintains the decoder state as a vector representation {\em per stabilizer}.
The model applies a neural network block (Fig.~\ref{fig:architecture}D) at each cycle to update the decoder state by incorporating the current cycle's stabilizers (Fig.~\ref{fig:architecture}C).
While decoders like MWPM typically accept sparse binary \emph{detection events}, a neural network allows for less restrictive inputs. We find better results and more stable training when providing both measurements and events, rather than events alone (\mmref{sec:SI-ablations}). In \cref{sec:beyond-Pauli} we will show that these inputs can be extended further to include probabilistic representations of I/Q readouts and leakage information.
Each stabilizer $i$ at cycle $n$ is represented by a stabilizer embedding vector $S_{ni}$, created by combining the inputs in the network of Fig.~\ref{fig:architecture}C, and directly added in to the decoder state (Fig.~\ref{fig:architecture}D).

The key processing operation is the {\em Syndrome Transformer} (Fig.~\ref{fig:architecture}E) which updates the decoder state by passing information between stabilizer representations in a learned, structured manner.
The Syndrome Transformer augments the multi-headed attention of a conventional Transformer~\cite{transformer} with an {\em attention bias} (\cref{fig:attention-bias-vis}) and spatial convolutions, each of which learn to modulate the information flow between stabilizer representations based on their physical relationship and type. At any stage in the computation, the decoder state can be processed by a readout network (Fig.~\ref{fig:architecture}F) to predict whether a logical error has occurred. In the readout network the decoder state is scattered to a 2D representation and projected to a per-data-qubit representation before pooling into a representation per-row or column of data qubits (representing logical $X_L$ or $Z_L$ observables, respectively), which a final residual network processes before predicting the probability of a logical error. We show the effectiveness of these architecture design decisions by ablation (\cref{fig:SI-ablations}).

\FloatBarrier
\subsection{Decoding the Sycamore memory experiment}\label{sec:MS2}

As a first demonstration, we apply our ML decoder to the \emph{Sycamore memory experiment dataset}, experimental data for surface code logical qubits on Google’s Sycamore device \cite{milestone2}. The experiment comprised both distance 3 and 5 surface codes; the $3\times 3$ code block executed at four separate locations (labelled \emph{north}, \emph{east}, \emph{south}, \emph{west}, `NESW'), within the Sycamore chip, and the $5\times 5$ code block executed at a single location. Both $X$ and $Z$ memory experiments were performed for up to 25 error correction cycles.
Within each experiment, stabilizer syndromes were measured over each cycle, followed by a final cycle of data qubit measurements, from which a final set of stabilizers in the experiment basis as well as the logical readout were computed.  50\,000 experiments were performed for each total cycle count $n\in \{1, 3, \ldots, 25\}$, and the resulting data split into \emph{even} and \emph{odd} subsets for 2-fold cross-validation.

\begin{figure}[t!]
    \centering
    \begin{subfigure}[b]{\textwidth}
    \caption{}
    \includegraphics[width=\textwidth]{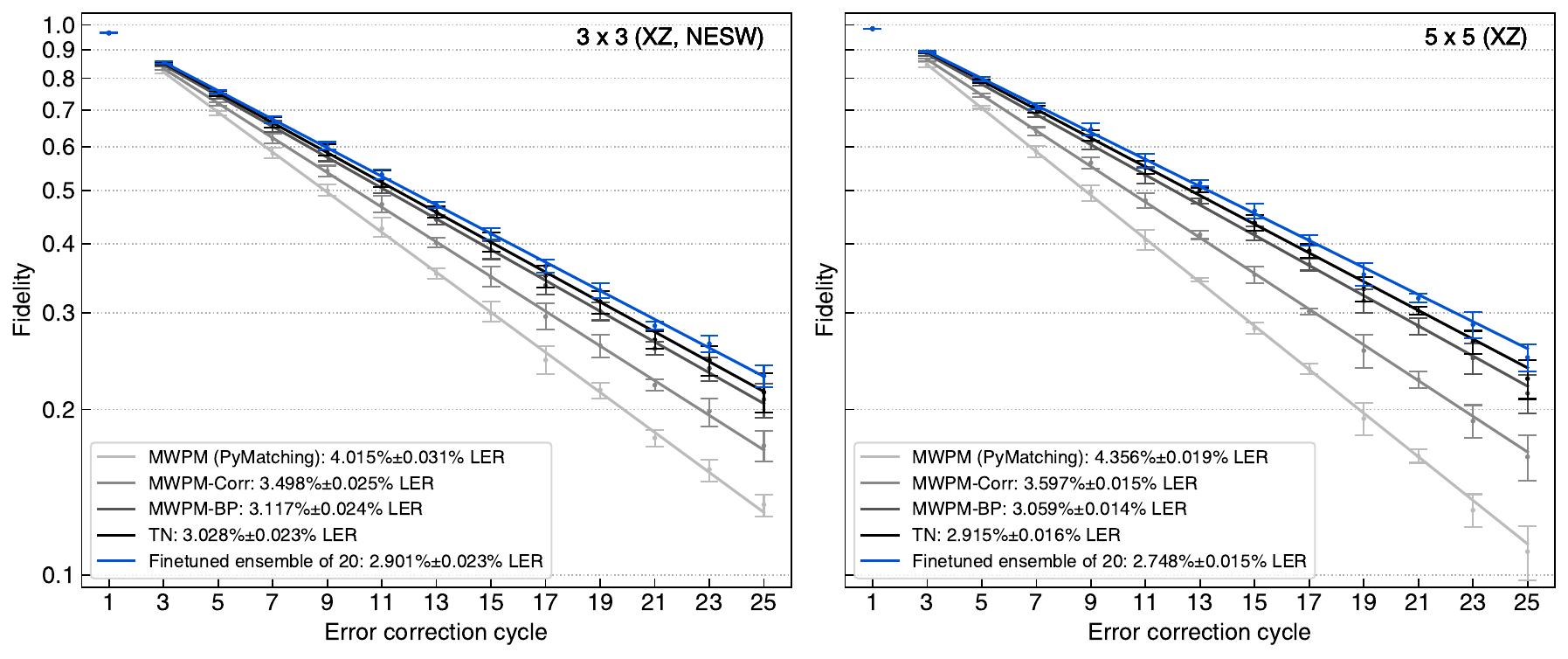}
    \label{fig:figure-1a}
    \end{subfigure}
    \begin{subfigure}[b]{\textwidth}
    \caption{}
    \includegraphics[width=\textwidth]{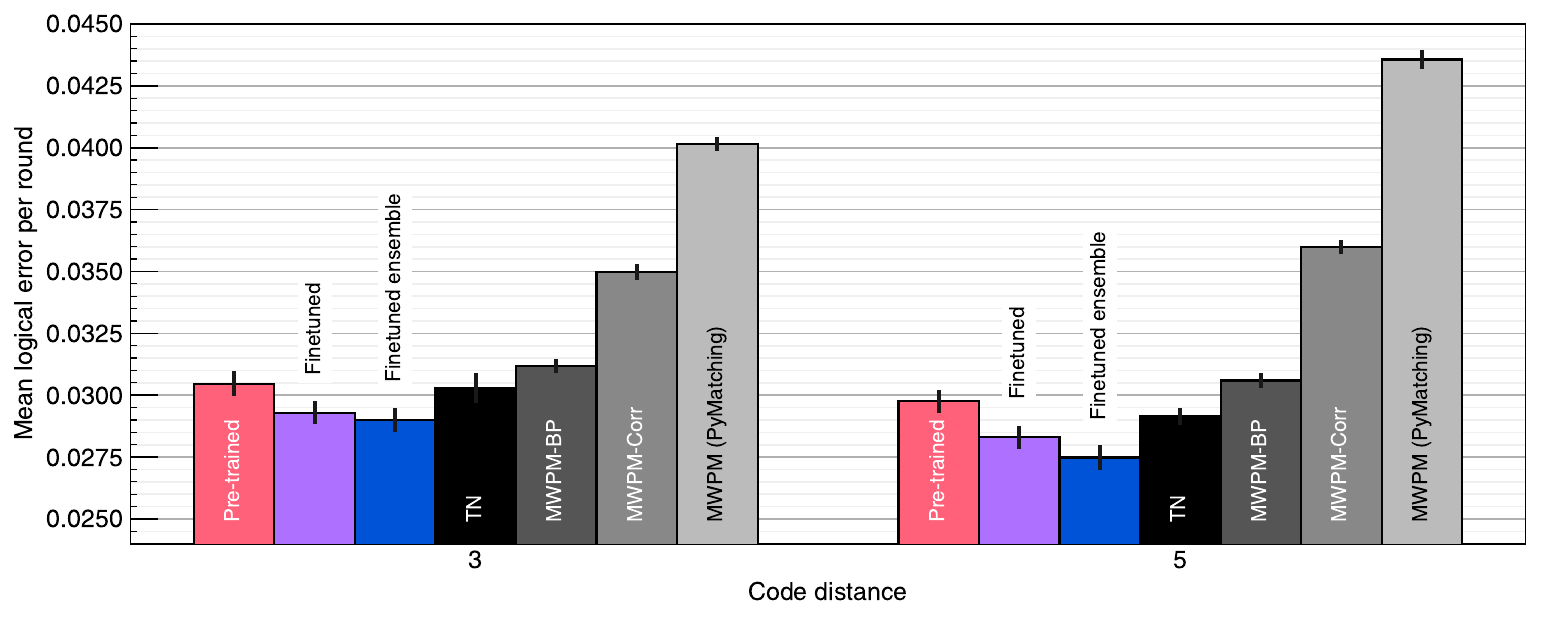}
    \label{fig:figure-1b}
    \end{subfigure}
    \caption{\textbf{ Accuracy of our machine-learning decoder and other leading decoders on the Sycamore experimental data.} All results are averaged across bases, even/odd cross-validation splits and, for the $3\times3$ experiments, the location (NESW). (\textbf{A}) Fidelity ($2 \times \mathrm{accuracy} - 1$) vs.\ error correction cycle for code distance 3 and 5 memory experiments in the Sycamore experimental dataset for the baseline tensor network decoder (black), our decoder (blue) and three variants of MWPM (shades of gray). In the legend, we show the logical error per round (LER), calculated from the slope of the fitted lines. 
    (\textbf{B}) Fitted logical error per round for different decoders for the $3\times 3$ and $5\times 5$ datasets. Variants of our ML decoder without ensembling or fine-tuning are also compared to show the advantage of the complete training procedure. }
    \label{fig:figure-1}
    
\end{figure}

Decoder performance is quantified by the \emph{logical error per round} (where round means the same as cycle, abbreviated LER), a measure of the fraction of experiments in which the decoder fails for each additional error correction cycle.
As in \cite{milestone2}, we calculate the LER by a linear regression of the log-fidelities for experiments of different numbers of cycles $n\in\{3,5,\ldots,25\}$ (\mmref{sec:SI-fit}, \cref{fig:figure-1}A).

We trained our decoder in two stages: pre-training and fine-tuning. 
In the pre-training stage, and e.g.\ for the even fold, we train on $2\times 10^9$ samples drawn from detector error noise models (DEMs, \cite{gidney2021stim}) fitted to the even experimental detection event distributions, and choose the model with the best validation LER, computed on the even experimental samples. 
For the fine-tuning stage we randomly partition the even experimental samples, training on $19\,880$ with weight decay relative to the pre-trained model to regularize the parameters. We choose the parameters giving the lowest LER on the remaining $5\,120$ samples (after $\approx 120$ passes through the data).  This procedure allows us to train a decoder to high accuracy with limited access to experimental data, while holding back the other (odd) fold as test set.
When the pre-training stage uses a more `standard' noise model such as circuit depolarizing noise, fine-tuning compensates for most of the mismatch (\mmref{sec:SI-pretraining-ablation}).

Our ML decoder achieves a lower LER ($2.901\pm 0.023\%$ at distance 3, and $2.748\pm0.015\%$ at distance 5, for a $\Lambda=1.056\pm0.010$) than the tensor network (TN) decoder ($3.028\pm0.023\%$ resp.\ $2.915\pm0.016\%$, $\Lambda=1.039\pm0.010$), the most accurate decoder hitherto reported for this experiment \cite{milestone2,bravyi2014efficient}. 
The accuracy gain becomes even more pronounced in comparison to state-of-the-art MWPM-based decoders, such as correlated matching (MWPM-Corr), matching with belief propagation (MWPM-BP), as well as PyMatching, an open-source implementation of MWPM \cite{higgott2022fragile,milestone2,pymatching} (\cref{fig:figure-1}A).
Fine-tuning on experimental data, as well as {\em ensembling} (training multiple models and combining their outputs, \mmref{sec:SI-Ensembling}) take our technique from parity with the TN decoder to a clear advantage (\cref{fig:figure-1}B). 

\FloatBarrier
\subsection{Beyond Pauli noise: leakage, cross-talk, I/Q noise}\label{sec:beyond-Pauli}

To achieve reliable quantum computation, the decoder must scale to higher code distances.  Since hardware implementations at the time of our study do not reach scales capable of running a surface-code experiment beyond distance 5, we explore the performance of our decoder for larger code distances using simulated data at error rates both comparable to and significantly lower than the Sycamore experimental data (\mmref{sec:SI-Training}, \cref{fig:event-densities}). 
The conventional circuit-noise model~\cite{o2017density,Baireuther_2018} is limited, failing to account for several crucial real-world effects, such as cross-talk, leakage and analog readout information. 
A decoder that recognizes these patterns can be especially valuable,  since correlated errors are highly damaging to fault-tolerance~\cite{preskill2006}. 
In this section, we use circuit simulators that can model and modulate the above effects independently at code distances $3, 5, 7, 9,$ and $11$.

\subsubsection{Analog readouts} 

Projective measurement is essential to extract information about errors from the surface code. Typically, each measurement is classified as a discrete outcome $\ket0$ or $\ket1$. However, in some quantum computing architectures, there is underlying analog data with information about uncertainty and leakage to non-computational states. This is the case for standard dispersive measurement of superconducting qubits, where a microwave pulse probes the frequency of a superconducting resonator that is coupled to a qubit in order to infer the qubit’s state \cite{blais2004cavity, wallraff2005approaching, jeffrey2014fast}. The returned microwave signal contains information in its amplitude and phase, traditionally represented in a two-dimensional space of in-phase (“I”) and quadrature (“Q”) amplitudes.
It is simple to provide this information to our ML decoder: for each stabilizer readout, instead of a binary value (0 if the most likely state is $\ket0$, 1 if $\ket1$), we provide the probability of being in state $\ket1$, based on a decay model, parameterized by the signal-to-noise ratio (SNR) and normalised measurement time $t = t_{\text{meas}}/T_1$.
This is provided as the network input at cycle $n$ along with a probabilistic analog to detection events (\mmref{sec:SI-soft-events}, \cref{fig:iq-pdfs}).

\begin{figure}[t]
    \centering
    \hspace*{-1mm}\mbox{
    \begin{subfigure}[b][][t]{0.36\textwidth}
    \caption{}  
    \hspace{-3mm}\includegraphics[trim={0 -1mm 0 0},clip,width=1.1\textwidth]{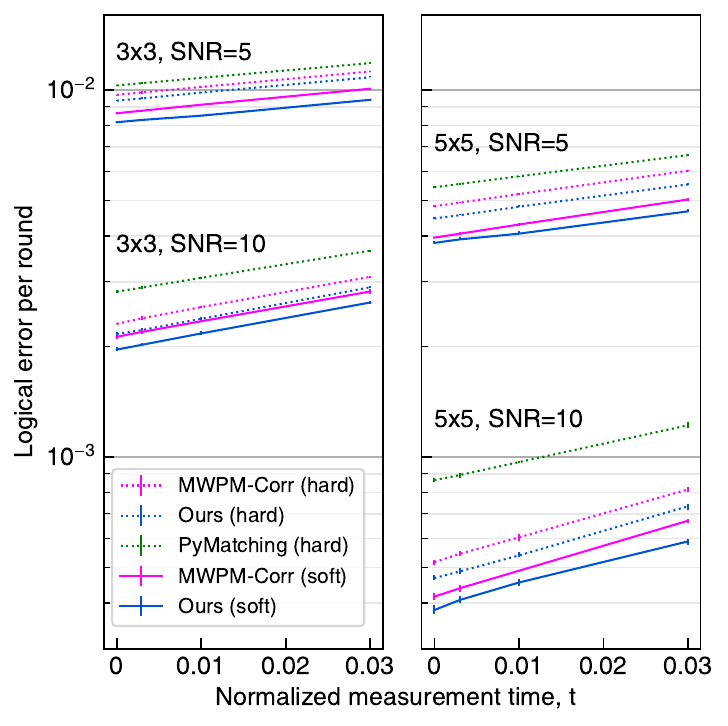}
    \end{subfigure}
    \hspace{0.2cm}
    \begin{subfigure}[b][][t]{0.344\textwidth}
    \caption{}
    \hspace{-2mm}\includegraphics[trim={2mm 0 0 0},clip,width=1.09\textwidth]{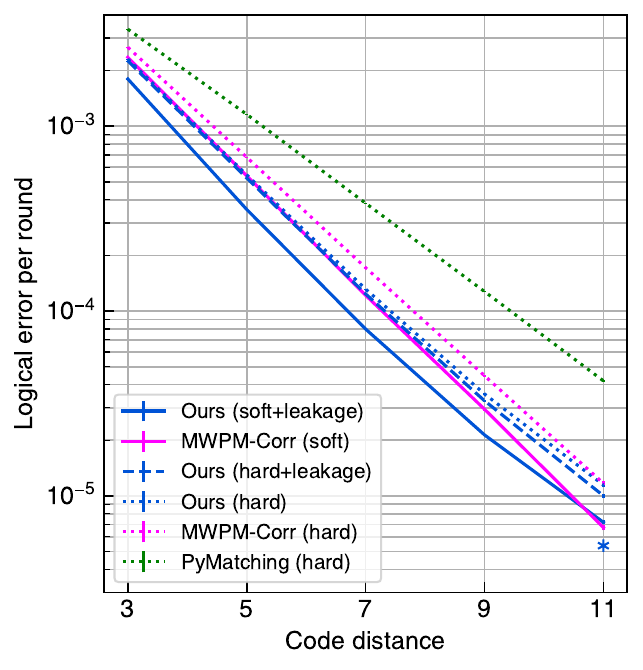}
    \end{subfigure}
    \hspace{0mm}
    \begin{subfigure}[b][][t]{0.266\textwidth}
    \caption{}
    \includegraphics[width=1.01\textwidth]{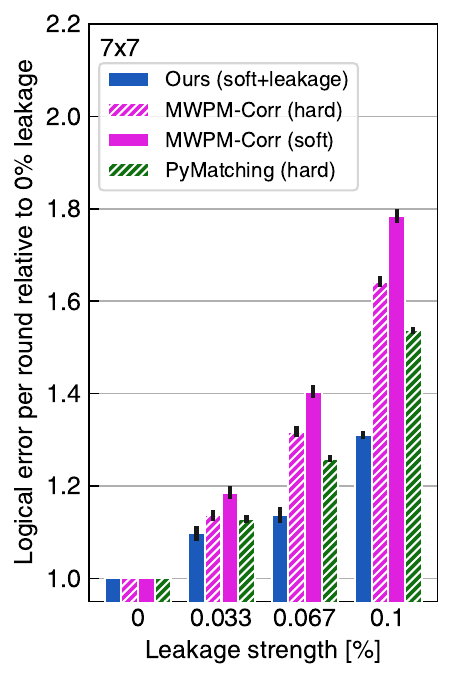}
    \end{subfigure}
    }
    \caption{\textbf{ Scaling the decoder and the effect of noise levels and input modality on decoder accuracy.} \textbf{(A)}~Decoder performances for SI1000 data generated with different I/Q noise parameter values. Line styles denote whether the model is provided with the richer, continuous input representation for I/Q readouts (``soft inputs'', continuous lines) or with thresholded binary values (``hard inputs'', dotted lines). Both the ML decoder (blue) and MWPM-Corr (magenta) benefit from the soft inputs but our ML-based technique is more accurate for a range of readout noise levels. \textbf{(B)}~Logical error per round for different decoders at different code distances, for the same Pauli+ noise model (SNR 10, $t=0.01$, 0.1\% leakage). Line styles represent whether decoder inputs are binary (hard) or soft and whether they contain information about leakage. Our ML decoder (blue), trained on $2\times 10^9$ samples, leads MWPM-Corr (magenta) by a margin that further improves when augmented with leakage information and probabilistic inputs. The additional point at distance $11$ extended training to $10^{10}$ training samples. \textbf{(C)} The relative impact of leakage on the performance of different decoders and input modalities at distance $7$. Our ML decoder is more robust to leakage than the MWPM-based decoders.}
    \label{fig:figure-2}

\end{figure}

Figure~\ref{fig:figure-2}A shows that ML can outperform correlated matching for a superconducting-inspired circuit depolarizing noise model (SI1000 noise, \mmref{sec:SI-SI1000,sec:SI-IQ+SI1000}, \cite{Gidney2021}), augmented with I/Q measurements (where we always threshold the final data qubits to prevent the logical measurement from leaking into the training inputs, \mmref{sec:SI-time-scalability}). If the network is provided with the richer probabilistic inputs derived from I/Q readouts (which our ML decoder can process without runtime penalty), the performance improves further, for both $3\times 3$ and $5\times 5$ codes. Incorporating probabilistic measurement inputs into MWPM-Corr \cite{pattison2021improved} also leads to an improvement (though this comes with a runtime penalty).

\subsubsection{Leakage and crosstalk}\label{sec:pauli_plus}

Training and evaluating on Pauli+ simulated data (\mmref{sec:SI-pauli_plus}), we show that our ML decoder is robust to crosstalk and leakage (\cref{fig:figure-2}B). For SNR $=10$, characteristic decay rate $t_{\text{meas}}/ T_1 = t=0.01$, and 0.1\% leakage (meaning $0.1\%$ leakage chance for stabilizer readouts and $\approx 0.3\%$ on the data qubit measurements, where leakage is removed after each round \cite{McEwen2021Leakage}), our ML-based decoder is more accurate than MWPM-Corr for all code distances 3--9 when trained on a consistent $2\times 10^9$ samples, and at distance 11 when training for $10^{10}$ samples.
We highlight that the number of network parameters is kept almost constant across code distances (\mmref{sec:SI-Speed}, \cref{tab:parameters}).

At distance 11, the ML decoder achieves a logical error per round of $5.37\pm 0.01 \times 10^{-6}$ (resp.\ $6.74\pm 0.02 \times 10^{-6}$ for MWPM-Corr).
While the shape of the ML decoder curves suggest a decrease in relative error suppression $\Lambda$ at larger code distances, training to convergence takes longer (cf.\ \cref{fig:training-examples}); the additional point at distance $11$ illustrates that at sufficient training samples an average error suppression factor from distance 3 to 11 of $\Lambda=4.28\pm0.02$ is possible, on par with MWPM-Corr ($\Lambda=4.33\pm0.04$, see \cref{tab:lambda-table}).
$\Lambda$ for matching is determined by the distribution of errors and the decoder’s noise model. For an ML decoder it also becomes a function of training: improved training improves the logical error per round, and this improvement versus the code distance is what determines $\Lambda$.

Hitherto, leakage states were usually ignored and classified as either $\ket0$ or $\ket1$, but can sometimes be distinguished in the I/Q representation~\cite{Sank2016Measurement, khezri2022measurementinduced}.
With traditional binary inputs and without leakage information, our machine-learning decoder outperforms MWPM-Corr at code distances 3--9, and achieves parity at distance 11.
Providing leakage information in the form of an additional binary input (leaked/not leaked) slightly increases the lead (\cref{fig:figure-2}B, blue dashed vs.\ dotted lines) beyond MWPW-Corr (dotted magenta line).
Yet, if we provide the model with richer soft information (a probability of being leaked, and a conditional probability of being in $\ket1$, given the measurement was not in a leaked state), the advantage against correlated matching becomes greater (\cref{fig:figure-2}B, solid blue line), even when MWPM-Corr incorporates probabilistic measurement inputs (\cref{fig:figure-2}B, solid magenta line).

For intermediate code distances, the relative penalty incurred from increased leakage as compared to the case of no leakage is much less pronounced for our machine learning decoder (\cref{fig:figure-2}C for code distance 7 and \cref{fig:SI-leakage-degradation} for 3,5,9 and 11).
E.g.\ compared to no leakage, when adding 0.1\% leakage, the ML decoder's logical error per round at distance $7$ increases by $\approx 30\%$ (blue bar), whereas leakage-unaware MWPM-Corr suffers a larger penalty of $\approx 60\%-80\%$ (magenta bar). While there have been proposals for incorporating leakage information into matching in idealized models \cite{suchara2015leakage}, our ML decoder architecture learns to mitigate the effects of leakage without the difficulty of directly characterizing them.

\begin{figure}[t!]
    \centering
    \begin{subfigure}[b][][t]{0.49\textwidth}
    \caption{}
    \includegraphics[width=\textwidth,trim={0 0 0 0},clip]{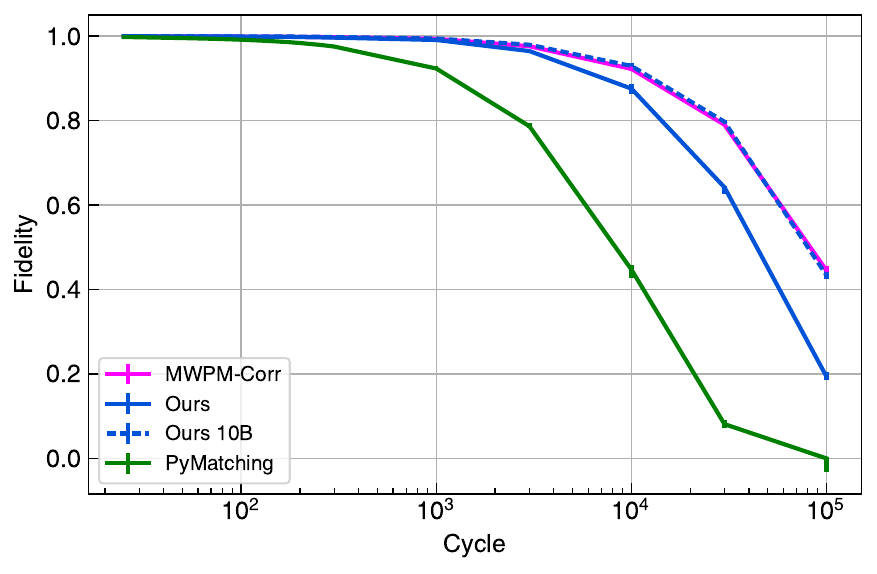}
    \end{subfigure}
    \begin{subfigure}[b][][t]{0.49\textwidth}
    \caption{}
    \includegraphics[width=\textwidth,trim={0 0 0 0},clip]{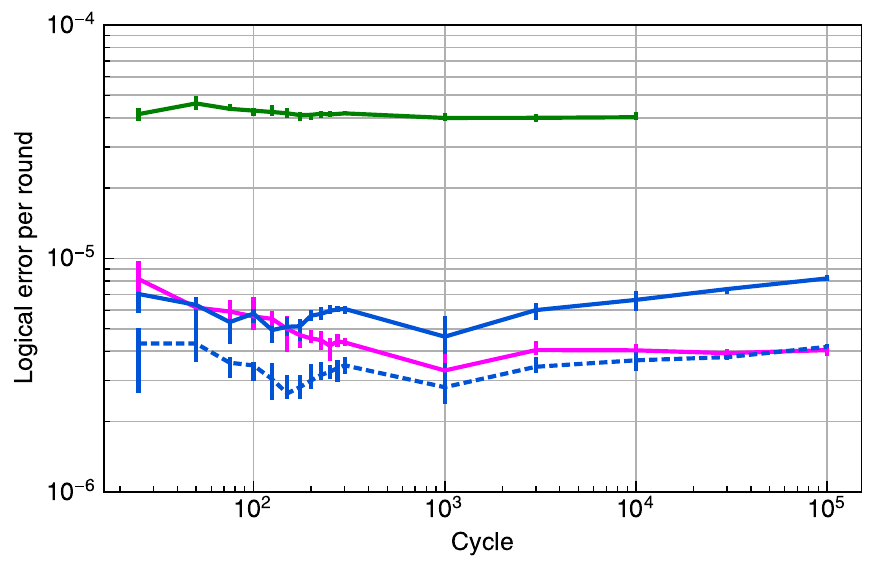}
    \end{subfigure}
    \caption{\textbf{ Generalization to larger number of error correction cycles at code distance 11.} Fidelity (2 $\times$ accuracy - 1) after up to 100,000 error-correction cycles (\textbf{A}) and the corresponding logical error per round (\textbf{B}) for PyMatching (green), MWPM-Corr (magenta) and for our ML decoder (blue) trained only on Pauli+ simulated experiments of up to 25 cycles, with 2B training samples (solid line) and 10B training samples (dashed line). Both training and test samples are Pauli+ (SNR=10, t = 0.01, 0.1\% leakage). We only plot LER values where the corresponding fidelity is greater than $0.1$. The data is generated from the same simulated experiments, stopped at different number of cycles.
    }
    \label{fig:300-rounds}
    
\end{figure}

We demonstrate that the advantage of our ML decoder against MWPM decoders persists even in the longest experiments we tried. We trained our decoder using Pauli+ simulated experiments of up to 25 error correcting cycles (soft inputs, SNR $=10$, $t=0.01$, and leakage 0.1\%), and found its performance to generalize to Pauli+ experiments of up to at least 100\,000 cycles (\cref{fig:300-rounds}, \mmref{sec:SI-time-scalability}). 

Ultimately, our ML decoder's flexibility in seamlessly learning from leakage readout, I/Q information, and cross-talk induced error patterns is an advantage over other state-of-the-art decoders, as evidenced by its improved accuracy.

\FloatBarrier
\subsection{Postselection: improving accuracy with confidence}\label{sec:confidence}

\begin{figure}[t!]
    \centering
    \begin{subfigure}[b][][t]{0.41\textwidth}
    \caption{}
    \includegraphics[width=\textwidth,trim={0 0 0 0},clip]{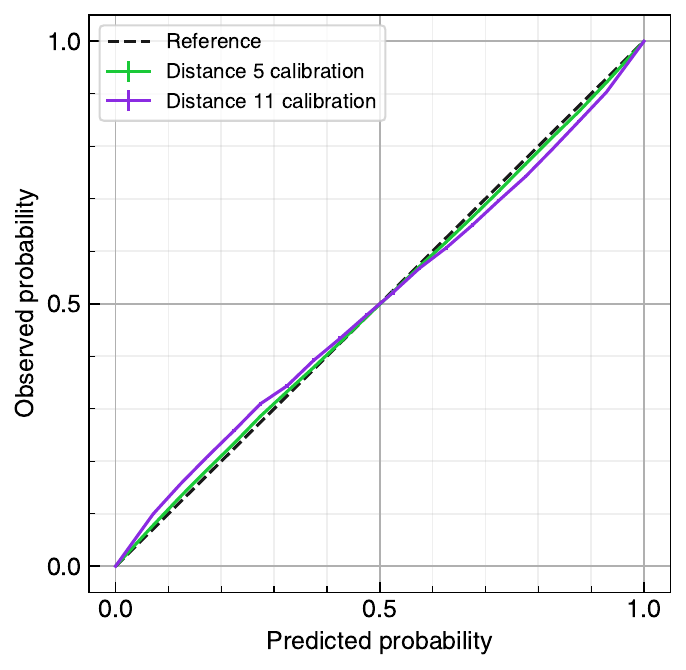}
    \end{subfigure}
    \begin{subfigure}[b][][t]{0.58\textwidth}
    \caption{}
    \includegraphics[width=\textwidth,trim={0 0 0 0},clip]{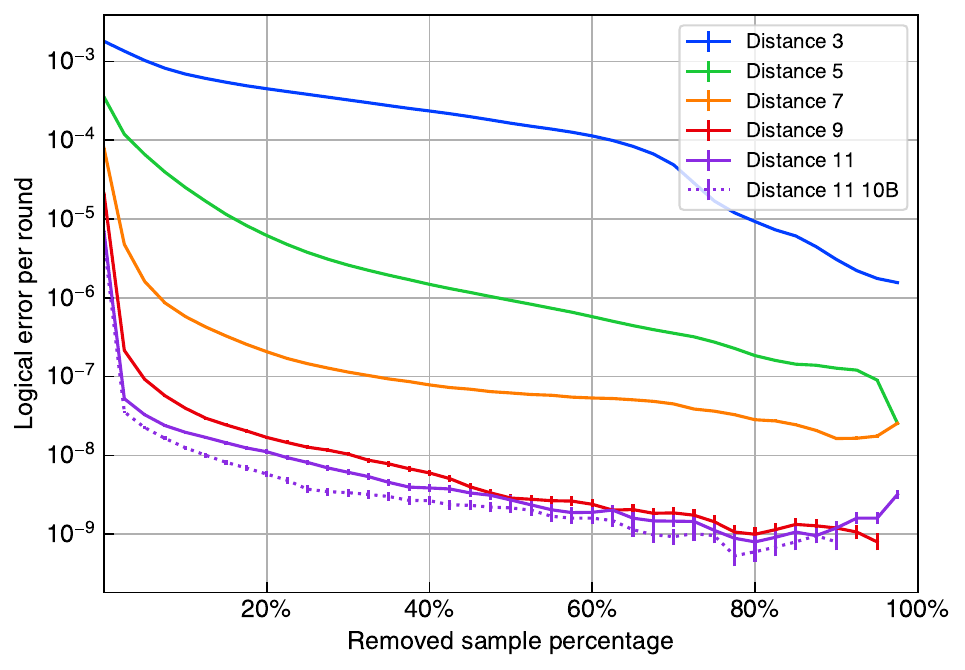}
    \end{subfigure}
    \caption{\textbf{ The use of the network's output as a confidence measure for post-selection.} For the ensembled model performance of our ML decoder, trained using 25-cycle Pauli+ simulated experiments (SNR $=10$, $t=0.01$, 0.1\% leakage), tested on $10^9$ samples. \textbf{(A)} Example calibration plot at distances 5 (green continuous line) and 11 (purple continuous line), with bootstrapped errors (small but present error bars). Black dashed line represents a perfectly calibrated classifier.
    \textbf{(B)} Logical error rate vs.\ the fraction of experiments discarded when discarding experiments with low confidence,  (error bars: standard error of mean from values in each bin, visible for a LER $\lesssim 10^{-8}$).}
    \label{fig:postselection}
    
\end{figure}

We trained the neural network with a logistic output, minimizing cross-entropy against binary labels.
As a result, its output can be interpreted as the probability that the true label is $\ket1$, a probability we found to be well-calibrated (\cref{fig:postselection}A, \ref{fig:individual-calibration-plots}).
For example, of samples with prediction probability 0.8, approximately 80\% indeed have the true label $\ket1$. Samples with a probability close to 0.5 are more likely to have been misclassified than samples with a probability closer to 0 or 1 (\cref{fig:individual-calibration-bins}).

The probabilistic output can be used as a confidence measure to discard the least confident samples (Fig.~\ref{fig:postselection}B).
On the same Pauli+ simulation (SNR=10, $t=0.01$, 0.1\% leakage) as in \cref{sec:pauli_plus}, and by rejecting only $0.2\%$ of the 25-cycle experiments at distance 11, we can achieve a postselected error rate a factor of $\approx 10$ lower ($10\%$ rejection gives a factor $\approx 250$), which could prove useful in repeat-until-success protocols \cite{Bravyi2005,Bravyi2012,Campbell2017,gidney2021factor,poulin2006optimal, pattison2023hierarchical}.

\FloatBarrier
\subsection{Discussion}
\subsubsection{Advantages}

In addition to improved error-suppression, our ML decoder inherently provides features desirable for a quantum error correction decoder.
\paragraph{Trainable on raw data.} The network can be directly trained on examples from real hardware. As such, its performance is less dependent on the availability or accuracy of a noise model characterizing the device. This reduces reliance on the ability to obtain an accurate noise profile of the device, which is often difficult~\cite{milestone2,o2017density}.

\paragraph{Ability to use rich inputs.} The network can process rich inputs, such as analog readout signals from the hardware \cite{jeffrey2014fast}, I/Q points, signatures of leakage \cite{Miao2022}, or potentially other side information such as temperature or calibration parameters~\cite{kelly2016,milestone2}. This leads to improved accuracy compared to processing stabilizer syndrome inputs alone (\cref{fig:architecture}C, \cref{fig:figure-2}).

\paragraph{Calibrated output.} Instead of a binary output, the decoder outputs a calibrated error probability that can be used as a confidence measure e.g.\ for post-selection.

\subsubsection{Throughput considerations}
To avoid backlog, any quantum error correction decoder needs to achieve a throughput (processing time per cycle) commensurate with the syndrome accumulation speed of the quantum chip---currently 1µs for superconducting qubits \cite{arute2019quantum,milestone2}, and $>\!\!1$ms for trapped ion devices \cite{Schafer2018,Ryan-Anderson2021}.
Latency (the delay between the final input to the decoder and its output) plays a secondary role in dictating the quantum computer's logical clock speed \cite{terhal2015quantum,overwater2022neural}.
Improving throughput and latency remains an important goal for both machine learning and matching-based decoders \cite{jeffrey2014fast,Liyanage2023, skoric2022parallel, tan2022scalable}.

While the model has not yet been optimized for inference speed, a host of techniques can be applied to speed up the model (\mmref{sec:SI-Speed}); yet even without those, when running our ML decoder on an accelerator and for code distances up to 25 (untrained for distances $>\!\!11$), its throughput is already within about one to two orders of magnitude of the target throughput rate of 1µs (see \cref{fig:timing-a}).
Note that, in contrast to graph-based decoders, our decoder throughput is independent of the noise rate by design \cite{varsamopoulos2019comparing}, avoiding slowdowns caused by spikes in noise rate. 
The recurrent architecture allows our architecture to decode an indefinite number of error correction cycles, and benefits from parallel processing (particularly GPUs and TPUs). 
Consequently, the ML decoder's inference time scales efficiently across code distance with a single, fixed-size architecture.

\subsubsection{Training data requirements}
\cref{fig:training-examples} shows the number of training samples needed to achieve LER parity with MWPM (PyMatching) and MWPM-Corr.
As observed before~\cite{varsamopoulos2019comparing,niu2020learnability}, the data requirements grow exponentially: a hundredfold increase to go from distance 3 to distance 11.
Indeed, the trend suggests that one might expect to pre-train a model for code distance 25 to reach parity with MWPM-Corr using $10^{13} - 10^{14}$ training examples.
However, this one-off pre-training on a generic noise model can be followed by fast fine-tuning on hardware-specific datasets to achieve the best performance, without training a completely new model from scratch.

\section{Conclusions and outlook}\label{sec:conclusions}
We have shown that a neural network decoder with inductive biases
motivated by the quantum error correction problem can learn to
decode the surface code with state-of-the-art error suppression.  On
experimental data, it outperforms the previous best-in-class tensor
network decoder, which takes orders of magnitude longer to run,
although achieving the real-time throughput rates of a superconducting architecture remains a challenge.

Our ML decoder's advantage in accuracy persists at scale, as we continue to
outperform correlated matching performance at distances up to 11.
This performance is maintained over numbers of error correction cycles
that far exceed the training regime. While the architecture itself can
be executed on even larger code distances with only a moderate
runtime increase, training the decoder to continue suppressing
error beyond distance 11 is a further challenge.

As a machine learning model, our decoder's greatest strengths come from its ability to learn from real experimental data.
Allowing it to seamlessly take advantage of rich inputs representing I/Q noise and leakage, these capabilities come without a human in the loop to design an explicit algorithm to use each novel feature.
This ability to use available experimental information showcases the strength of machine learning in a scientific context.

While we anticipate other decoding techniques will continue to improve, we believe that our work provides evidence that machine learning decoders may achieve the necessary error suppression and speed to enable practical quantum computing.

\bibliographystyle{abbrv}
\bibliography{main}

\section*{Acknowledgments}
The Google DeepMind team would like to thank
Jonas Adler,
Charlie	Beattie,
Sebastian Bodenstein,
Craig Donner,
Pavol Drot\'ar,
Fabian Fuchs,
Alex Gaunt,
Ingrid von Glehn,
James Kirkpatrick,
Clemens Meyer,
Shibl Mourad,
Sebastian Nowozin,
Ivo Penchev,
Nick Sukhanov, and
Richard Tanburn
for helpful discussions and other contributions to the project.
The Google Quantum AI team would like to thank
Austin Fowler,
Thomas O'Brien,
and Noah Shutty
for their feedback on the manuscript.

\section*{Authors contributions}
J.A.\ and D.K.\ developed models and wrote software for modeling realistic noise in superconducting processors.
J.B.\ conceptualized and supervised the research, and contributed to project administration, data curation, investigation, formal analysis, validation and visualization of results, as well as to the writing of the paper.
S.J.B.\ supported the investigation, resource provision, and software development aspects of the experiments.
S.B.\ provided project supervision, software tools and coordination, and direction of priorities for scalable decoding.
A.D.\ helped conceptualize and supervise the research, and contributed to data curation as well as the investigation, methodology, and writing of the paper.
T.E.\ contributed to data curation, resource provision and methodology, as well as the investigation and methodology, formal analysis, validation and visualization of results, and to the writing of the paper.
M.Y.N.\ contributed to the research conceptualization, the investigation and methodology, crosstalk error modeling in superconducting processors, and the writing of the paper.
C.G.\ contributed knowledge about the theory of decoders and configuring noise models, as well as software support.
D.H.\ contributed to the research conceptualization and supervision, and sponsored the research.
F.J.H.H.\ helped conceptualize the research, and contributed to data curation, investigation, methodology, formal analysis, validation and visualization of the results, as well as to the writing of the paper.
G.H.\ provided project administration, and supported the conceptualization of research direction.
C.J.\ provided project supervision, software tools and coordination, and direction of priorities for scalable decoding.
P.K.\ contributed to the research conceptualization and supervision, and sponsored the research.
H.N.\ contributed to the research conceptualization and supervision, and sponsored the research.
M.N.\ helped conceptualize the research, and contributed to the investigation and methodology, project supervision, knowledge about the theory of decoders, and analysis and validation of results, as well as to the writing of the paper.
M.Y.N.\ contributed to the investigation and methodology, helped in validating the result, contributed knowledge about the theory of decoders and configuring noise models, and helped write the paper.
K.S.\ contributed experimental knowledge on leakage, measurement, soft information, analyzing soft information for decoding, and helped write the paper.
A.W.S.\ led the investigation and methodology, helped conceptualize and supervise the research, and contributed to data curation, formal analysis, validation and visualization of the results, and to the writing of the paper.

\clearpage
\appendix
\setcounter{figure}{0}
\setcounter{table}{0}
\renewcommand\thefigure{S\arabic{figure}}    
\renewcommand\thetable{S\arabic{table}}

\section{Materials and methods}
\subsection{Datasets and noise models}\label{sec:SI-Datasets}
\FloatBarrier\subsubsection{Memory experiments}

A memory experiment is the most basic error correction experiment to run, but is representative of the difficulty of fault-tolerant computation with the surface code \cite{milestone2}. We encode a known single (logical) qubit state $\ket{\psi_i}$ as $\rho_i =\ketbra{\psi_i}$, perform multiple error correction cycles (i.e. stabilizer qubit measurements), and finally perform a (logical) measurement on the resulting state $\rho_f$. We declare success if the decoded logical outcome matches the initial logical encoding.

In a real-world experiment, all operations on the physical qubits are noisy. Several noise models capturing various levels of noisiness can be studied, corresponding to various levels of abstraction in modelling real-world noise,  e.g.\ the noiseless case; the code capacity case (noiseless readouts); or the phenomenological case (noise on physical qubits and readouts). Beyond this qualitative classification of noise types to study, the actual noise model that is used to describe the various operations on the physical qubits can vary tremendously in accuracy, from simple bit- or phase-flip noise, to a full simulation of the Master equation of the quantum system.\footnote{And even then, how accurately the Master equations describe a real-world system can vary significantly.}

\FloatBarrier\subsubsection{The rotated surface code}\label{sec:SI-surfacecode}

\begin{figure}[t]
    \centering\mbox{
    \hspace{-5mm}
    \begin{quantikz}[row sep={0.7cm,between origins},column sep=0.15cm,align equals at=1]
    \lstick{(a)}     & \qw     & \qw     & \qw     & \ctrl4  & \qw \\ 
    \qw              & \qw     & \qw     & \ctrl3  & \qw     & \qw \\ 
    \qw              & \qw     & \ctrl2  & \qw     & \qw     & \qw \\ 
    \qw              & \ctrl1  & \qw     & \qw     & \qw     & \qw \\
    \lstick{$\ket0$} & \targ{} & \targ{} & \targ{} & \targ{} & \meter{}
    \end{quantikz}
    \hspace{3mm}
    \begin{quantikz}[row sep={0.7cm,between origins},column sep=0.15cm,align equals at=1]
    \lstick{(b)}     & \qw     & \qw       & \qw       & \qw       & \targ{}   & \qw     & \qw \\ 
    \qw              & \qw     & \qw       & \qw       & \targ{}   & \qw       & \qw     & \qw \\ 
    \qw              & \qw     & \qw       & \targ{}   & \qw       & \qw       & \qw     & \qw \\ 
    \qw              & \qw     & \targ{}   & \qw       & \qw       & \qw       & \qw     & \qw \\
    \lstick{$\ket0$} & \gate H & \ctrl{-1} & \ctrl{-2} & \ctrl{-3} & \ctrl{-4} & \gate H & \meter{}
    \end{quantikz}
    \hspace{3mm}
    \begin{quantikz}[row sep={0.7cm,between origins},column sep=0.15cm,align equals at=1]
    \lstick{(c)}     & \qw     & \qw       & \gate H & \qw       & \qw       & \gate H & \ctrl{}   & \qw     & \qw \\ 
    \qw              & \qw     & \qw       & \gate H & \qw       & \ctrl{}   & \gate H & \qw       & \qw     & \qw \\ 
    \qw              & \qw     & \qw       & \gate H & \ctrl{}   & \qw       & \gate H & \qw       & \qw     & \qw \\ 
    \qw              & \qw     & \ctrl{}   & \gate H & \qw       & \qw       & \gate H & \qw       & \qw     & \qw \\
    \lstick{$\ket0$} & \gate H & \ctrl{-1} & \qw     & \ctrl{-2} & \ctrl{-3} & \qw     & \ctrl{-4} & \gate H & \meter{}
    \end{quantikz}}
    \caption{$X$ (a) and $Z$ (b) stabilizer readouts for the rotated surface code \cite{Versluis2017}. (c) Common $X$ and $Z$ stabilizer readout for the XZZX code \cite{Bonilla_Ataides2021}.}
    \label{fig:stabilizer-readouts}
\end{figure}
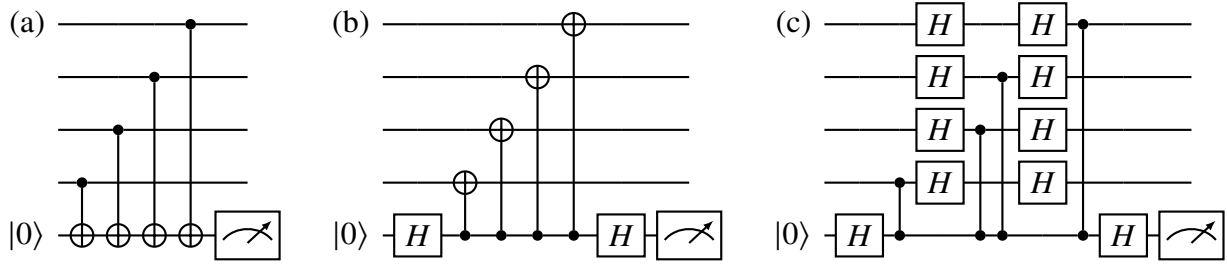

Here we study the memory experiment for a rotated surface code. Introduced in \cite{Bombin2007-np}, it is a variant of the surface code, which itself is a variant of Kitaev's toric code \cite{Kitaev1997}.
In the rotated surface code, stabilizers are interspersed in a checkerboard-like pattern in a 2D grid of data qubits (\cref{fig:figure-1}A). Stabilizer readouts are performed via stabilizer ancilla qubits, in a circuit as given in \cref{fig:stabilizer-readouts}, depending on the code variant.

\FloatBarrier\subsubsection{Sycamore memory experiment dataset}
\label{sec:SI-Datasets-Experimental}
This is the publicly-released dataset \cite{milestone2data} accompanying Google's Sycamore surface code experiment from \cite{milestone2}, comprising:
\begin{enumerate}
    \item Four areas at code size $3\times 3$, dubbed \emph{north}, \emph{south}, \emph{east}, and \emph{west}; as well as one area for code size $5\times 5$.
    \item Both $X$ and $Z$ memory experiment bases.
    \item Individual experiments at $1, 3, 5, \ldots, 25$ error correction cycles, at $50\,000$ shots each.
\end{enumerate}
Each dataset was split into an \emph{even} and \emph{odd} subset for two-fold cross-validation, and accompanied by a detector error model fitted to the respective subset, to be used for decoding the other fold, respectively.

\FloatBarrier\subsubsection{Detector error model (DEM)}
\label{sec:SI-dem}
A detector error model (DEM) \cite{gidney2021stim} can be thought of as an error hypergraph, where stochastic error mechanisms are hyperedges connecting the clusters of detectors they trigger. These mechanisms are independent and have an associated error probability. The DEMs we use were previously fitted \cite{milestone2} to each experimental set using a generalisation of the $p_{ij}$ method \cite{google2021exponential}.

We use the open-source program Stim \cite{gidney2021stim} to generate samples using the DEMs. This is necessary for pre-training our ML decoders, as the limited quantity of experimental data available makes training with only experimental data unfeasible (see Appendix \ref{sec:SI-Training}).  

\FloatBarrier\subsubsection{Circuit depolarizing noise: SD6 and SI1000}\label{sec:SI-SI1000}

\newcommand{\includesc}[1]{
    \includegraphics[width=1.7cm]{figures/circuits/circuit-sc--#1_svg-raw.pdf}
}
\newcommand{\includescv}[1]{
    \includegraphics[width=1.7cm]{figures/circuits/circuit-sc-xzzx--#1_svg-raw.pdf}
}
\begin{figure}[t]
    1. Initialization:\\[2mm]
    \includesc{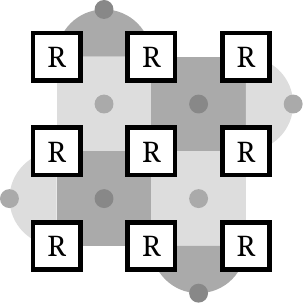}
    \includesc{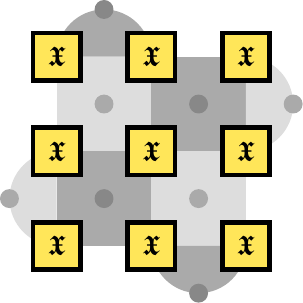}
    \includesc{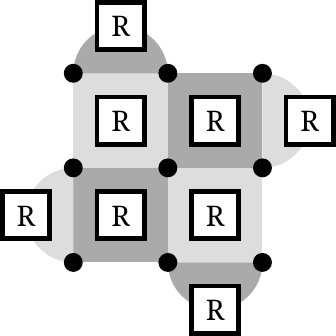}
    \includesc{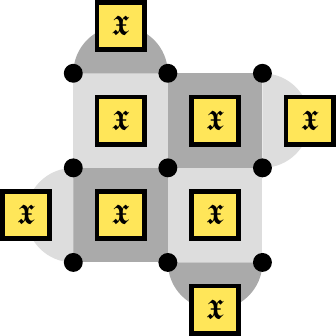}\\
    2a. Stabilizer measurement cycle:\\[2mm]
    \includesc{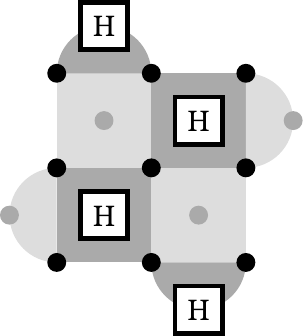}
    \includesc{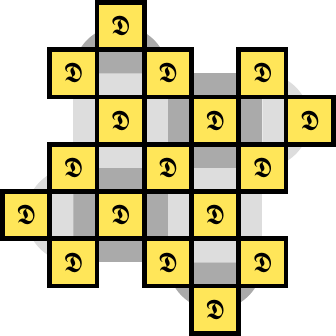}\\
    \includesc{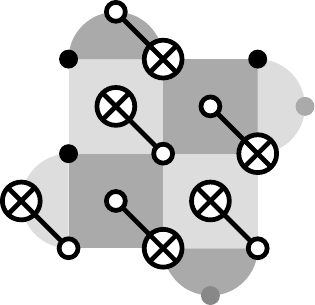}
    \includesc{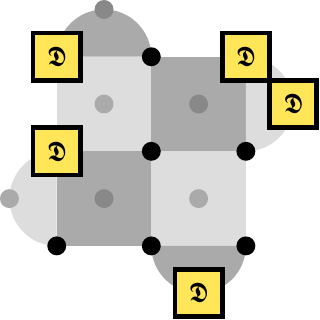}
    \includesc{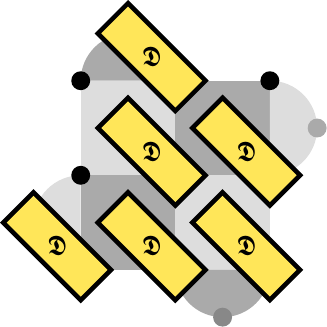}\hspace{1cm}
    \includesc{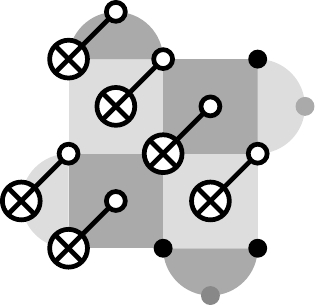}
    \includesc{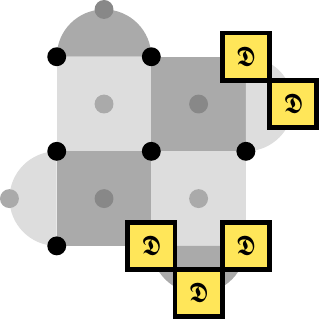}
    \includesc{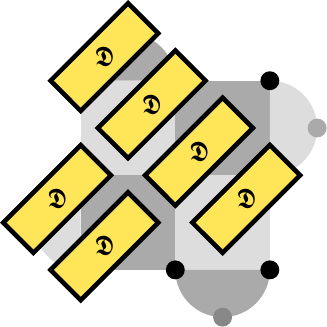}\\
    \includesc{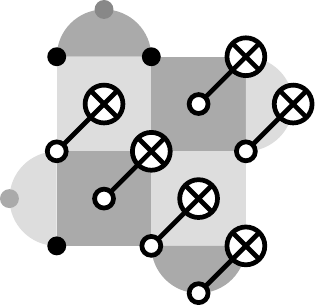}
    \includesc{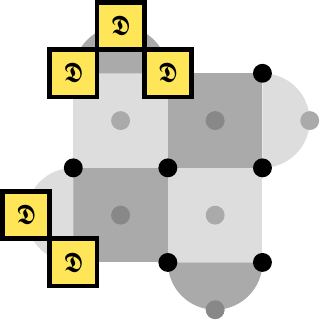}
    \includesc{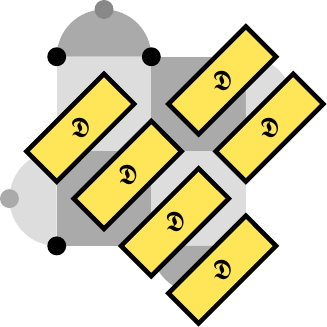}\hspace{1cm}
    \includesc{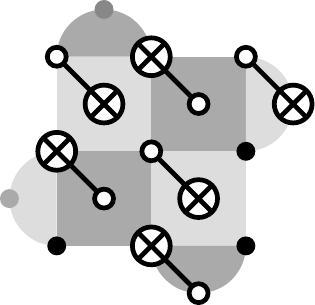}
    \includesc{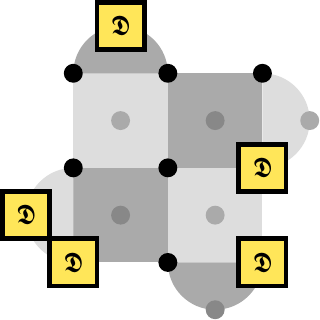}
    \includesc{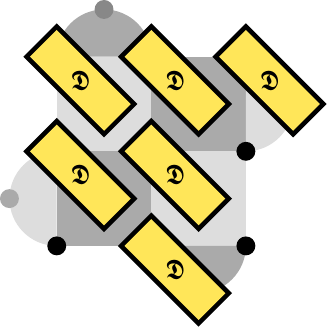}\\
    \includesc{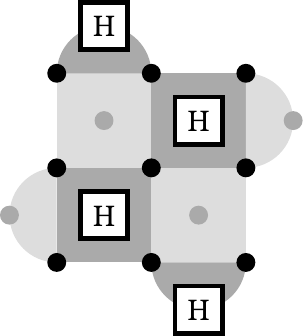}
    \includesc{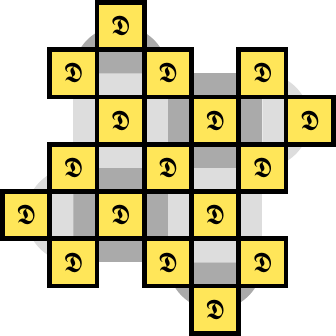}\\
    2b. Stabilizer qubit readouts and reset:\\[2mm]
    \includesc{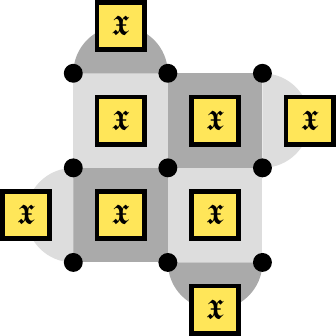}
    \includesc{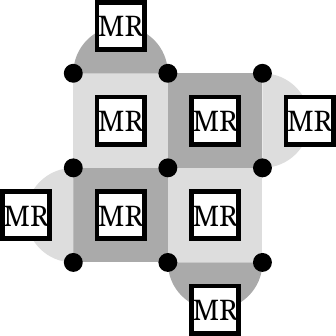}
    \includesc{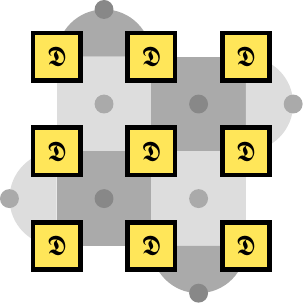}
    \includesc{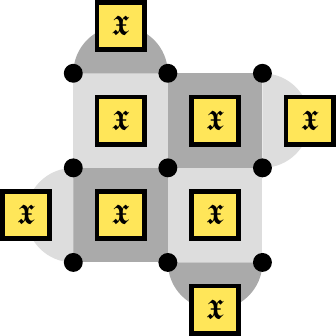}\\
    3. Data qubit readouts:\\[2mm]
    \includesc{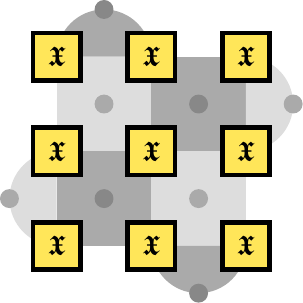}
    \includesc{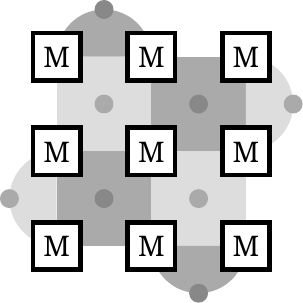}
    \caption{\textbf{Circuit depolarizing noise gate and error schema for a $3\times3$ rotated surface code in the $Z$ basis.} Black dots indicate data qubits, gray dots indicate $X$/$Z$ stabilizer qubits, as detailed in \cref{fig:figure-1}B.
    $\mathfrak D$ labels single- or two-qubit depolarizing noise, and $\mathfrak X$ labels a bit flip channel. $\mathrm M$, $\mathrm R$, and $\mathrm{MR}$ are measurements, reset, and combined measurement and reset in the $Z$ basis. $\mathrm H$ is a Hadamard gate, and $\mathrm{CNOT}$ gates are indicated by their standard circuit symbol.}
    \label{fig:sc}
    
\end{figure}

\begin{figure}[t]
    1. Initialization:\\[2mm]
    \includescv{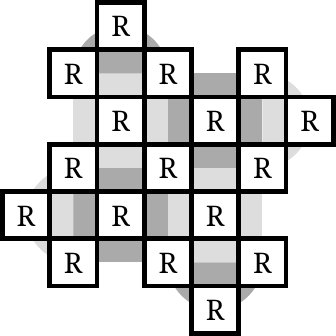}
    \includescv{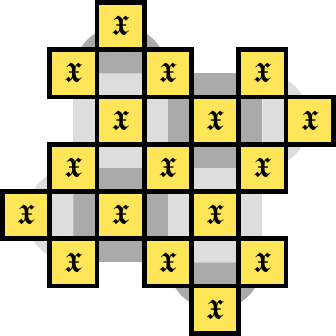}\\
    2a. Stabilizer measurement cycle:\\[2mm]
    \includescv{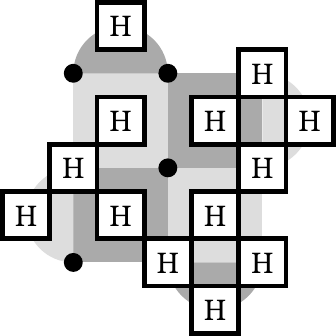}
    \includescv{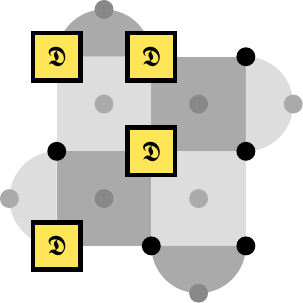}
    \includescv{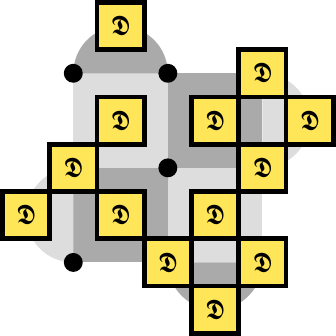}\hspace{1cm}
    \includescv{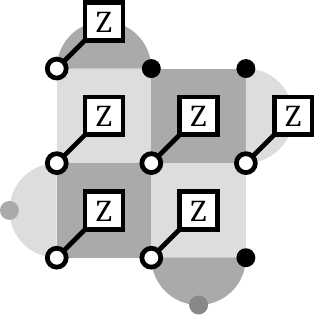}
    \includescv{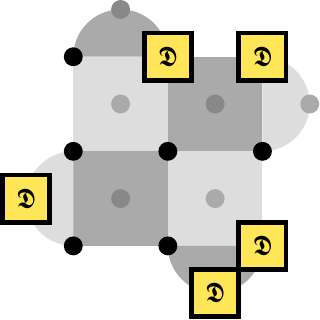}
    \includescv{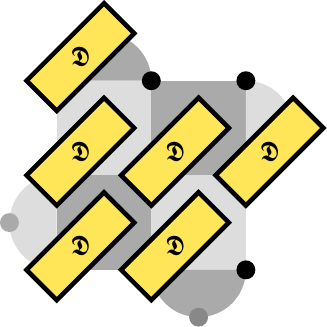}\\
    \includescv{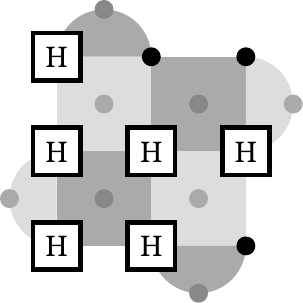}
    \includescv{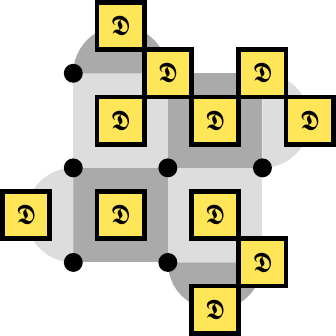}
    \includescv{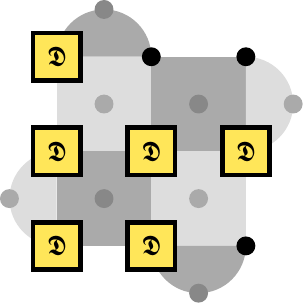}\hspace{1cm}
    \includescv{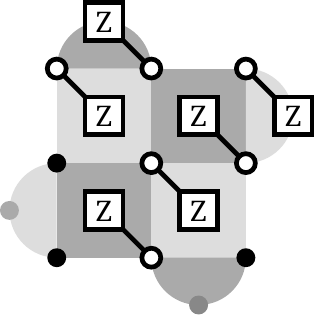}
    \includescv{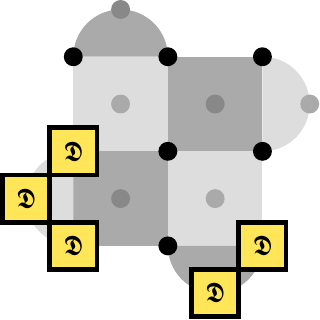}
    \includescv{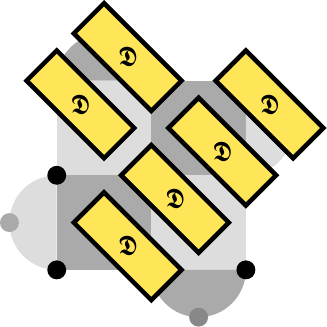}\\
    \includescv{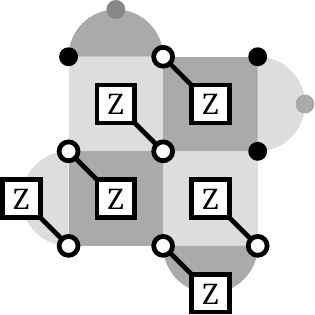}
    \includescv{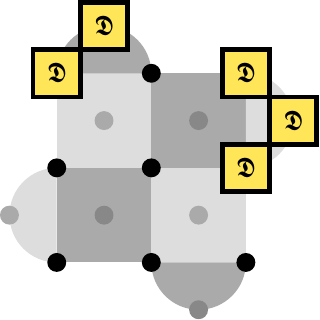}
    \includescv{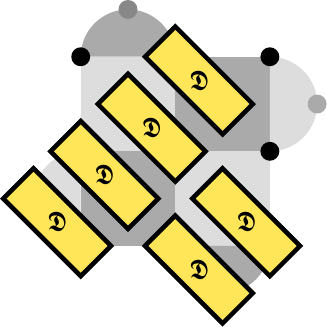}\hspace{1cm}
    \includescv{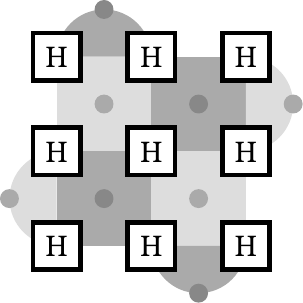}
    \includescv{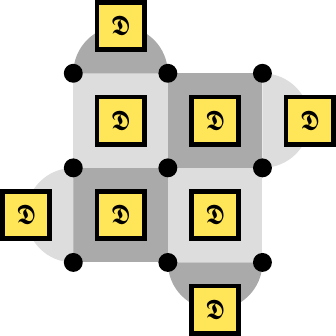}
    \includescv{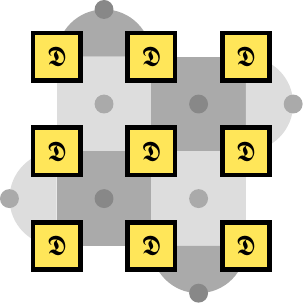}\\
    \includescv{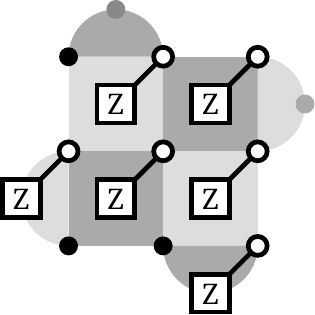}
    \includescv{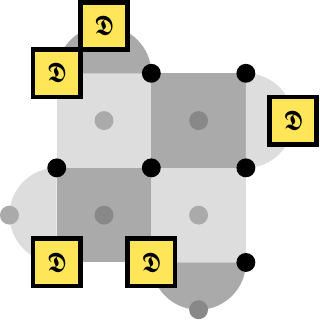}
    \includescv{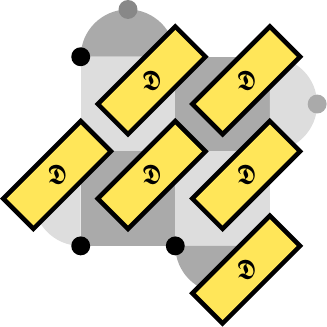}\hspace{1cm}
    \includescv{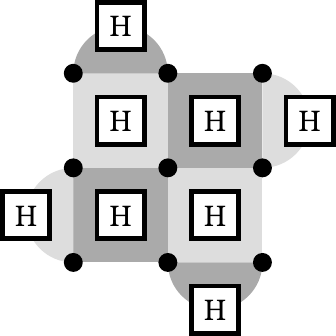}
    \includescv{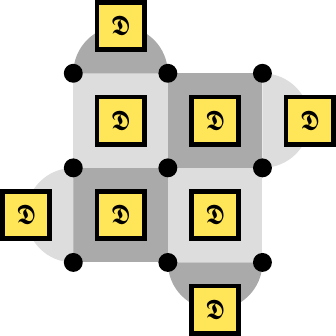}\\
    2b. Stabilizer qubit readouts and reset:\\[2mm]
    \includescv{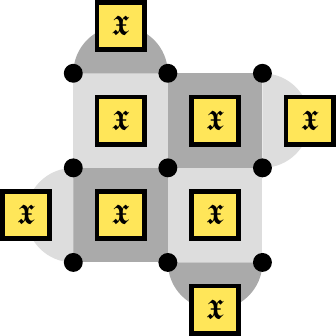}
    \includescv{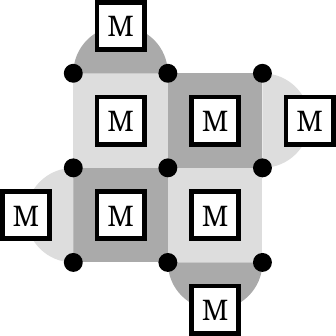}
    \includescv{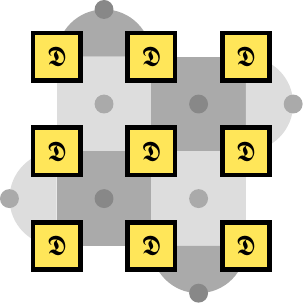}\hspace{1cm}
    \includescv{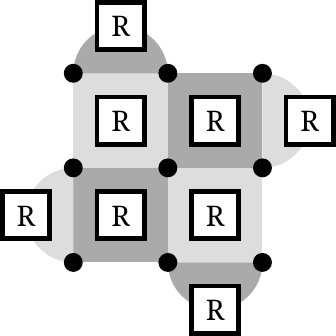}
    \includescv{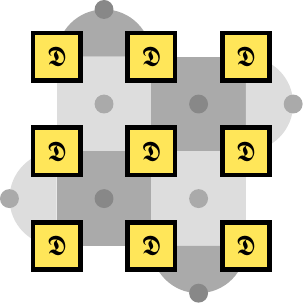}
    \includescv{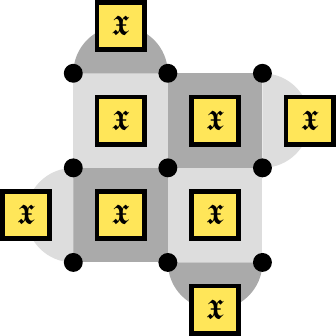}\\
    3. Data qubit readouts:\\[2mm]
    \includescv{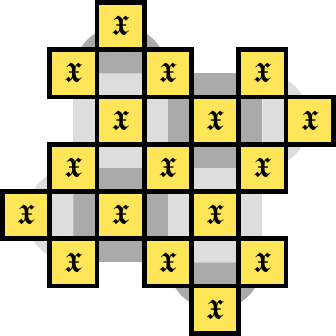}
    \includescv{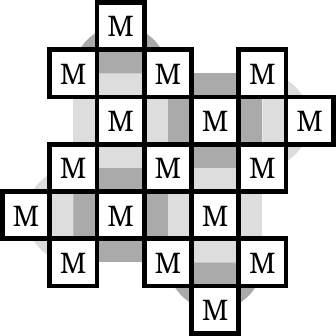}
    \caption{\textbf{Circuit depolarizing noise gate and error schema for a $3\times3$ XZZX rotated surface code in the $Z$ basis.} Labels are as in \cref{fig:sc}, and $\mathrm{CZ}$ gates are indicated by their standard circuit symbol.}
    \label{fig:sc-xzzx}
    
\end{figure}

As error syndromes cannot be read out directly with a single measurement, a stabilizer readout circuit has to be applied to deduce the $X$ and $Z$ stabilizers, as shown in \cref{fig:stabilizer-readouts}. The entire sequence of circuit depolarizing noise for a memory experiment of the surface code is shown in \cref{fig:sc}, and in \cref{fig:sc-xzzx} for an XZZX variant of the rotated surface code \cite{Bonilla_Ataides2021}.

\paragraph{SD6 noise.}
SD6 (standard depolarizing 6-step cycle) is a circuit depolarizing noise model \cite{Gidney2021}.

\paragraph{SI1000 noise.}
Introduced in \cite{Gidney2021}, SI1000 (superconducting-inspired $1000$ ns cycle) is a circuit depolarizing noise model comprising Pauli errors of non-uniform strengths, which approximate the relative noisiness of the various circuit processes in superconducting circuits: e.g.\ as measurements remain a major source of errors in superconductors, its relative weight with respect to the noise parameter $p$ is $5p$.
In contrast, single qubit gates and idling introduce only a very small amount of noise, hence their relative strength is $p/10$.

\begin{table}[t]
    \centering
    \begin{tabular}{rll}
        \toprule
        Operation & Noise Type & Strength \\
        \midrule
        Z (X) Measurement & Preceded by bitflip (phaseflip) channel & $5p$ \\
        Z (X) Reset & Resets to $\ket0$ ($\ket+$) and applies a bitflip (phaseflip) channel & $2p$ \\
        Resonator Idle & Applies 1q depolarizing channel to qubits not measured/reset & $2p$ \\
        2q Clifford & 2q depolarizing channel & $p$ \\
        1q Clifford & 1q depolarizing channel & $p/10$ \\
        Idle & 1q depolarizing channel & $p/10$ \\
        \bottomrule
    \end{tabular}
    \caption{Noisy operations in an SI1000 circuit depolarizing noise model.}
    \label{tab:si1000}
\end{table}

\FloatBarrier\subsubsection{Measurement noise}\label{sec:SI-IQ}
In each error-correction cycle, we projectively measure many of the qubits, allowing us to extract information about errors that have occurred. Consider measuring a self-adjoint operator $A$ with discrete eigenvalues $\lambda_i$. Let $P_i$ be the projector into the subspace with eigenvalue $\lambda_i$. Then the probability of observing $\lambda_i$ in a measurement is given by Born’s rule, $p_i = \textrm{Tr}(\rho P_i)$, and in that case the resulting state is projected into $P_i \rho P_i / p_i$. For the case of a single qubit measured in the computational basis $\{\ket0, \ket1\}$, $p_{\ket0} = \bra0\rho\ket0$ and $p_{\ket1} = \bra1\rho\ket1$. In an error-correction cycle, we only measure a subset of the qubits, but the projective nature of the measurement causes the entangled state of the data qubits to remain an eigenstate of all the stabilizer operators.

In practice, measurement is a challenging engineering problem: ordinarily, we want qubits isolated from their environment to allow coherent operations, but measurement necessitates interaction with the environment.  Additionally, we need to immediately re-use measured qubits for the next error-correction cycle, which requires either a ``non-demolition” measurement (where the qubit is faithfully projected into $\ket0$ or $\ket1$ corresponding to the measurement outcome) or other state preparation, such as unconditional reset to $\ket0$ following measurement. Unconditional reset also provides an opportunity to remove leakage from the system \cite{McEwen2021Leakage}. Measurement can cause other problems like state transitions and unwanted dephasing, which must be carefully avoided \cite{Sank2016Measurement,khezri2022measurementinduced}.

Implementations vary between physical platforms. For example, in standard dispersive measurement of superconducting qubits, a linear resonator (or series of resonators) serves as an intermediary between the qubit and the outside world \cite{blais2004cavity}. The measurement is implemented as a microwave scattering experiment to probe the resonator’s frequency, which depends on the qubit state due to coupling with non-linear Josephson elements \cite{wallraff2005approaching}. The scattered microwave pulse is amplified and digitized to determine its amplitude and phase, which encodes information about the qubit state \cite{jeffrey2014fast}.

The resulting amplitude and phase is traditionally represented in a two-dimensional space of in-phase (“I”) and quadrature (“Q”) amplitudes, (I, Q). Ideally, there is a distinct point in (I, Q) space associated with each qubit state ($\ket0$, $\ket1$, and potentially leakage states like $\ket2$). However, the measured signals are obfuscated with noise from sources like transmission loss, amplifiers, and electronics, manifesting as a spread or ``cloud’’ of points in (I, Q) space associated with each state. Additionally, qubits can exhibit unwanted transitions between states during the measurement, like decaying from $\ket1$ to $\ket0$, which would result in an average point between the $\ket0$ and $\ket1$ centers in (I, Q) space \cite{Sank2014}. Ordinarily, a measured (I, Q) value is classified to $\ket0$ or $\ket1$ (or in some cases $\ket2$), and this discrete measurement outcome is given to the decoder. However, a neural network decoder can use the raw (I, Q) value instead, giving it access to richer information without further preprocessing.

\begin{figure}[t]
    \centering
    \includegraphics[width=\textwidth]{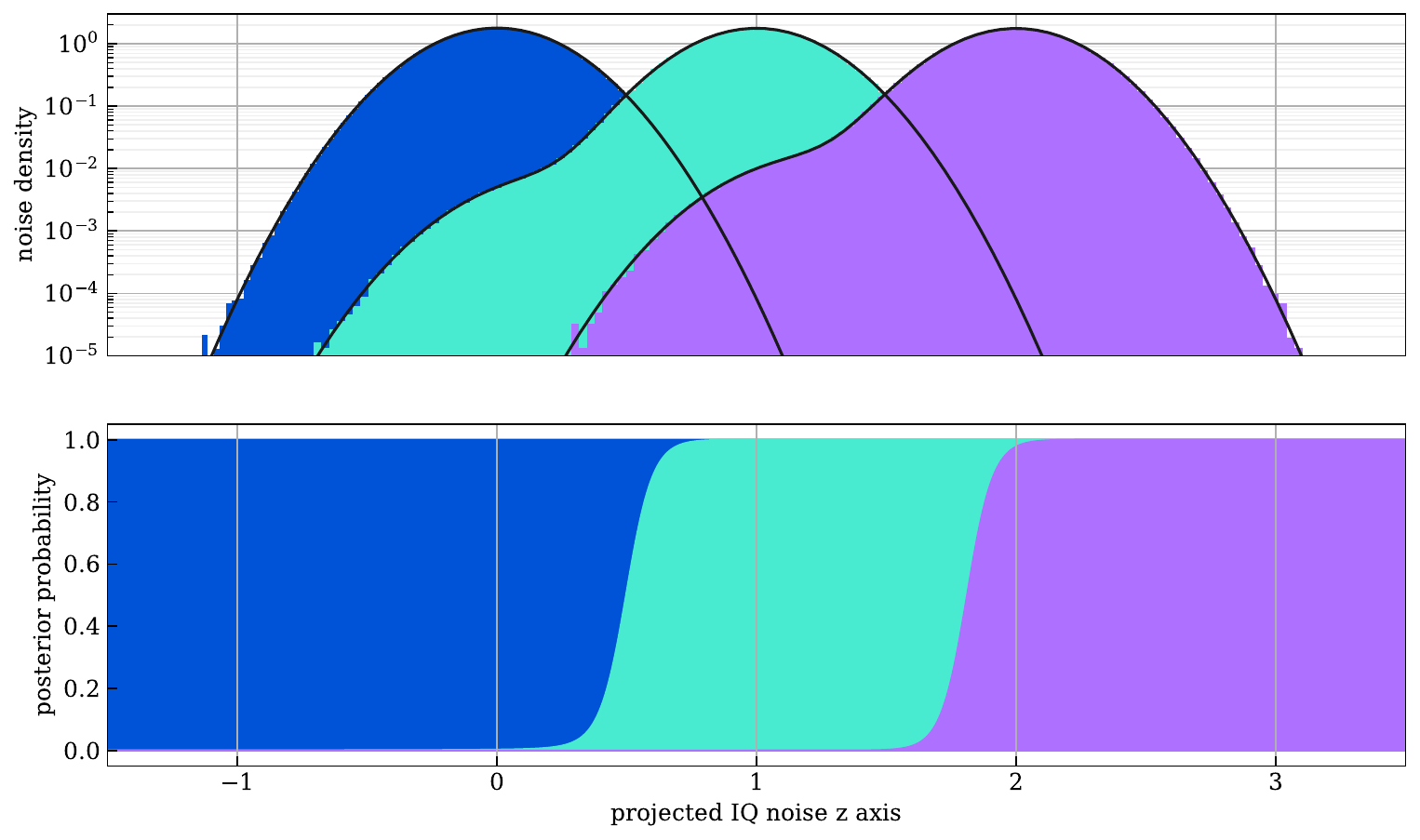}
    \caption{\textbf{Simplified I/Q noise with signal-to-noise ratio $\mathbf{\text{SNR}=10}$ and normalised measurement time $\mathbf{t=0.01}$}. Top plot: point spread functions when projected from its in-phase, quadrature, and time components onto a one-dimensional $z$ axis. Shown are the sampled point spread functions for $\ket0$ (blue), $\ket1$ (green), and a leaked higher-excited state $\ket2$ (violet). Bottom plot: posterior sampling probability for the three measurement states, for prior weights $w_2=0.5\%, w_0=w_1=49.75\%$.}
    \label{fig:iq-pdfs}
\end{figure}

To simulate this process, we run a simulation with noiseless measurements and then add noise after the fact. This can be as simple as discrete assignment error (for example, flip each measurement outcome with some probability), or we can emulate the richer (I, Q) signals. For our simulations, we consider a simplified one-dimensional space for our analog readout signal, with probability density functions $P_i$ for $\ket0$, $\ket1$, and $\ket2$ centered around $z=0, 1, 2$, respectively, shown in \cref{fig:iq-pdfs}. 

These probability distributions are parameterized by a signal-to-noise ratio, SNR, and a dimensionless measurement duration, $t = t_\textrm{meas}/T_1$, the ratio of the measurement duration to the qubit lifetime, $T_1$.
The distribution for $\ket0$, $P_0(z, \textrm{SNR})$, is simply a Gaussian distribution centered at $z=0$. For $\ket1$, we center at $z=1$ and add the effect of decay from $\ket1$ to $\ket0$.
For $\ket2$, we center at $z=2$ and assume the decay from $\ket2$ to $\ket1$ occurs roughly twice as frequently as the $\ket1$ to $\ket0$ case (in reality qubits can deviate from this based on details in the qubit $T_1$ spectra).
For simplicity, we do not include the second-order process of decaying from $\ket2$ to $\ket1$ to $\ket0$, though that does happen experimentally.

In this simplified single-parameter picture, we can thus write
\begin{align*}
  P_0(z, \text{SNR}) &= \sqrt{\frac{\text{SNR}}{\pi}}\exp \left(-\text{SNR}\cdot z^2 \right) \\
  P_1(z, \text{SNR}, t) &=\frac{t}{2} \exp \left( -t \left( z - \frac{t}{4 \, \text{SNR}} \right) \right) \\
  & \times \left[
    \text{Erf}
      \left(
        \sqrt{\text{SNR}} \left( z - \frac{t}{2 \, \text{SNR}} \right)
      \right)
    + \text{Erf}
      \left(
        \sqrt{\text{SNR}} \left( 1 - z + \frac{t}{2 \, \text{SNR}} \right)
      \right)
  \right] \\
& + e^{-t} \sqrt{\frac{\text{SNR}}{\pi}}
  \exp \left(
    - \text{SNR} \left( z - 1 \right)^2
  \right) \\
  P_2(z, \text{SNR}, t) &= P_1(z-1, \text{SNR}, 2t).
\end{align*}
For each measurement outcome from the simulation of state $\ket i$, we sample an ``observed’’ value of $z$ according to the associated probability density function. This ``observed’’ $z$ can then be processed using a prior probability distribution and the known probability density functions to determine a posterior probability for each state. For example, we may have a prior distribution that leakage occurs with probability 0.01 and we split the remaining 0.99 evenly between $\ket0$ and $\ket1$. More generally, we express these posterior probabilities as

\begin{align*}
    \text{post}_1 :=&\ \prob\!\left( \ket 1 \, \middle| \, \lnot\ket 2 \right)\\
    =&\ \frac{\prob(\ket 1 \land \lnot \ket 2)}{\prob(\lnot\ket 2)} = \frac{\prob(\ket 1)}{\prob(\lnot\ket 2)} = \frac{P_1(z, \text{SNR})}{(\bar w_0 / \bar w_1) P_0(z, \text{SNR}) + P_1(z, \text{SNR}, t)} \\
\intertext{and}
    \text{post}_2 :=&\ \prob(\ket 2) = \frac{P_2(z, \text{SNR}, t)}{w_0 / w_2 P_0(z, \text{SNR}) + w_1 / w_2 P_1(z, \text{SNR}, t) + P_2(z, \text{SNR}, t)},
\end{align*}
where $w_0 + w_1 + w_2 = 1$ are the prior probabilities of the three measurement outcomes, and $\bar w_0 := w_0 / (w_0 + w_1)$, $\bar w_1 := w_1 / (w_0 + w_1)$ are the marginal prior probabilities, conditioned on not having seen leakage.

This means we provide the network with two inputs:
\paragraph{post\textsubscript1.} The posterior probability of observing a $\ket 1$ state, conditioned that the state was not leaked (i.e., not in $\ket 2$). Once thresholded, this is the traditional measurement output from which syndrome or data qubit measurements, and subsequent detection events, are derived.

Note that due to our ordering of the states $\ket0$, $\ket1$, and $\ket2$ along the $z$ axis (see \cref{fig:iq-pdfs}), if a state was leaked it is most likely attributed to $\ket 1$, which is a valid choice of mapping an observed leaked state to a $\ket0$ or $\ket1$ measurement outcome.
For matching-based decoders, this assignment is a valid choice of  assignment of a leaked state to the $\{ \ket0, \ket1 \}$ subspace, and as good as e.g.\ a random mapping.
Indeed, for a decoder unable to process leakage information, a leaked state is ``lost information'', and thus an assignment to $\ket1$ will create a detection event in $\approx 50\%$ of cases.

\paragraph{post\textsubscript2.}
This is the probability of having seen leakage.
Due to the usually very low prior probability of seeing leakage in first place (commonly $<1\%$), the posterior distributions are skewed against $\ket 2$, as can be seen in \cref{fig:iq-pdfs}: even though the psf for the state $\ket 2$ is centered around $z=2$, has the same width as the other two distributions, and additionally has a higher decay tail towards $z=1$ due to its twice-as-high normalized measurement time $t$, the prior weight shifts its posterior to only give a significant chance of interpreting a measurement outcome as $\ket 2$ at a $z$ value already very close to $z=2$.

\FloatBarrier\subsubsection{Soft measurement inputs vs.\ soft event inputs}\label{sec:SI-soft-events}
\newcommand\xor\oplus
\newcommand\Ber{\mathrm{Ber}}
For our model, we have found that directly providing stabilizer measurements as inputs instead of stabilizer detection events is beneficial (\cref{fig:SI-ablations}).
Traditionally we have binary stabilizer readouts $s_{i,n} \in \{ 0, 1 \}$, where $i$ indexes the stabilizer qubit in the surface code, and $n$ the error correction cycle.
A detection event is then derived simply as the \emph{change} of a stabilizer measurement across EC cycles, $d_{i,n} := s_{i,n} \xor s_{i,n-1}$, which itself is a binary variable $\in \{0, 1\}$. This is the quantity that is traditionally used by most decoders, e.g. MWPM.

The XOR operation used to compute the change in stabilizer measurements results in a $1:1$ correspondence of information encapsulated in the events input vs.\ the measurements input; indeed, given detection events $d_{i,n}$, we can---up to a possibly unknown initial measurement frame---obtain back the stabilizer measurements $s_{i,n} \equiv \sum_{m=0}^n d_{i,n} \pmod 2$, where $d_{i,0}$ is the first event frame.\footnote{The first event frame was derived by either XOR'ing with an assumed zero frame prior to the first measurement (e.g.\ for those stabilizers corresponding to the memory experiment basis, cf.\ \cref{fig:figure-2}B), or was set to zero to remove an initially random stabilizer frame that does not allow extraction of more information about a first detection event (e.g.\ for the stabilizers orthogonal to the memory experiment basis)}.

This bijection allows us to present a comparison of measurement and event inputs, as they both contain the same amount of information for the decoder (possibly up to the first frame, as aforementioned; however for Pauli noise we take the initial off-basis stabilizers and XOR them onto the stabilizers anyhow, so that this assumed initial random frame is precisely zero as well).

If the measurements are transformed into posterior probabilities for each stabilizer measurement, we can assume that each such posterior
\begin{equation}
  p_{i,n} := \textbf{post}_1(i,n) = \prob\!\left(\ket 1_{i,n}\, \middle|\, \lnot\ket2_{i,n}\right)
\end{equation}
parameterizes a Bernoulli random variable $M_{i,n} \sim \Ber(p_{i,n})$.
As those are also boolean-valued random quantities, we can then transform pairs of measurement variables into corresponding ``detection events'', $E_{i,n} := M_{i,n} \xor M_{i,n-1}$, completely analogous to the binary measurement case. This means that
\[
  E_{i,n} \sim \Ber(q_{i,n})\quad\text{where}\quad
  q_{i,n} := p_{i,n} \big(1-p_{i,n-1}\big) + \big(1-p_{i,n}\big) p_{i,n-1}
\]
is parameterized by the probability that exactly one of $M_{i,n}$ and $M_{i,n-1}$ is $1$.

Analogously to before, this ``soft XOR'' defines a linear recurrence on the detection events that can be integrated to obtain back the posterior measurement probabilities from the soft detection events:
\[
  p_{i,-1} = 0
  \quad\text{and}\quad
  p_{i,n} = \frac{q_{i,n} - p_{i,n-1}}{1 - 2p_{i,n-1}}
\]
It is clear from the above that the special case of complete uncertainty ($p_{i,n} = 1/2$ for some $n$) is non-invertible, as all information is lost in that case.

\newcommand\sxor{\mathrm{SoftXOR}}
By induction one can also show that for a series of measurements (e.g.\ along an edge of data qubits in the surface code grid), thresholding the measurements against $1/2$ and then XOR'ing the set is equivalent to performing an iterative ``soft XOR'', and then thresholding.
To show this, let us simplify notation and drop the multiindex; our soft measurement probabilities are $p_1,\ldots,p_n$ such that all $p_i \neq 1/2$, and the corresponding thresholded Boolean values are $z_i := p_i > 1/2$. We denote with $\sxor(p_1,\ldots,p_n)$ the soft XOR defined above, and want to show $\sxor(p_1, \ldots, p_n) > 1/2 \Leftrightarrow z_1 \xor \ldots \xor z_n$.
The induction start is then immediate from the definition. Let us thus assume the hypothesis holds up to some value $m-1$. Then
\begin{align*}
    \sxor(p_1, \ldots, p_m) &= p_m \Big[  1 - \sxor(p_1, \ldots, p_{m-1}) \Big] + (1-p_m)\sxor(p_1, \ldots, p_{m-1}) \\
    &=: p_m\left(1-b\right) + (1-p_m)b \\
    &> 1/2\quad\Longleftrightarrow\quad b(1-2p_m) > \frac12 - p_m.
\end{align*}
Now if $p_m<1/2$, we have $1-2p_m > 0$ and thus $b>1/2$; otherwise if $p_m>1/2$ we have $b<1/2$. Thus
\[
    \sxor(p_1,\ldots,p_m) > 1/2\quad\Longleftrightarrow\quad \left(b < \frac12 \land p_m > \frac12\right) \lor \left( b>\frac12 \land p_m<\frac12\right).
\]
The two cases then translate to
\begin{align*}
    b < \frac12 \land p_m > \frac12\quad&\Longleftrightarrow\quad \lnot(z_1 \xor\ldots\xor z_{m-1}) \land z_m =: A \\
    b > \frac12 \land p_m < \frac12 \quad&\Longleftrightarrow\quad z_1\xor\ldots\xor z_{m-1} \land \lnot z_m := B,
\end{align*}
and $A \lor B = z_1\xor\ldots\xor z_m$.

\FloatBarrier\subsubsection{Pitfalls for training on soft information}\label{sec:pitfalls}
A crucial safeguard in all machine learning models is to never leak the label (i.e.\ the value to be predicted) into the input of the model.
For a distance-$d$ rotated-surface-code experiment with binary stabilizer labels, there exist exactly $d^2 - 1$ bits of information that are extracted at each error correction cycle; and it is impossible to deduce, from these measurements alone, the logical state of the qubit in the experiment basis (that information is contained in the remaining dimension $\cong \mathbb C^2$).

Naturally, this also holds true in the final cycle of a memory experiment, when we measure all data qubits in the experiment's basis---$d^2$ bits---and compute the same-basis stabilizer measurements from them---$(d^2 - 1) / 2$ many for an odd-distance surface code patch. As those same-basis stabilizers are a strict subset of the full $d^2 - 1$ stabilizers derived in previous cycles, the same argument applies: no information about the logical state of the surface code qubit can be leaked, as all stabilizers commute with the logical operators of the code.

However, when reading $d^2$ data qubits with soft information, and then re-computing the stabilizers from them via SoftXOR, there is a map of $d^2$ floating point values to $d^2-1$ floating point values. We found that this gives the model the ability to deduce the label from the inputs.
For this reason, we always threshold the data qubit measurements used for computing stabilizer measurements in the last memory experiment round (and the leakage data as well) , irrespective of whether we were providing the model with soft or hard inputs in previous rounds. In this way we ensure that there is precisely the same amount of information derived from the data qubits as in a standard ``non-soft readouts'' memory experiment, i.e.\ $d^2$ bits, which in turn are mapped to $(d^2 - 1) / 2$ stabilizer measurements in the experiment's basis. This makes it impossible for any decoder to discern the logical measurement label from its inputs.

\FloatBarrier\subsubsection{Pauli+ model for simulations with leakage}
\label{sec:SI-pauli_plus}
Realistic device noise was implemented in a manner similar to the Pauli+ model described in supplementary material of Ref.~\cite{milestone2}.  This model was updated by scaling noise strengths down from the near-threshold regime in that work to realize approximately $\Lambda = 4$ in surface code performance, where $\Lambda$ is the ratio of logical error rates between two surface codes of distance $d$ and $d+2$, as in the supplementary material of Ref.~\cite{Miao2022}.  Moreover, the simulator was modified from the ``stabilizer tableau'' representation to use the Pauli-frame representation, which yields indistinguishable results since transitions between stabilizer states are Pauli channels.

We briefly review the Pauli+ model here, though details are described in supplementary material of Ref.~\cite{milestone2}.  The Pauli+ model extends a Pauli frame simulator with leakage states.  These include transitions to and from leaked states, as well as error channels where a two-qubit gate applied to a qubit pair where one input is leaked is replaced by a noise channel on the non-leaked qubit.  For a noise channel in the simulation (in general, a Kraus channel), the qubit subspace is Pauli twirled, and transitions to/from leaked states are converted to stochastic transitions.  A Pauli-frame simulator is extended such that in addition to a Pauli operator at each qubit, leaked states can be tracked.  For example, the possible states of one qubit with leaked excited states could be $\{I, X, Y, Z, L2, L3\}$, where $L2$ and $L3$ are states of the Pauli frame simulator that represent quantum states $|2\rangle$ and $|3\rangle$.  

The noise strength is adjusted to what might be achievable in superconducting quantum processors in the medium term, several years from time of this study.  Each gate in the simulation is associated with a baseline amount of depolarizing noise.  The strength of depolarizing noise for each operation was informed by recent device characterization~\cite{milestone2} and an estimate of how noise might improve in future devices, as described in supplementary material of Ref.~\cite{Miao2022}.  In addition to this conventional Pauli-channel noise, there is a model for coherent crosstalk that accounts for interactions between pairs of CZ gates; this model is Pauli twirled to produce Pauli channels that are correlated on groups of qubits up to size four, and the unitary calculated includes leakage levels~\cite{milestone2}.  Leakage is introduced in three ways.  There is a probability of leakage introduced by dephasing during the CZ gate, a ``heating rate'' of leakage that is a function of gate duration, and leakage terms that arise from the crosstalk unitary described above.  The leakage rates were adjusted from the values in supplementary material of Ref.~\cite{milestone2} such that CZ dephasing and crosstalk were reduced to 0.25 of the previous values (e.g. CZ dephasing was $2 \times 10^{-4}$ instead of $8 \times 10^{-4}$), but the heating rate was unchanged at $1/(700~\mu s)$.  When leakage is scaled in this work, it is these three rates that are scaled together. Leakage is removed from the system by multi-level reset gates applied after measurement, by data-qubit leakage-removal~\cite{Miao2022} applied to code qubits every syndrome cycle, and by a passive decay rate that is proportional to $1/T_1$.

\subsection{Metrics}

\FloatBarrier\subsubsection{Logical error per round}\label{sec:SI-fit}

\begin{figure}[t!]
    \centering
    \includegraphics[width=\textwidth]{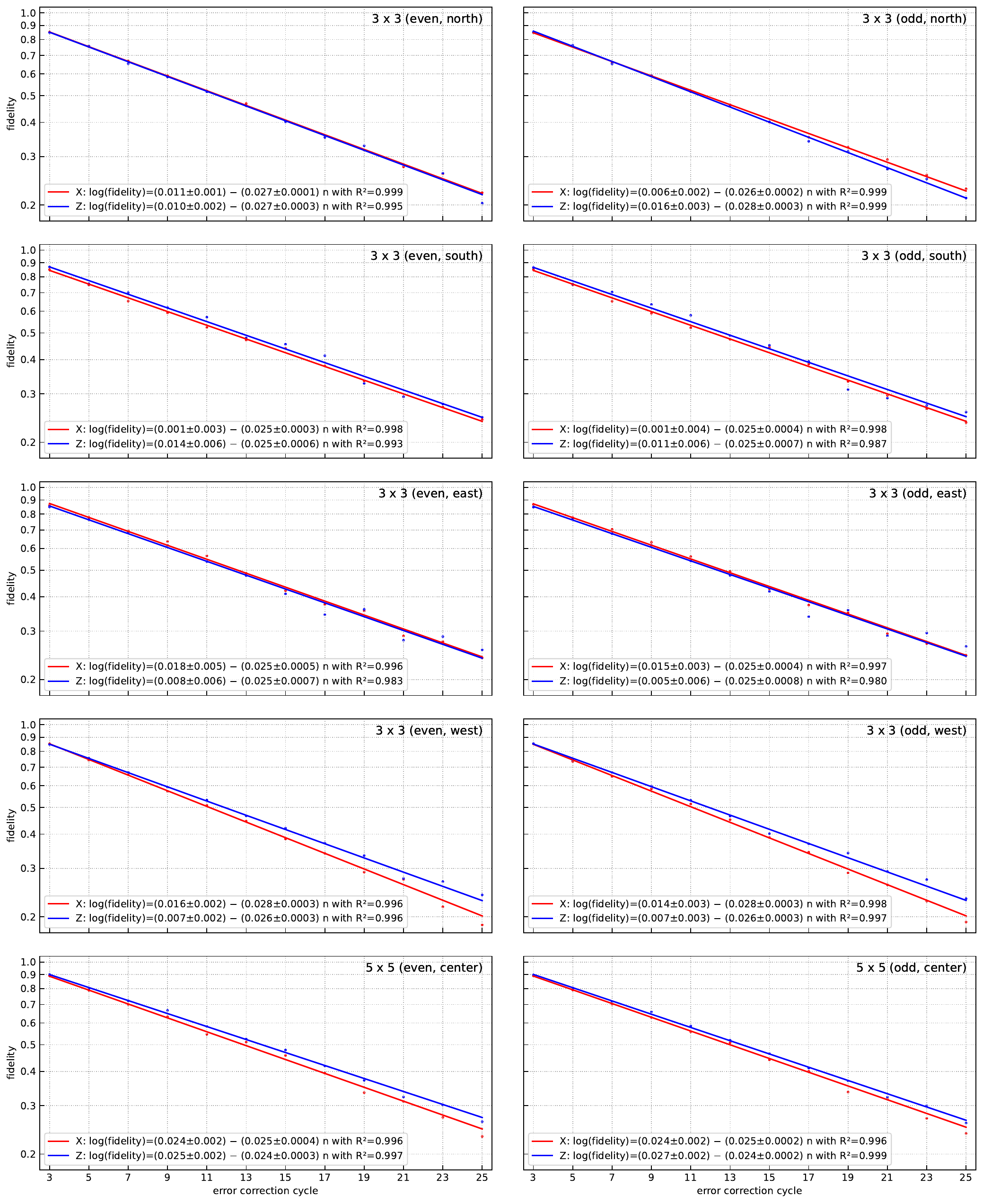}
    \caption{\textbf{Individual fits of logical error per round for $\mathbf{3\times3}$ and $\mathbf{5\times5}$ memory experiments.}}
    \label{fig:exploded-dem-fits}
\end{figure}

If $E(n)$ denotes the decoder error rate (computed as the erroneous fraction of  logical error predictions) at stabilizer measurement cycle $n$, we can make an ansatz for its functional dependence on $n$ via
\begin{equation}\label{eq:error-from-ler}
  E(n) = \frac12\left( 1 - (1-2\epsilon)^n\right),
\end{equation}
as derived in \cite[supp.\ mat., eq.\ 3]{google2021exponential}. For \cref{eq:error-from-ler}, we can see that $E(0) = 0$ (i.e., we assume no error at cycle $n=0$), $E(1)=\epsilon$, and $E(n)$ approaches $1/2$ for larger $n$.
In this context, the quantity $\epsilon$ is called the \emph{logical error per round}; indeed, it describes the exponential decay of the decoder's fidelity $F(n) := 1 - 2E(n)$, via the formula
\begin{equation}
    F(n + 1) = (1 - 2 \epsilon) F(n) \quad\Longrightarrow\quad F(n) = (1 - 2 \epsilon)^n,
\end{equation}
where we used $F(0) = 1$.

How we obtain the logical error per round $\epsilon$ from the error rates depends on whether we consider results after different number of cycles or not. The two ways of deriving the metric are compatible, in the sense that performing a fit on a single round experiment (with the added constraint of setting $F(0) = 1$ explicitly) yields exactly the same logical error per round as inverting the error $E(n)$ directly via \cref{eq:ler}.

\subsubsection{Experiment at a fixed number of cycles} 

For an experiment at a fixed number of cycles (e.g.\ $n=25$) we simply invert \cref{eq:error-from-ler}, and obtain
\begin{equation}\label{eq:ler}
  \epsilon = \frac12\left( 1 - \sqrt[n]{1 - 2 E(n)} \right).
\end{equation}

\subsubsection{Experiment across multiple rounds} 

We determine $\epsilon$ via a linear fit of the log fidelity,
\begin{equation}
\log F(n) = \log F_0  + n \log (1 - 2 \epsilon).
\end{equation}
To assess the fit's quality, we use the goodness of fit $R^2$.
In addition, we expect $F_0$ to be close to $F(0) = 1$, so we consider significant departures of  $\log F_0$ from 0 to indicate a bad fit (See \cref{sec:SI-Termination}).
As shown in \cref{fig:exploded-dem-fits}, all our fits for the $3\times3$ and $5\times5$ memory experiments exhibit a very high $R^2\ge0.98$, and $F_0\ge 1$.

As done in \cite{milestone2}, we exclude the point $(1, E(1))$ from our fits due to a strong time boundary effect which yields a much stronger error suppression per round at the first error correction cycle.

\FloatBarrier\subsection{Note on statistics}

\begin{figure}[t]
    \centering
    \begin{subfigure}[b]{\textwidth}
    \caption{}
    \includegraphics[width=\textwidth]{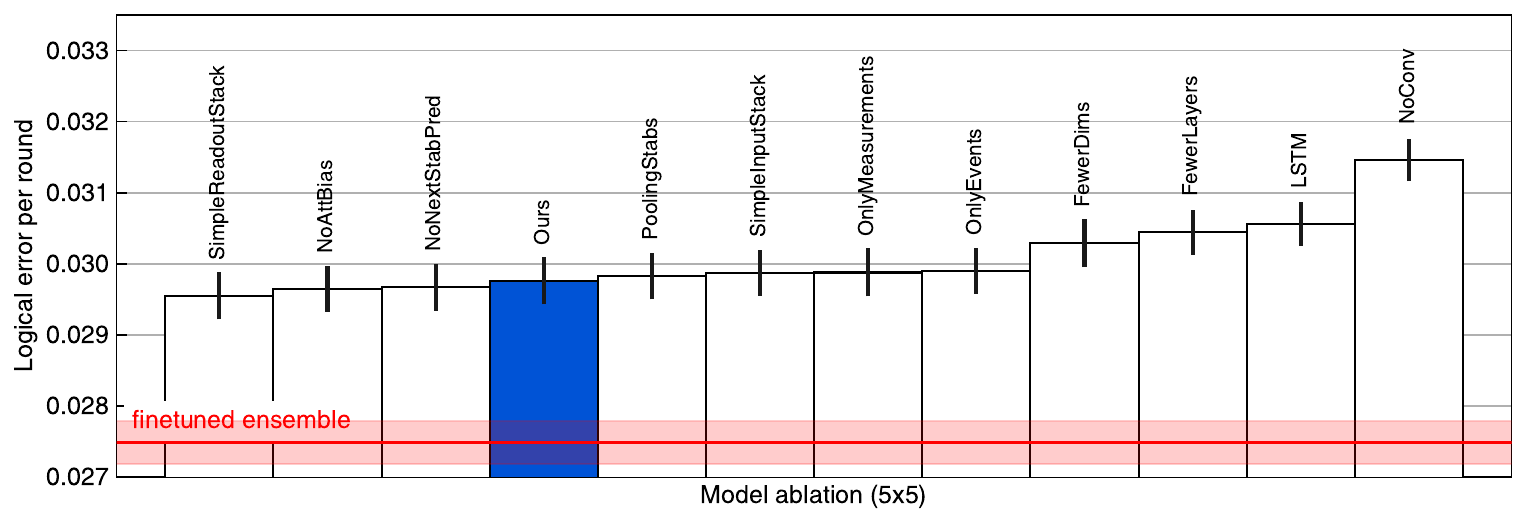}
    \end{subfigure}
    \begin{subfigure}[b]{\textwidth}
    \caption{}
    \includegraphics[width=\textwidth]{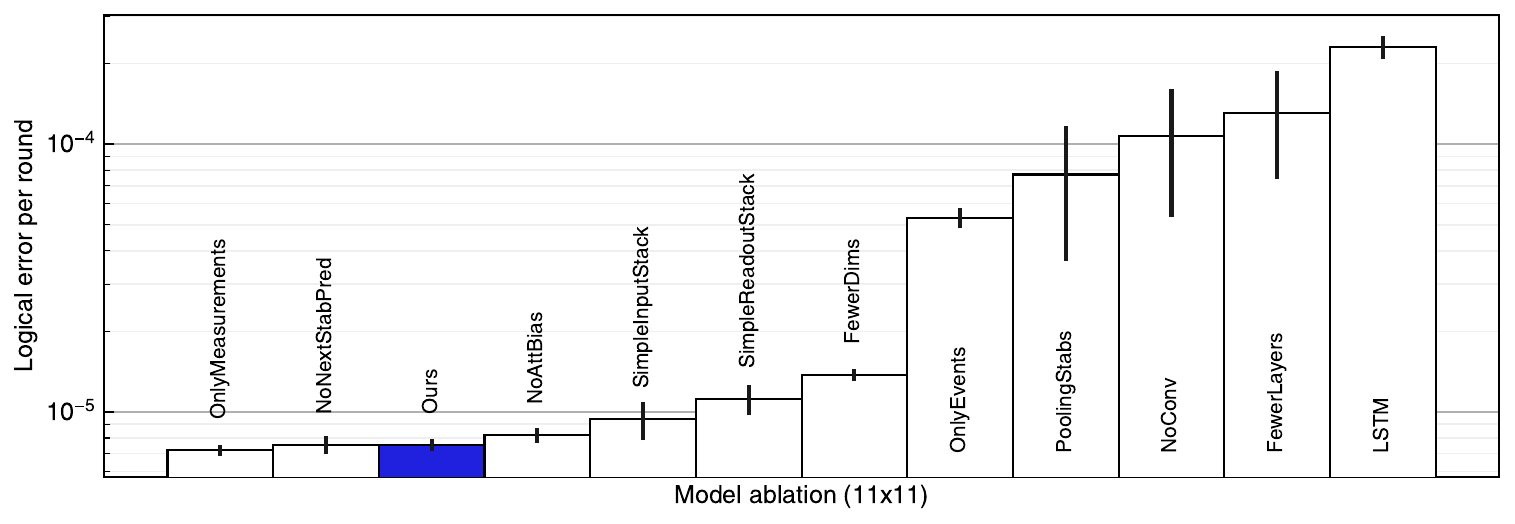}
    \end{subfigure}
        \caption{\textbf{Ablations: The effect of removing or simplifying decoder architecture elements.} Decoder performance under ablations (white bars) compared to the non-ablated decoder (blue bar), ordered by mean LER. (\textbf{A}) For $5\times5$ DEM training, averaged across bases ($X$ and $Z$) and the two cross-validation folds. The red horizontal line represents the performance of the fine-tuned ensemble. Error bars represent bootstrapping-estimated errors from individual fidelities. (\textbf{B}) For $11\times11$ Pauli+ training. Error bars represent estimated mean error estimated from 5 runs.} 
    \label{fig:SI-ablations}
\end{figure}

\paragraph{Combining different datasets.}
\renewcommand\dd{\mathrm{d}}
In the experimental datasets (e.g. \cref{fig:figure-2}B, \cref{fig:SI-ablations}A), we have 16 (for code distance 3) or 4 (for 5) distinct datasets per aggregated model performance---the combination of $X$\&$Z$ bases, even/odd subsets, and the different device regions. As we do not expect performance on the different datasets to be the same, we purposefully exclude the spread across datasets from our error estimation. Our error estimation is derived exclusively from the bootstrap estimation of the individual fidelity points. As in \cite{milestone2}, we propagate them by Gaussian error propagation; i.e., we sum two quantities $e_1 \pm \dd e_1$, $e_2 \pm \dd e_2$ via $e = (e_1 + e_2) \pm \sqrt{\dd e_1^2 + \dd e_2^2}$, discarding the spread between the quantities.
\paragraph{Combining different seeds.}
When we have multiple seeds but only one dataset (e.g.\ the ablation for Pauli+ data, \cref{fig:SI-ablations}B), we exclusively consider the spread across datasets, discarding the bootstrapped variance of the individual samples.

\subsection{Model details}
\label{sec:SI-Model}

Our ML decoder is a neural network designed to decode the surface code for a range of code-distances and for experiments of arbitrary duration.  Here we describe features of the architecture, particularly those that are adapted to the QEC problem.  \cref{sec:SI-ablations} shows that several of these become more important for the more complex decoding problem at larger code distances. 

The recurrent architecture design (Fig.~\ref{fig:architecture}B) reflects the time-invariant nature of the problem with the Syndrome Transformer maintaining the decoder state which represents the information from previous stabilizers relevant to deciding an experiment outcome. The decoder state has the potential to store some information for a window of arbitrary duration and is not limited to a fixed window. 

\FloatBarrier\subsubsection{Input representation}
\label{sec:SI-Inputs}

The network is provided with between one and four inputs per stabilizer. 
In the simplest case, analogous to MWPM, we provide binary {\em events} which are the temporal differences of binary {\em measurements} of the stabilizer state. Although these contain the same information, in practice we find that measurement inputs lead to better results than event inputs but provide both (see \cref{fig:SI-ablations}). When simulating I/Q noise, we provide probabilistic measurements and events as described in section~\ref{sec:SI-soft-events}.
For experiments with leakage, we also provide the leakage probability and the temporal-difference analog. 

A representation is built up for each input stabilizer as shown in Figure~\ref{fig:architecture}C, by summing linear projections of each of the input features. To allow the transformer to distinguish between the stabilizers, we also add a learned input embedding of the stabilizer index $i$.  Since the final-round stabilizers are not measured but computed from the data-qubits, we use a separate embedding for the final round, with a single learned embedding for all the undefined off-basis stabilizers and a separate final-round linear projection for the on-basis computed stabilizers. 
Each stabilizer representation is independently passed through a two-layer residual network to derive the stabilizer representation, $S_{ni}$, provided to the recurrent neural network. ($S'_{Ni}$ for the final stabilizers.) 

At each error-correction cycle, the stabilizer representations are added to the corresponding decoder state vectors and then scaled by a factor 0.707 to control the magnitude (Fig.~\ref{fig:architecture}D).

\FloatBarrier\subsubsection{Syndrome Transformer}
\label{sec:SI-syndrometransformer}

We designed the computation block  (Fig.~\ref{fig:architecture}D) to match the QEC task. At the heart of our RNN block architecture is the Syndrome Transformer, a self-attention architecture based on the Transformer~\cite{transformer} which has seen recent success in a variety of problems~\cite{brown2020language,jumper2021highly}. Transformers consist of multi-head attention followed by a gated dense block~\cite{shazeer2020glu}. Here we augment this architecture with an attention bias and 2-dimensional convolutions (Fig.~\ref{fig:architecture}E). 

The Syndrome Transformer updates the per-stabilizer decoder state representation by incorporating information from other stabilizers based upon their location. While previous decoders have exploited symmetries of the toric code  \cite{egorov2023end} and used convolutional neural networks for processing the surface code~\cite{gicev2021scalable} we argue that the boundary conditions of the surface code together with non-uniformity of real physical devices mean that a model that goes beyond rigid spatial-invariance can deliver benefits. Much of the information passing can be local, to handle local spatial correlations, and can be modelled with 2D convolutions. Longer-range interactions are partially supported by dilated convolutions but dense all-to-all attention enables the model to reason about all possible stabilizer pairs, for instance to reason about matching-like event pairings. 

For each transformer layer we apply three dilated convolutions after first scattering the stabilizer representations to their 2D spatial layout in a $(d+1) \times(d+1)$ grid, with a learned padding vector for locations where there is no 
stabilizer.

\subsubsection{Attention bias}
\label{sec:SI-attention-bias}

The attention allows information exchange between all pairs of stabilizers, but since the stabilizers have a predetermined spatial connectivity, we learn an attention bias~\cite{jumper2021highly} which modulates the attention between stabilizer $i$ and $j$, learned separately for each head in each transformer layer. 

The attention bias embeds fixed information about the layout and connectivity of the stabilizers constructing a learned embedding for each stabilizer pair $i,j$. This embedding is independent of the decoder state and at each transformer layer is projected down to a bias per head to be added to the conventional content-based attention logits. 

The attention bias embedding is a $(d^2-1) \times (d^2-1) \times 48$ tensor  constructed by adding learned embeddings of discrete features for each stabilizer pair $i,j$ based on their spatial layout.
\begin{enumerate}
    \item The spatial coordinates of stabilizer $i$ and stabilizer $j$.
    \item The signed spatial offset of stabilizer $i$ from stabilizer $j$.
    \item The Manhattan distance between $i$ and $j$.
    \item A bit to indicate if the bases for $i$ and $j$ are the same or not.
\end{enumerate}
These learned embeddings are then independently passed through a residual network to form the final embedding. While the embedding is learned, after training the attention bias embedding is constant and can be precomputed. 

To provide a simple further bias, at each round the current and previous stabilizers are used to compute indicator features for spatial and time-space event correlations \cite{milestone2} for each $i,j$ pair. At cycle $n$ these are the products:
\begin{enumerate}
    \item $\mathrm{event}_{ni} \times \mathrm{event}_{nj}$ (spatial)
    \item $\mathrm{event}_{ni} \times \mathrm{event}_{(n-1)j}$ (time-space)
    \item $\mathrm{event}_{(n-1)i} \times \mathrm{event}_{nj}$ (time-space)
    \item $\mathrm{event}_{(n-1)i} \times \mathrm{event}_{(n-1)j}$ (spatial),
\end{enumerate}
as well as the diagonals of these (7 features since two diagonals are identical).

These features are concatenated to the attention bias embedding and directly projected to the attention bias with a learned projection.

\begin{figure}[tp!]
    \centering
    \begin{subfigure}[b]{0.49\textwidth}
    \caption{}
    \includegraphics[width=\textwidth]{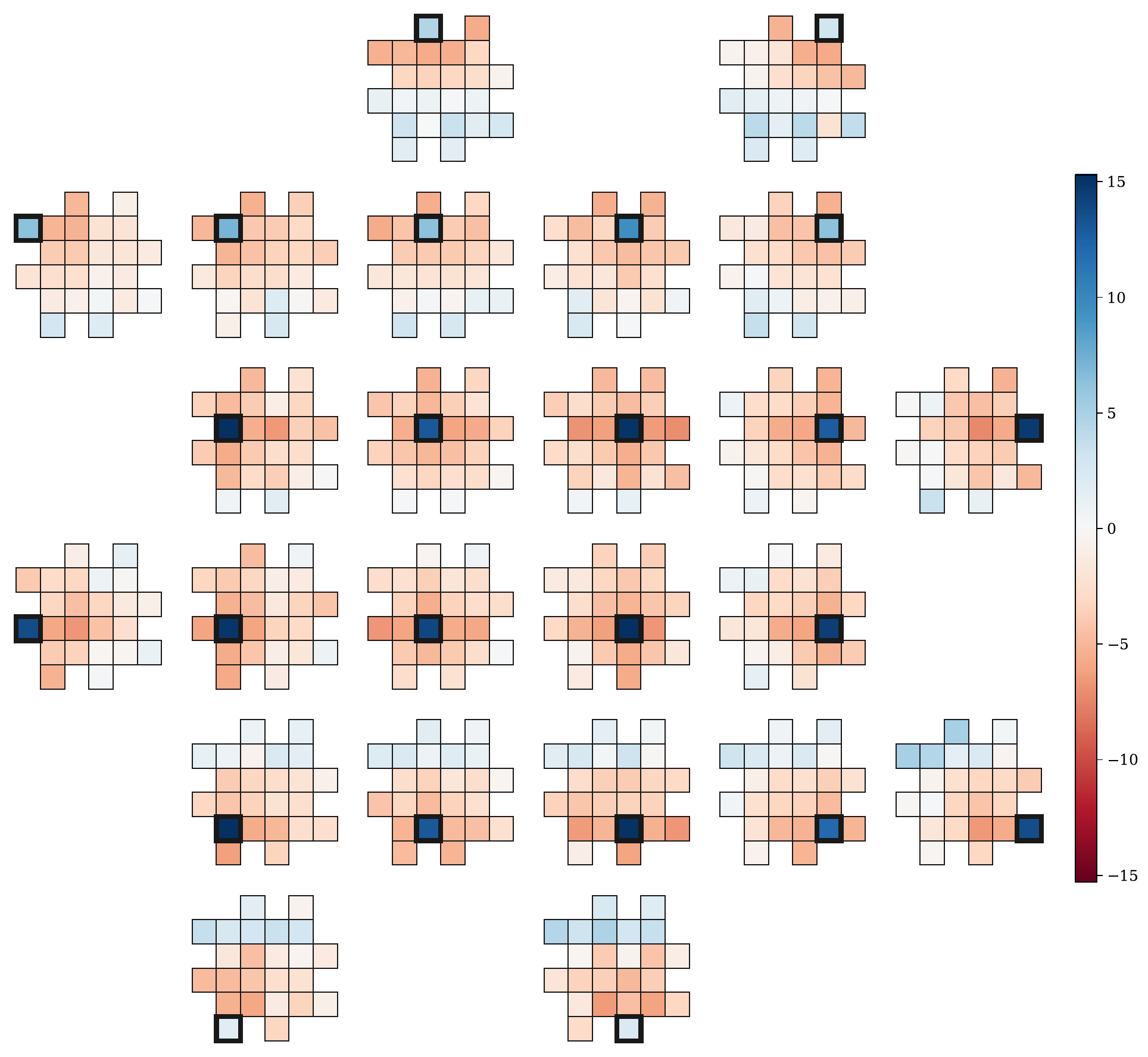}
    \end{subfigure}
    \hfill
    \begin{subfigure}[b]{0.49\textwidth}
    \caption{}
    \includegraphics[width=\textwidth]{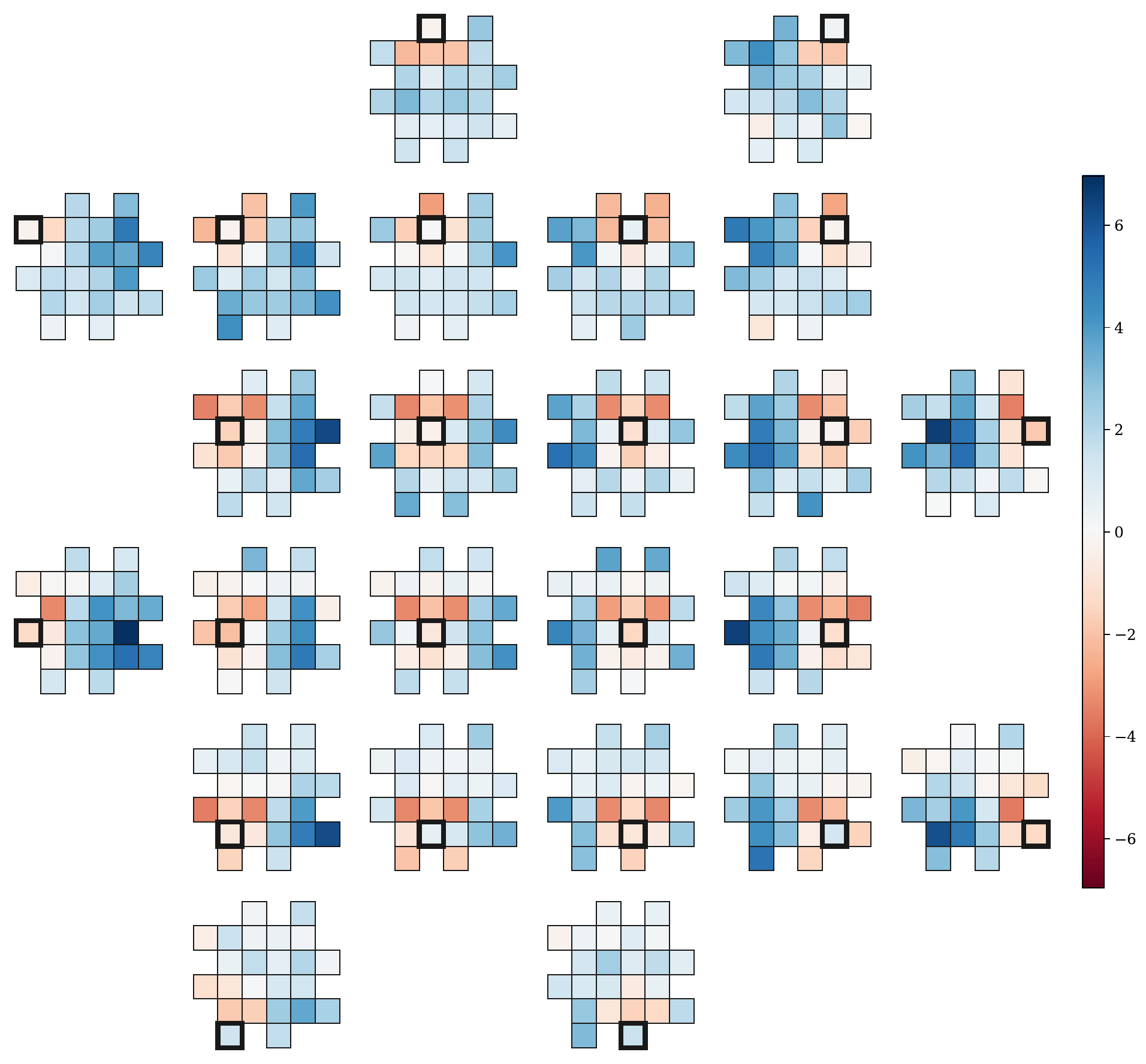}
    \end{subfigure}
    \newline
    \begin{subfigure}[b]{0.49\textwidth}
    \caption{}
    \includegraphics[width=\textwidth]{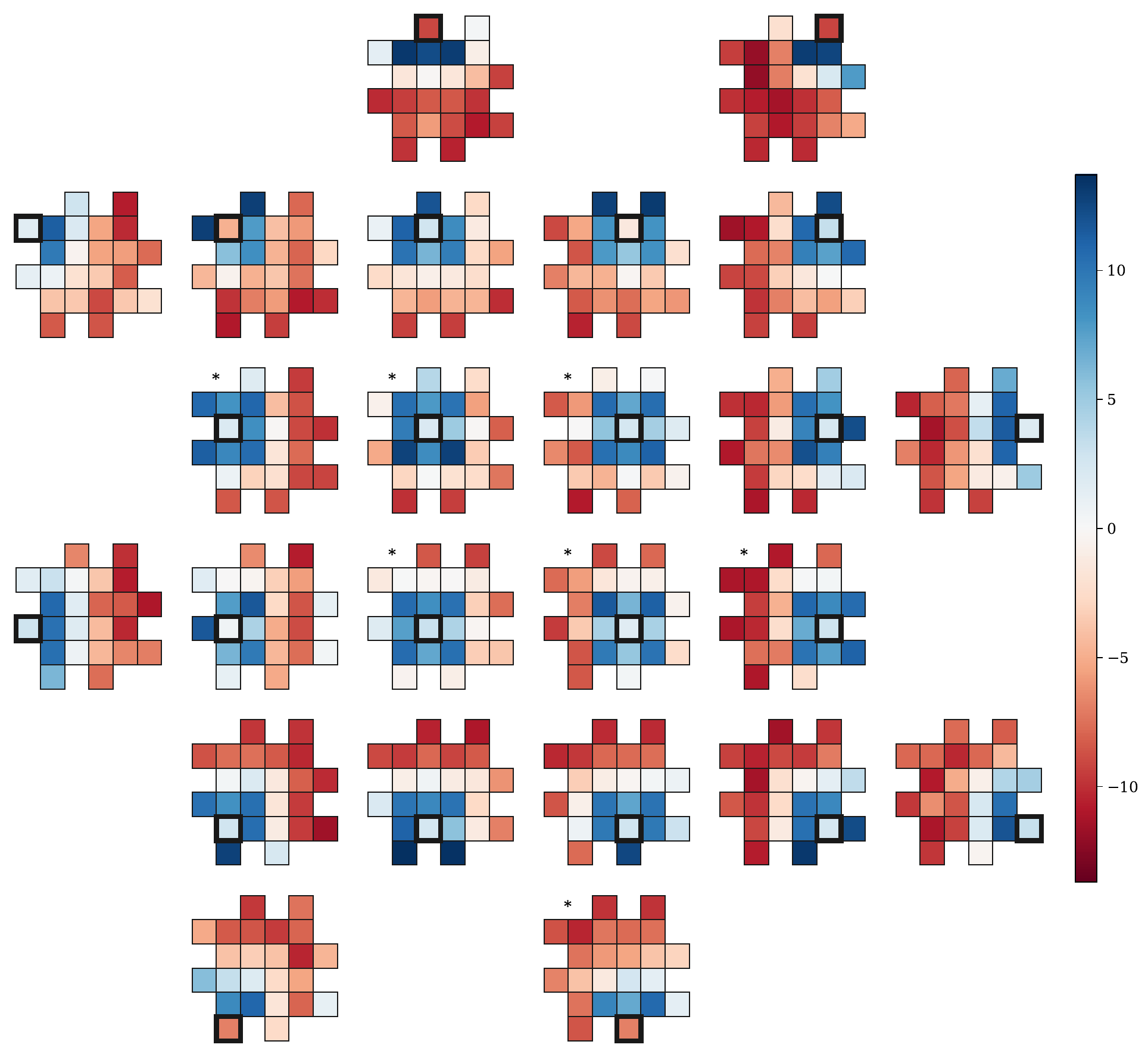}
    \end{subfigure}
    \hfill
    \begin{subfigure}[b]{0.49\textwidth}
    \caption{}
    \includegraphics[width=\textwidth]{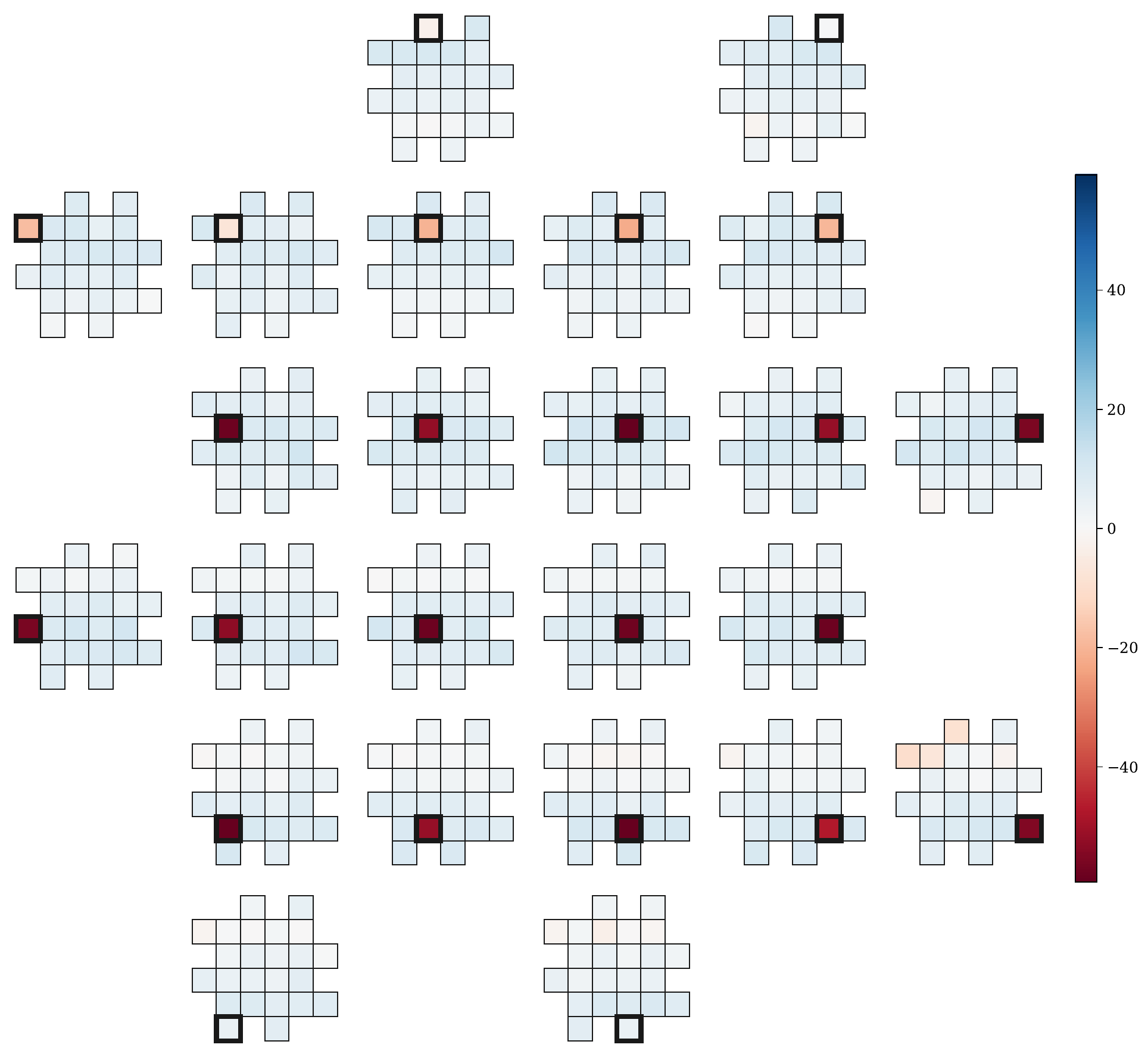}
    \end{subfigure}
    \caption{\textbf{Attention bias visualisation.} Attention bias logits of the four heads of the first Syndrome Transformer layer of our decoder model pre-trained on 5x5 DEM in the Z basis. We obtain the logits by combining the learned attention bias embedding with all-zero stabilizer values. The $24\times24$ attention logit matrices are each visualized as one grid per stabilizer, laid out according to the physical layout of the attending stabilizer qubits. Each grid shows the logits for the attention to each stabilizer, with self-attention highlighted with a black square.}
    \label{fig:attention-bias-vis}
\end{figure}

\paragraph{Attention bias visualisations.} 
In order to investigate whether the attention bias learns an interpretable representation, we visualise its logits in Figure \ref{fig:attention-bias-vis}. For each of the four attention heads for the first transformer layer of one ($5\times5$) DEM-trained model, we plot the attention logits for each stabilizer in a physical layout. It clearly shows that the different attention bias heads perform distinct functions. The first head modulates the attention towards the same stabilizer and stabilizers far away in the surface code. The second head discourages attention to immediate neighbouring stabilizers (even more-so to the diagonally-adjacent stabilizers in the same basis) while encouraging attention to non-neighbouring stabilizers. The third head instead does the opposite, strongly encouraging local attention while discouraging attention to stabilizers further away. Additionally, the third head seems to show patterns of higher attention biases for same-basis stabilizers than off-basis. This is visible in the attention maps marked with~$\ast$. Lastly, the final head predominantly discourages attention to the same stabilizer while being slightly encouraging towards attention for non-same stabilizers. We observed similar patterns of local and non-local attention bias for other models, however it did not always show as clearly and in some models the attention bias offered no obvious interpretation.

\FloatBarrier\subsubsection{Readout network}
After the RNN has processed the final stabilizers from cycle $N$ to create the $\text{decoder state}_N$ representation a readout network (Fig.~\ref{fig:architecture}F) processes the state to make a final prediction, again using the spatial distribution of the stabilizers. In the readout network, we first transform the per-stabilizer representation to a per-data-qubit representation by a scatter operation which arranges the decoder state representation according to the stabilizers' spatial layout and then applies a $2\times2$ convolution which combines information from the four stabilizer neighbours of each data qubit. We then apply a dimensionality reduction and pooling along rows or columns of the data qubits  perpendicular to the logical observable rows or columns of qubits (depending on the measurement basis), to arrive at a vector representation per equivalent logical observable.
This representation is then processed by a residual network to make the final label prediction.  We can compute one label for each of the $d$ rows or columns corresponding to equivalent choices of logical observables in the experiment basis (cf.\ \cref{fig:architecture}A) and average the loss for all of these if all the labels are available (as they are for the scaling experiment simulations). Only the first line is used at inference time.  The network was designed to pool \emph{along} logical observables to give a prediction per line, but in practice we found better results pooling \emph{perpendicular} to them.

\FloatBarrier\subsubsection{Auxiliary tasks}
\label{sec:SI-Auxiliary}
Often, training neural networks to make predictions other than those required for the main machine learning task, known as  {\em auxiliary tasks}, can lead to improved training or better performance on that main task. Here we ask the network to make a prediction of the next stabilizers, by a linear projection and logistic output unit from each stabilizer's representation.  
Figures~\ref{fig:SI-ablations} and \ref{fig:SI-ablations-2d}) shows that this auxiliary task seems to detract slightly from the network performance, but leads to slightly faster training.

\FloatBarrier\subsubsection{Efficient training for variable-duration experiments}
\label{sec:SI-variable-duration}
\begin{figure}[t!]
    \centering
    \includegraphics[width=0.8\textwidth]{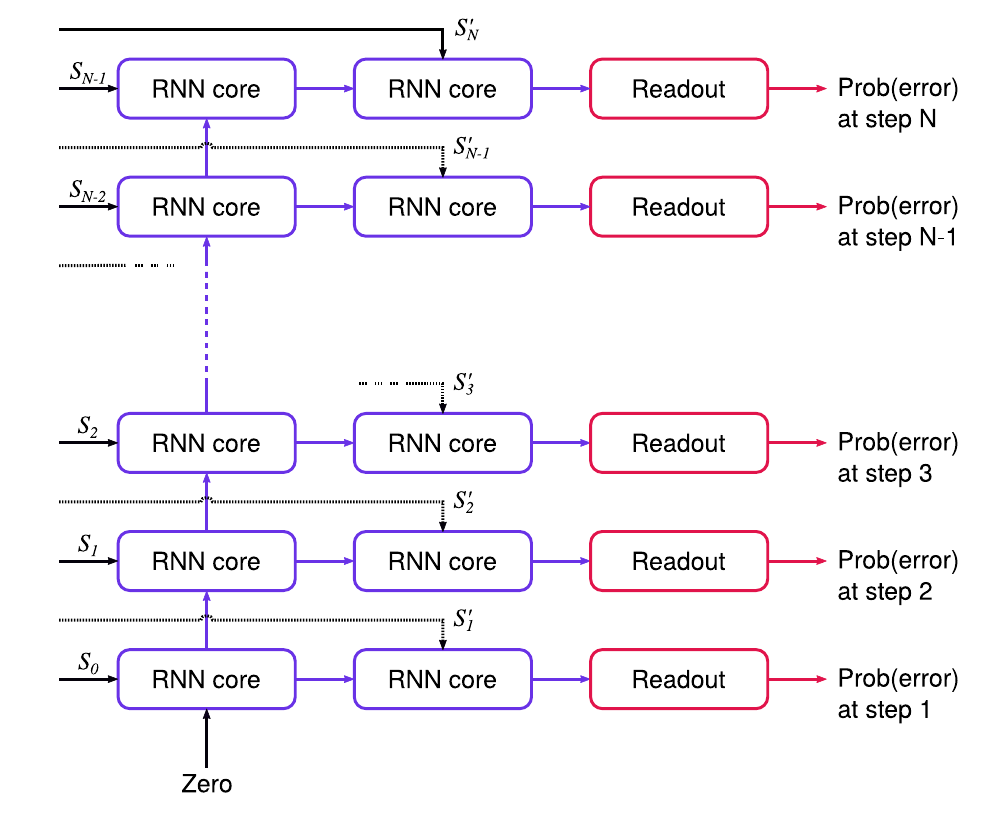}
    \caption{\textbf{Architecture of the network when predicting labels at every round.} $S_n$ are the stabilizer representations as explained in \cref{fig:architecture}B, where the primed quantities $S'_n$ indicate the embeddings are computed using different embedding parameters and based only on the stabilizers in the experiment basis computed from the final data qubit measurements (\cref{sec:SI-Inputs}).}
    \label{fig:architecture-with-intermediate-labels}
\end{figure}

Since the computation could be terminated at any cycle, the network could be asked to make a prediction at any cycle. This means that we might require a label for any cycle. By calling the simulator repeatedly with the same seed for increasing numbers of cycles, or through access to the internal simulator state, we are able to make a multi-round experiment with a shared bulk and a label for each round. Due to the special nature of the final stabilizers being computed from the measured data qubits there is a set of final stabilizers for round $n$ which are different from the bulk stabilizers for round $n$ of experiments which last longer. With such simulated data, when training, we can share computation for these experiments of different lengths as shown in Fig.~\ref{fig:architecture-with-intermediate-labels}. For $N$ cycles, $N$ labels can be trained with $2N$ applications of the embedding and RNN core ($N$ for the bulk and $N$ for the final stabilizers for each duration), and $N$ Readout computations. (vs. $N(N+1)/2$ applications of the embedding and RNN with $N$ Readout computations for training on $N$ separate examples of durations $1,\ldots,N$.)

\subsection{Implementation details}
Our ML decoder architecture is implemented and trained using the JAX, Haiku and JAXline~\cite{deepmind2020jax} machine-learning frameworks.

\FloatBarrier\subsection{Training details}
\label{sec:SI-Training}

\subsubsection{Sycamore data}\label{sec:SI-milestone2}

\paragraph{Cross-validation.}
For the data from Google's Sycamore memory experiment paper~\cite{milestone2}, we consider the two disjoint sets of odd- and even-indexed experiments (see \ref{sec:SI-Datasets-Experimental}) to perform 2-fold cross-validation. 

\paragraph{Pre-training.}
Because of the limited amount of experimental data captured in the Sycamore experiment, compared to the number of training examples required for an ML decoder, we pre-train on samples from a DEM fitted on one half of the data (25\ 000 samples per experiment length, see \ref{sec:SI-dem}). Examples are uniformly sampled from the lengths $\{1, 3, \ldots , 25\}$. Accuracy is measured using the actual samples from that same half, with early stopping using the parameters giving the best fitted LER up to 2 billion samples. 

We use a ``noise curriculum'' in which we show data with lower noise strength at the beginning of the training, continuously transitioning to higher noise strength during the training process. More precisely, we consider several replications of the DEM with error event probabilities scaled with factors $f = 0.5, 0.6 \ldots 1.0$. Each dataset is sampled with a probability proportional to
\begin{equation}
p_f(t) \propto 1 + w_{\text{c}} G(f_{\text{c}}(t), \sigma_{\text{c}}, f)
\end{equation}
where $G(\mu, \sigma; x)$ is the standard un-normalized Gaussian function
and $f_c(t)$ is the ``peak scale factor'':
\begin{equation}
f_c(t) = f_{c, min} + \frac{1 - f_{c, min}}{1 + \exp \left( -s_{\text{c}}(t / t_{\text{c}} - 1) \right) }
\end{equation}
that transitions from the minimum peak scale factor, $f_{c, min}$ to 1 at a number of training steps $t = t_{\text{c}}$. The values of the noise curriculum parameters can be found in Table \ref{tab:noise_curriculum}.

\begin{table}[t!]
    \centering
    \begin{tabular}{rll}
        \toprule
        Parameter name & Symbol & Value \\
        \midrule
        Curriculum weight & $w_{\text{c}}$ & 12 \\
        Weight std & $\sigma_{\text{c}}$ &  0.1 \\
        Initial weight peak  & $f_{\text{c, min}}$ & 0 \\
        Number of training steps before transition & $t_{\text{c}}$ & 50\,000 \\
        Slope of transition & $s_{\text{c}}$ & 1 \\
        \bottomrule
    \end{tabular}
    \caption{Noise curriculum parameters used in pre-training for the Sycamore experiment.}
    \label{tab:noise_curriculum}
\end{table}

\paragraph{Fine-tuning.}
We then fine-tune the model using the first 19\ 880 samples of the experimental dataset with early stopping keeping the parameters giving the best fitted LER on the remaining 5\ 120 samples (up to 30\,000 steps). The final model is evaluated on the other half of the data---the 25\,000 held out samples not used for training nor early stopping.

\subsubsection{SI1000 and I/Q noise}\label{sec:SI-IQ+SI1000}
We train and evaluate the model on SI1000 data generated using Stim with added I/Q noise. The hyperparameters used for these models are the same as before. We train models for 5 seeds each. Examples all have 25 error detection cycles. We train for a total of 2 billion samples and then use the parameters giving the best fitted LER on the dev set. SI1000 $p=0.2\%$. For I/Q noise we trained models with either hard or soft inputs for SNR $\in \{5, 10, 20\}$ and $t\in\{0, 0.003, 0.01, 0.03\}$. We use the same hyperparameters as for scaling investigation (see Table \ref{tab:hyperparameters} and \ref{tab:dilations}).

\subsubsection{Pauli+}

For the scaling investigation, we trained models using samples from the Pauli+ simulator (see \ref{sec:SI-pauli_plus}) using 5 different seeds each for 4 different leakage levels (corresponding to approximately 0.0\%, 0.033\%, 0.067\% and 0.1\% chance of leakage in the stabilizer measurements). We furthermore trained with either soft or hard inputs, with I/Q hyperparameters: SNR $=10$, $t=0.01$. To aid training, we used an auxiliary task of predicting intermediate logical labels as well as predicting the next stabilizers. After termination, the model parameters obtaining the highest development set LER were chosen. For each code distance in 3,5,7,9 and 11, we generated 990 million training samples from the Pauli+ simulator, which were then augmented with I/Q noise in post-processing. The models were trained for a total of 2 billion samples such that each individual Pauli+ training sample was seen roughly twice by the model, but with different I/Q noise added.

\begin{figure}[t!]
    \centering
    \includegraphics[width=0.9\textwidth]{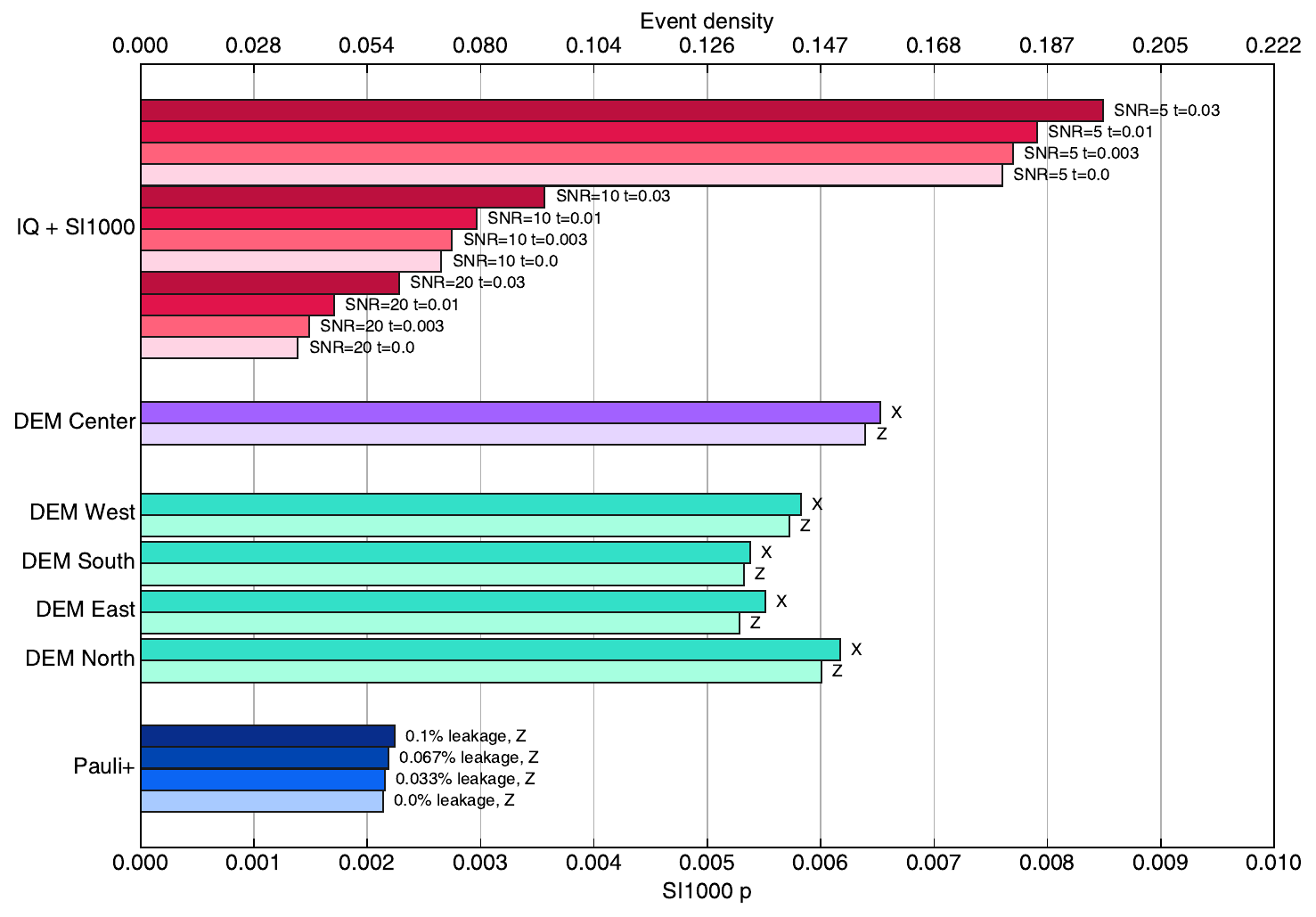}
    \caption{\textbf{Event densities for different datasets and the corresponding SI1000 p-value.} We indicate the event density of the different datasets in the top x-axis, with a non-linear scale. The detector error models are fitted to the Sycamore surface code experiment \cite{milestone2}, and are for an XZZX variant (with CZ gates for the stabilizer readout, cf.\ \cref{sec:SI-surfacecode,fig:stabilizer-readouts}) of the surface code; the same code variant is used for the Pauli+ data. The I/Q + SI1000 datasets use a traditional surface code variant (with CNOT gates). As we are never compiling between gatesets, there is no implied noise overhead; these are the final event densities observed when sampling from the respective datasets. Note that the I/Q + SI1000 dataset \emph{replaces} the binary channel readout noise with I/Q noise at the same value of $p=0.2\%$, which explains why for an SNR=20 (i.e., very low readout noise) the event density can be lower than the $p=0.2\%$ reference value. For datasets with soft I/Q noise, the plots above show the average \emph{soft} event density as explained in \cref{sec:SI-IQ}.}
    \label{fig:event-densities}
    
\end{figure}

\FloatBarrier\subsubsection{Loss}

We train the model using cross-entropy objectives with binary targets. For the scaling experiments, where we have a label for each logical observable for each experiment duration, all these losses are averaged.

As an auxiliary loss, we use next stabilizer prediction cross-entropy loss (Sec.~\ref{sec:SI-Auxiliary}) averaged across all cycles and all stabilizers and then down-weighted relative to the error prediction loss (Tab.~\ref{tab:hyperparameters}). 

\FloatBarrier\subsubsection{Loss minimization}

We minimize loss using stochastic gradient descent. We use the Lamb \cite{lamb} and Lion \cite{lion} optimizers for experimental and scaling datasets respectively. We use weight decay (L2 norm on non-bias parameters) everywhere, either relative to zero (for pre-training) or relative to pre-trained parameters (for fine-tuning, using a stronger weight decay).

The learning rate is piecewise constant after an initial linear warm-up of 10\,000 steps, with reductions by a factor 0.7 at specified numbers of steps. The batch size used for all experiments is 256, quadrupled once during training. 

\FloatBarrier\subsubsection{Termination}
\label{sec:SI-Termination}
Training is terminated after 2 billion examples. Model parameters are accumulated with an exponential moving average and regularly evaluated on development data to compute the LER (by fitting across cycles $3, 5, \dots 25$ for the Sycamore data, or by computing for 25 cycles for the scaling data). The set of parameters with the lowest development set LER is retained. With noisy fidelity estimates, particularly early on in training, we found that LER could be overestimated (see \ref{sec:SI-fit}), so we exclude parameter sets for which the fit has $R^2 \le 0.9$  or an intercept $\le \max ( -0.02, -\sigma)$ where $\sigma$ is the estimated standard deviation for the intercept of the fit line. 

\label{sec:10b}
We observed that training is not fully converged by 2 billion training examples for the larger code distances, resulting in a higher LER on the validation set than might be expected. To investigate the magnitude of this effect, we continued training the $11\times 11$ model ($0.1\%$ leakage, with soft inputs) up to 10 billion examples, resulting in a lower LER (the $*$ on Fig.~\ref{fig:figure-2}). Table~\ref{tab:lambda-table} shows the improvement in Lambda from the further training. Fig.~\ref{fig:training-examples}A shows the validation set LER vs number of training examples for these models. For the continued training, we removed the ``next stabilizer prediction'' loss which we found to be counterproductive (\cref{sec:SI-ablations}). 

\FloatBarrier\subsubsection{Hyperparameters}
\label{sec:SI-Hyperparameters}
For the Sycamore paper, we tuned the hyperparameters of the network by training models  
on the $5\times 5$ DEM, using samples from the same DEM as validation data for hyperparameter selection. We used the same hyperparameters for $3\times 3$ except for learning-rate ($\times \sqrt 2$) and using dilation 1 convolutions (See Table~\ref{tab:dilations}).

For the scaling investigation, the same base model was used, with some hyperparameter values further tuned to minimize validation set LER for the $11\times11$ code. Again the same model is used for all other code distances except for adjusting the learning rate and choosing the convolution dilations (Table~\ref{tab:dilations}).

\begin{table}[t]
    \centering
    \begin{tabular}{lp{6cm}|cc}
    \toprule
    Module & Hyperparameter & \multicolumn{2}{c}{Value}\\
    & & Sycamore & Scaling \\
    \midrule
Optimizer & Method & Lamb~\cite{lamb} & Lion ~\cite{lion}\\
& Weight decay & $10^{-5}$ & $10^{-7}$ \\
& Fine-tuning weight decay & 0.08 & -- \\
  & b2 & \multicolumn{2}{c}{0.95}\\
  & Initial batch size & \multicolumn{2}{c}{256}\\
  & Final batch size & \multicolumn{2}{c}{1024}\\
  & Batch size change step & $4\times10^6$ & $8\times10^5$ \\
  & Learning rate decay factor &  \multicolumn{2}{c}{0.7} \\
  & Learning rate decay steps $\times10^5$ & $\{0.8,2,4,10,20\}$ & $\{4,8,16\}$ \\
  & Next stabilizer prediction loss weight & 0.01 & 0.02 \\
  & Parameter exponential moving average constant & \multicolumn{2}{c}{0.0001} \\
    \midrule
    Feature embedding & ResNet layers & \multicolumn{2}{c}{2}\\
    \midrule
  Syndrome transformer & Layers & \multicolumn{2}{c}{3}\\
  & Dimensions per stabilizer & 320 & 256 \\
  & Heads &\multicolumn{2}{c}{4} \\ 
  & Key size  &\multicolumn{2}{c}{32} \\  
  & Convolution layers &\multicolumn{2}{c}{3} \\ 
  & Convolution dimensions & 160 & 128 \\ 
  & Dense block dimension widening & \multicolumn{2}{c}{5} \\
    \midrule
  Attention bias & Dimensions & \multicolumn{2}{c}{48} \\
  & Residual layers & \multicolumn{2}{c}{8}\\
  & Indicator features & \multicolumn{2}{c}{7} \\ 
    \midrule
  Readout ResNet & Layers & \multicolumn{2}{c}{16} \\
  & Dimensions & 64 & 48 \\
    \bottomrule
    \end{tabular}
    \caption{Hyperparameters of the network.}
    \label{tab:hyperparameters}
\end{table}
\begin{table}[t!]
    \centering
    \begin{tabular}{ll|cc}
    \toprule
    Code distance & Dilations & \multicolumn{2}{c}{Learning rate}\\
    & & Sycamore & Scaling \\
    \midrule
3 & 1, 1, 1 & $3.46\times 10^{-4}$ &  $1.3\times 10^{-4}$\\
5 & 1, 1, 2 & $2.45\times 10^{-4}$ &  $1.15\times 10^{-4}$\\
7 & 1, 2, 4 & -- &  $1\times 10^{-4}$ \\
9 & 1, 2, 4 & -- &  $7\times 10^{-5}$\\
11 & 1, 2, 4 & -- & $5\times 10^{-5}$\\
    \bottomrule
    \end{tabular}
    \caption{The dilations of the three $3\times3$ convolutions in each syndrome transformer layer and the experiment learning rates are determined by the code-distance of the experiment.}
    \label{tab:dilations}
\end{table}

\FloatBarrier\subsubsection{Parameters}
\label{sec:SI-parameters}

\begin{table}[t]
    \centering
    \begin{tabular}{cc}
    \toprule
    Code size & Parameters \\
    \midrule
$3\times3$ & 5\,444\,674 \\
         $5\times5$ & 5\,453\,826\\
         $7\times7$ & 5\,467\,074\\
         $9\times9$ & 5\,484\,418\\
         $11\times11$ & 5\,505\,858\\
    \bottomrule
    \end{tabular}
    \caption{The number of parameters for the model scaling over code distance. Since the core of the model is essentially the same, the number of parameters is constant apart from the index embedding parameters.}
    \label{tab:parameters}
\end{table}
Since the architecture used for all code distances is the same, the number of parameters (the weights of the neural network) is constant except for additional stabilizer index embedding parameters needed for larger code distances.  All the convolutions are $3\times 3$ albeit that the dilations are varied with the code distance.  Table~\ref{tab:parameters} shows the actual number of parameters in each model.

\subsection{Ensembling}
\label{sec:SI-Ensembling}
It is possible to combine multiple classifiers to obtain a more accurate prediction (ensembling, e.g. \cite{breiman2001random}). Since the models can be run independently in parallel, ensembling does not change the computation speed, but does require more computational resources.  We apply ensembles of multiple models for both the Sycamore and scaling experiments. 

We train multiple models with identical hyperparameters, but different random seeds leading to different parameter initializations and training on different sequences of examples. For the Sycamore data we use 20 seeds, and 5 for the scaling experiments. We average the logits from the different networks (computing a geometric mean of the predicted error probabilities). Since there is a range of accuracy between seeds it might be possible to achieve greater performance by ensembling only the models with the best validation set performance, but here we ensemble all the models trained.

\subsection{Decoding speed}
\label{sec:SI-Speed}

\begin{figure}[t!]
    \centering
    \includegraphics[width=0.6\textwidth]{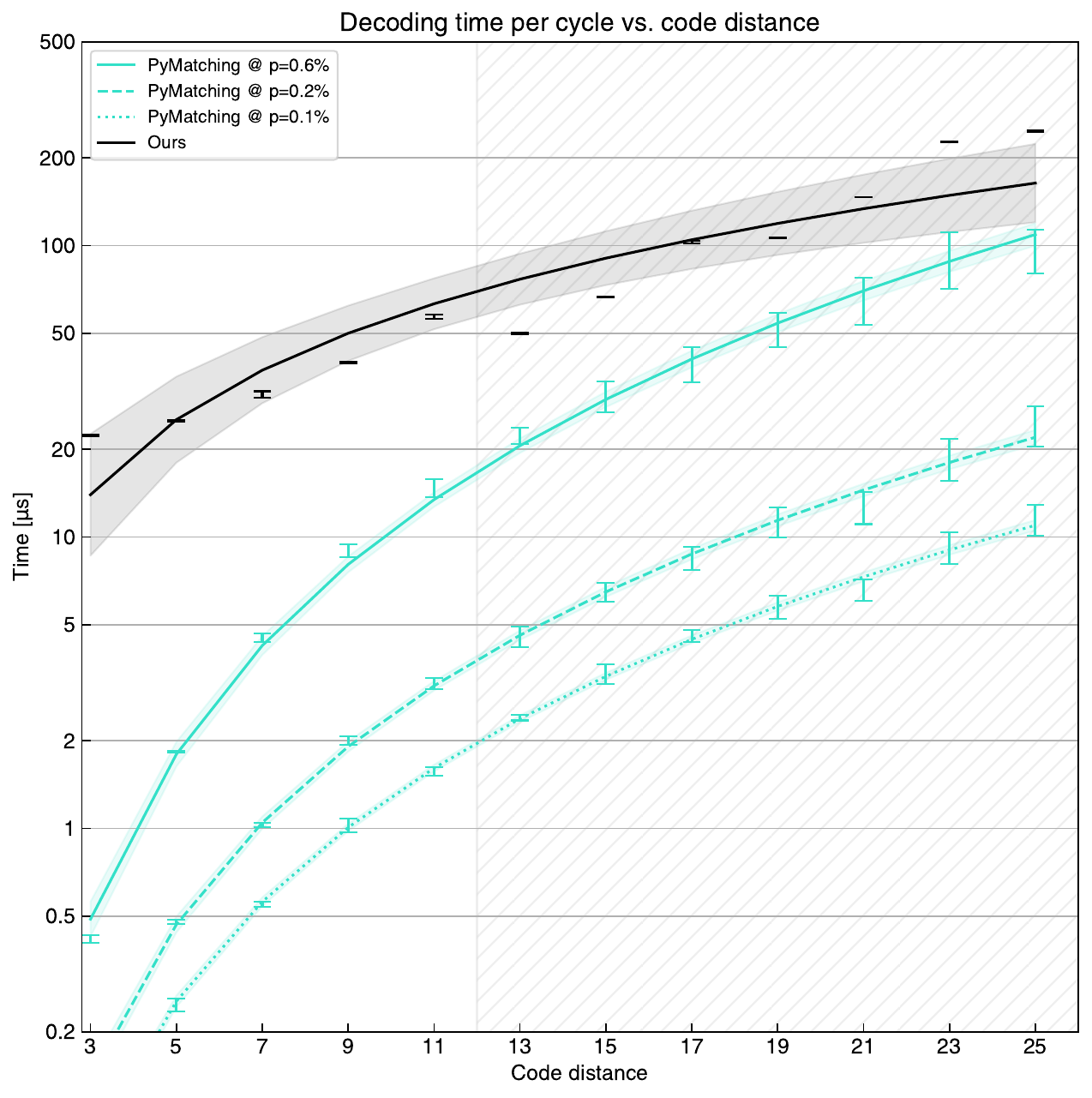}
    \caption{\textbf{Decoding time per QEC cycle vs.\ code distance.} The hatched region for $d>11$ indicates that while the ML decoder is the same for all code-distances, it has not been trained or shown to work beyond $d=11$. The line is a least squares fit to $a\times d^\mathrm{exponent}$, and the shaded region marks a 95\% CI interval. (a) Timing of uncorrelated matching (PyMatching). Times for PyMatching use a current CPU (Intel Xeon Platinum 8173M) and for the ML decoder use current TPU hardware with batch size 1.}
    \label{fig:timing-a}
\end{figure}

We designed and tuned the current model to deliver the best error suppression possible up to $11\times11$ codes, with the only speed consideration being to keep training time manageable for experiments. So far, we put no effort into optimizing the design for inference speed.
Figure~\ref{fig:timing-a} compares the throughput of our decoder with that of the open-source PyMatching library which itself was recently optimized with a $>100$-fold speed increase \cite{Higgott2023}.  This measures only the time required for computation per-round, ignoring the {\em latency}---the time required to deliver a final answer after receiving the final round's stabilizers. While our decoder is slower on current hardware than needed for a practical quantum computer, at moderate noise levels the scaling vs. code-distance is better than that for PyMatching.

We believe that our decoder can be sped up significantly using a number of well-known techniques, beyond rigorous efficient implementation.  First, the ablations (\cref{fig:SI-ablations}) show that at a given code distance, certain features of the network may be less effective and tuning the network to a particular code-distance may allow us to remove or scale-back slower parts of the network with little or no effect on error suppression. Secondly, techniques such as distillation~\cite{hinton2015distilling}, lower precision inference and weight pruning can be applied to achieve similar performance with less computation. Finally, hardware specific implementation, possibly using custom hardware such as ASICs or FPGAs can deliver further speed improvements. We note that many of the operations of the network can be carried out in parallel, and some computation (e.g. the embedding of stabilizers) can be pipelined---increasing throughput by increasing the latency.

\FloatBarrier\subsubsection{Decoding speed scaling considerations}

By design, the model runtime is independent of the physical noise level (and hence the error syndrome densities), whereas matching is slower the greater the noise. The fixed runtime of neural network decoders is considered to be a practical advantage \cite{varsamopoulos2019comparing}.
Moreover, while the computational cost of the ML decoder's components does scale with the code distance (the worst-case layer being the attention blocks, with a computational cost per cycle that scales as $\sim d^4$ in the code distance $d$), they are all intrinsically parallelizable. 
The impact of parallelization  on scaling depends on the hardware on which the decoder is executed. We found that on generally-available ML accelerators (GPUs, TPUs) and to the code sizes we investigate ($d=25$), a sub-quadratic scaling in $d$ is achievable (\cref{sec:SI-Speed}, \ref{fig:timing-a}), suggesting a crossover point with the computational cost of single-threaded MWPM at larger code distances and as hardware improves, even though that is a high bar to cross given the heavy optimization that has gone into accelerating matching-based algorithms themselves, often side-stepping having to execute the costly Blossom algorithm itself \cite{Higgott2023}.
While parallel processing has also been explored for MWPM \cite{Liyanage2023}, there is potential room for improvement on the ML side, e.g.\ switching to sparse attention would lower the overall parallel compute requirement to $\sim d^2$.
We also note that, in principle, it is possible to achieve unbounded throughput by decomposing the matching decoding problem in an embarrassingly parallel fashion \cite{skoric2022parallel, tan2022scalable}.
We expect similar ideas might be applied to an ML decoder.

\FloatBarrier\subsection{Postselection and calibration}

\begin{figure}[t]
    \centering
    \includegraphics[width=\textwidth]{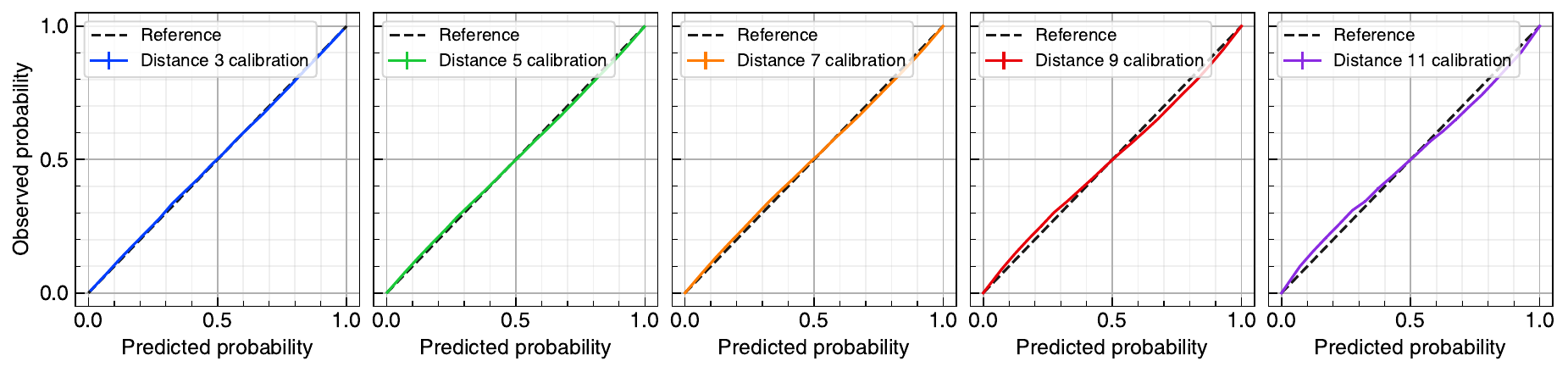}
    \caption{\textbf{Calibration of our ML decoder's outputs for code-distances 3--11.} For Pauli+ generated data with SNR $=10$, $t=0.01$, and $0.1\%$ stabilizer qubit leakage chance.}
    \label{fig:individual-calibration-plots}
\end{figure}

\begin{figure}[t]
    \centering
    \includegraphics[width=0.9\textwidth]{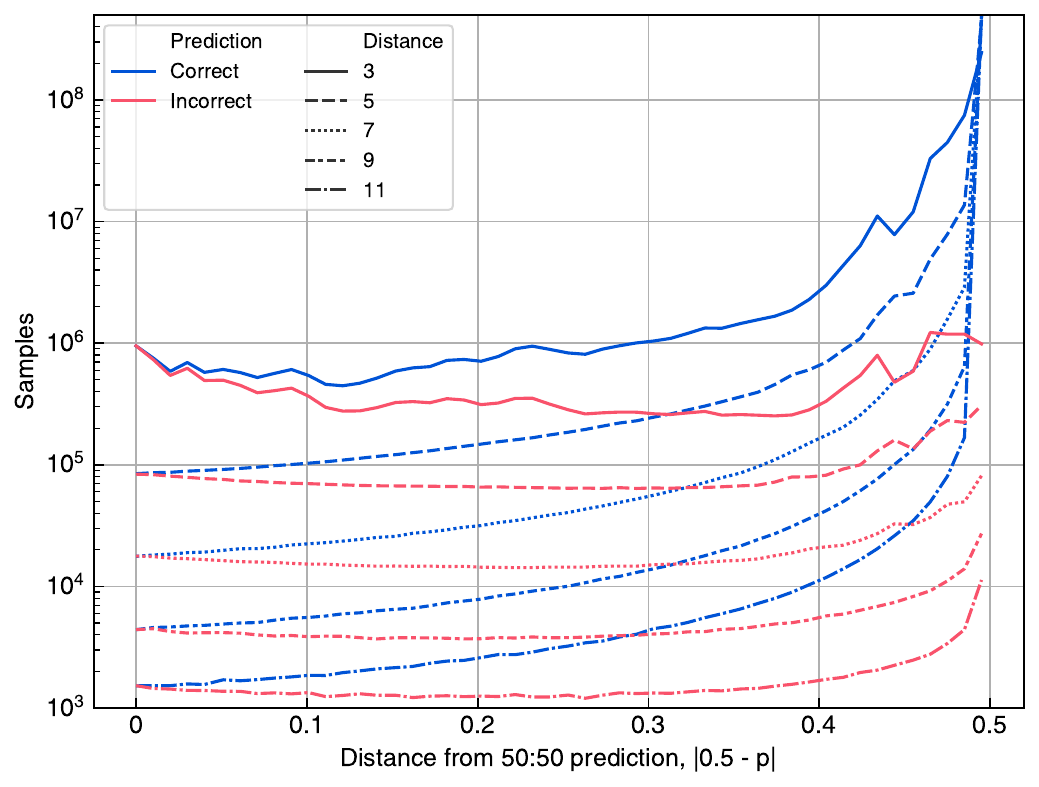}
    \caption{\textbf{Calibration histogram of predicted probabilities.} The predictions are grouped into correct (blue) and incorrect~(red) before building the histogram, and then binned into ``certainty'' bins depending on their distance from a $50:50$ prediction, i.e.\ by $|1/2 - p|$ for a predicted probability $p$. For all code distances, wrong predictions have a lower certainty. Correct predictions concentrate around the interval edges, i.e.\ at probabilities $0$ and $1$, resulting in a high certainty.}
    \label{fig:individual-calibration-bins}
\end{figure}

\FloatBarrier\section{Supplementary text}
\begin{table}[t!]
    \centering
    \begin{tabular}{r r r | l l }
    \toprule
        decoder & I/Q inputs & leakage inputs & $\Lambda_{3/5}$ from $d=3$ to $5$ &$\Lambda_{3/11}$ from $d=3$ to $11$ \\
    \midrule
     ML (10B) &   yes &   yes & --                        &  $4.28\pm0.02^*$ \\
           ML &   yes &   yes & $5.08\pm0.01$            &  $3.98\pm0.02$ \\
           ML &    no &   yes & $4.31\pm0.02$   &  $3.88\pm0.03$ \\
           ML &    no &    no & $4.23\pm0.02$            &  $3.77\pm0.03$ \\
    \midrule
   PyMatching &    no &    no & $ 2.89\pm0.01$ &  $2.99\pm0.01$ \\
    \midrule
    MWPM-Corr &   yes &    no & $4.38\pm0.02$ &  $4.33\pm0.04$ \\
    MWPM-Corr &    no &    no & $3.94\pm0.02$ &  $3.89\pm0.03$ \\
    \bottomrule
    \end{tabular}
    \caption{\textbf{Average error suppression factors $\mathbf\Lambda$ for Pauli+ experiment.} The error suppression factor is computed from the data in \cref{fig:figure-2}B, via $\Lambda_{3/11} = (\epsilon_3 / \epsilon_{11})^{1/4}$, for a logical error per round $\epsilon_3$ at code distance $3$, and $\epsilon_{11}$ at distance 11, respectively. $^*$ As the model is well-converged at distance 3 we compute $\Lambda_{3/11}$ for the 10B run (\cref{sec:10b}) with $\epsilon_{11}$ from the 10B run, vs.\ $\epsilon_3$ from the 2B run.}
    \label{tab:lambda-table}
\end{table}

\subsection{Further details of scaling experiments}

In \cref{tab:lambda-table}, we collect error suppression factors $\Lambda$ from distance 3 to 5, and 3 to 11, for various decoders and input modalities. The strongest error suppression through distance 11 is achievable with our ML decoder, at a LER of $5.37\pm0.01$.

\begin{figure}[t]
    \centering
    \includegraphics[width=0.5\textwidth]{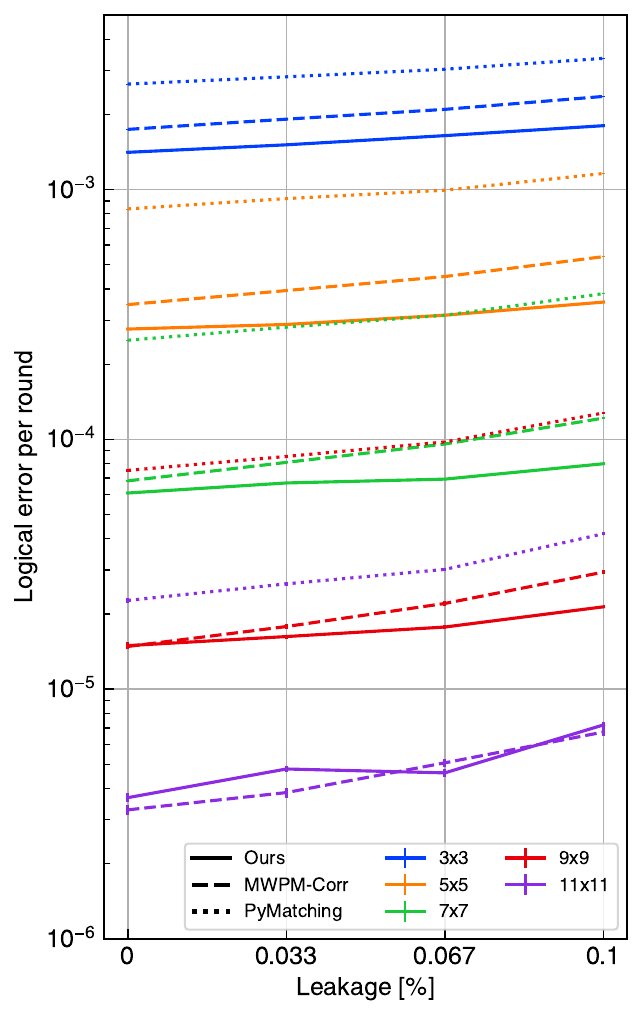}
    \caption{\textbf{Detailed view of logical error per round vs.\ leakage strength.} Evaluated in 25-cycle Pauli+ simulated experiments with SNR $=10$ and $t=0.01$, and for code distances 3 to 11.}
    \label{fig:ler-leakage-details}
\end{figure}

In \cref{fig:ler-leakage-details}, we show the effect of leakage on decoder performance, for our decoder, PyMatching, and MWPM-Corr.
As collated in \cref{fig:figure-2}c, our ML decoder's performance degrades significantly less with increased leakage inputs.

\begin{figure}[t]
    \centering
    \hspace*{-1mm}\mbox{
    \begin{subfigure}[b][][t]{0.19\textwidth}
    \caption{}  
    \hspace{-3mm}\includegraphics[trim={0 -1mm 0 0},clip,width=1.\textwidth]{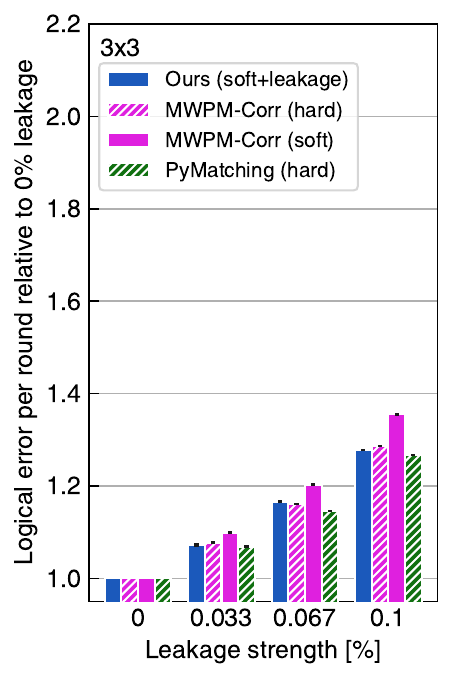}
    \end{subfigure}
    \hspace{0.cm}
    \begin{subfigure}[b][][t]{0.19\textwidth}
    \caption{}
    \hspace{-2mm}\includegraphics[trim={2mm 0 0 0},clip,width=1.\textwidth]{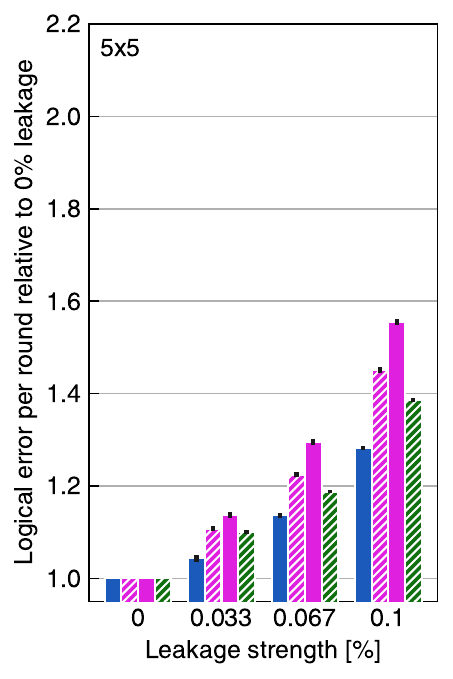}
    \end{subfigure}
    \hspace{0mm}
    \begin{subfigure}[b][][t]{0.19\textwidth}
    \caption{}
    \hspace{-2mm}\includegraphics[trim={2mm 0 0 0},clip,width=1.\textwidth]{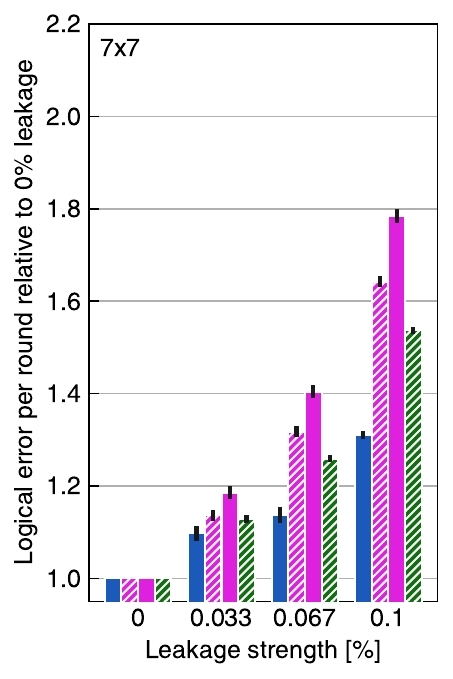}
    \end{subfigure}
        \hspace{0mm}
    \begin{subfigure}[b][][t]{0.19\textwidth}
    \caption{}
    \hspace{-2mm}\includegraphics[width=1.\textwidth]{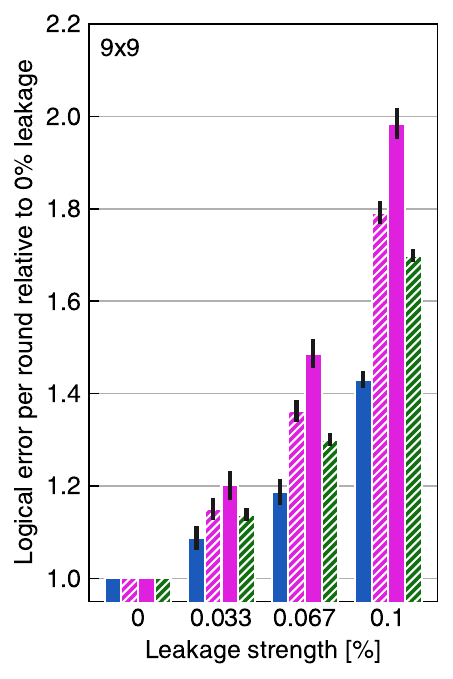}
    \end{subfigure}
    \hspace{0mm}
    \begin{subfigure}[b][][t]{0.19\textwidth}
    \caption{}
    \hspace{-2mm}\includegraphics[width=1.\textwidth]{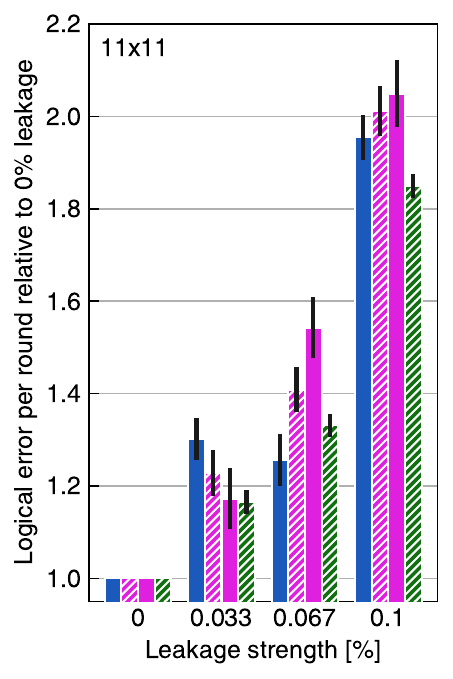}
    \end{subfigure}
    }
    \caption{The relative impact of leakage on the performance of different decoders and input modalities at code distances from $3$ to $11$.}
    \label{fig:SI-leakage-degradation}
\end{figure}

\FloatBarrier\subsection{Time Scalability}
\label{sec:SI-time-scalability}

\begin{figure}[t!]
    \centering
    \includegraphics[width=\textwidth]{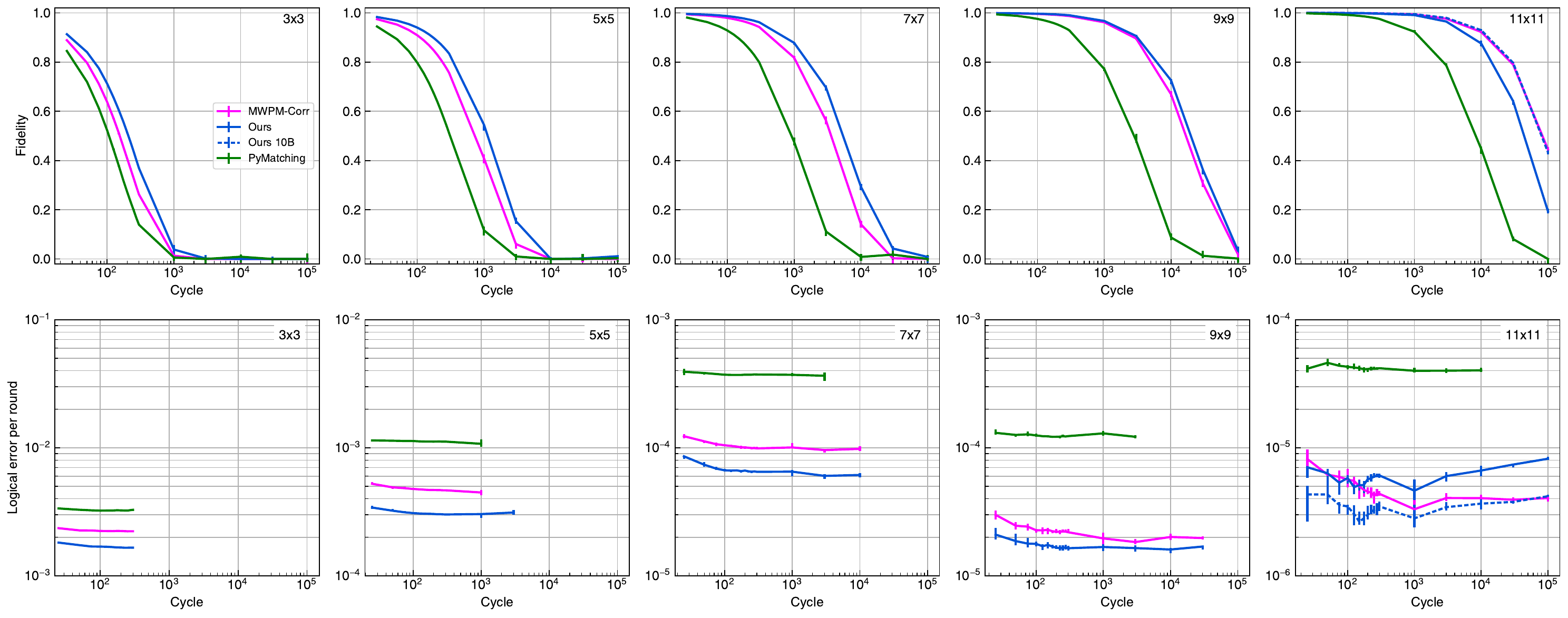}
    \caption{\textbf{Generalization to longer experiments.} Fidelities and logical error rates for networks trained on the scaling dataset for up to 25 error detection cycles but applied to decoding experiments of longer durations. We only plot logical error rates where the corresponding fidelities are greater than $0.1$. The data is generated from the same simulated experiments, stopped at different number of cycles.}
    \label{fig:SI-300rounds}
\end{figure}

Ultimately, in order to perform fault-tolerant quantum computation at arbitrary circuit depths, a decoder needs to be able to keep its decoding performance for an arbitrary number of error correction cycles. Figure~\ref{fig:SI-300rounds} shows the performance of networks trained up to 25 cycles and that they maintain their performance when applied to much longer experiments, up to $100\,000$ cycles or until the decoder fidelity drops below $10\%$.

We note there appears to be a systematic decrease in LER for the matching decoder, as well as the ML decoders on smaller distances. 
We hypothesize that this is caused by the relatively high measurement noise in our simulations (inspired by a superconducting noise profile). 
This measurement noise is particularly damaging in the terminal round, where data qubit measurements can cause space-like separated detection event pairs which tend to be more damaging than time-like separated detection event pairs. 
This is further exacerbated by withholding I/Q information about these measurements from the ML and MWPM-Corr decoders (cf. \cref{sec:pitfalls}).
While we have chosen to train on 25 cycles to mirror the experiments in \cite{milestone2}, it would be interesting to quantify the effect of extending the training to more cycles, or fine-tuning for extended performance.

We also note that while the recurrent architecture of our decoder allows us to perform this experiment without significant resource overhead on the decoder side---i.e.\ an error correction step at round $60\,661$ takes exactly the same amount of time and memory as one at round $7$---in its current implementation, PyMatching would take the entire matching graph as input, which incurs a significant memory overhead.
We leave comparisons to streaming decoder implementations to future work.

\FloatBarrier\subsection{Soft matching} 
The minimum-weight perfect matching (MWPM) decoder can also be augmented to use soft information \cite{pattison2021improved}.
For each I/Q point, we can compute the posterior probability that the sample was drawn from either the $\ket{0}$-outcome distribution or $\ket{1}$-outcome distribution---see Section ~\ref{sec:SI-IQ}.
Note that this can sometimes classify $\ket{2}$-outcomes as highly confident $\ket{1}$-outcomes.

We can threshold these posterior probabilities to obtain binary measurement outcomes that are used to compute detection events.
Then, the probability of the opposite outcome can be interpreted as a measurement error, which contributes a probability to one of the error edges in the error graph that instantiates the MWPM decoder.
The probabilities of these edges are reassigned, replacing the average measurement probability contribution with these instance-specific probabilities.
This change can be further propagated to the correlation reweighting step \cite{fowler2013optimal}.
The posterior probabilities for data qubit measurements are withheld to compare fairly with the ML decoder, from which these values are also withheld.
We leave comparison to a leakage-aware matching decoder, which reweights edges based on leakage detections \cite{suchara2015leakage}, to future work.

\FloatBarrier\subsection{Ablations}
\label{sec:SI-ablations}

\begin{figure}[t]
    \centering
    \includegraphics[width=\textwidth]{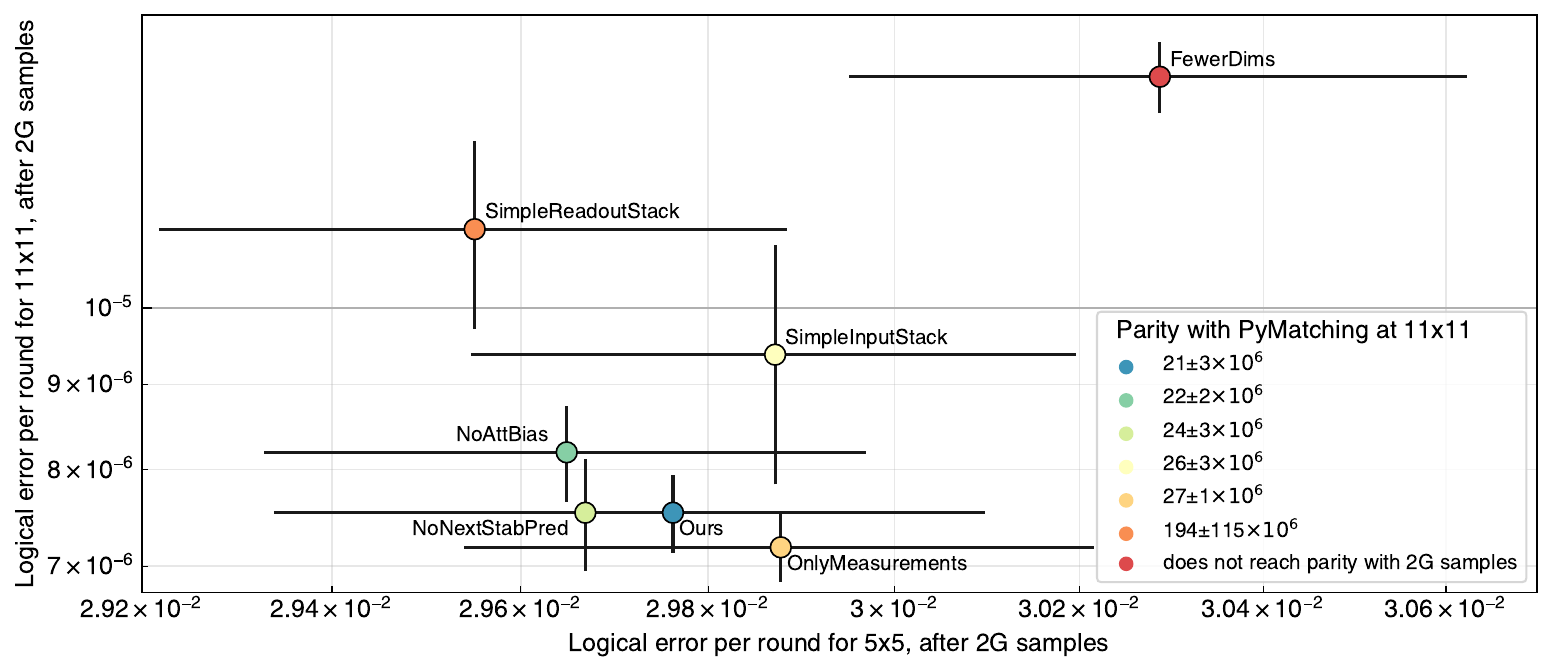}
    \caption{\textbf{Effect of ablations in performance and data efficiency.} (A) Decoding performance of the best performing subset of ablations in $5\times5$ DEM and $11\times11$ Pauli+. Colors indicate the number of training samples required for reaching performance parity with PyMatching in 5x5 DEM. 
    }
    \label{fig:SI-ablations-2d}
    
\end{figure}

To understand the effect of different components of the architecture we conduct \emph{ablations}, by training networks where we remove or simplify one aspect of the main design and seeing the effect.  For each scenario (described in the following sections) we trained 5 models with different random seeds and compare the mean test set LER in Figures~\ref{fig:SI-ablations}A and B for $5\times5$ Sycamore DEM pre-training and $11\times11$ Pauli+ training respectively.  For the former,  Fig.~\ref{fig:SI-ablations-2d} also shows the effect on training speed. In each case, other hyperparameters were not changed, and it is possible that lost performance could be recovered by compensating with other changes. 

While many of the ablations have only a small effect on the performance at $5\times5$, at $11\times11$ the effects are more marked.  

\subsubsection{Model Ablations}

\paragraph{LSTM.}  We substitute the whole recurrent core with a stack of 6 LSTMs, as implemented in Haiku~\cite{haiku2020github}. To keep the number of parameters roughly constant upon scaling (as our ML decoder does), we make the width of the LSTM hidden layers dependent on the code distance $d$, and equal to $ 64 \times (25 - 1) / (d^2 - 1) $. As the LSTM uses dense layers, which lack any spatial equivariance, we also remove the scatter and pooling operations in the readout.
\paragraph{NoConv.} We remove all the convolutional elements in the Syndrome Transformer.
\paragraph{SimpleReadoutStack.} We reduce the number of layers in the Readout ResNet from 16 to 1 (see \cref{tab:hyperparameters}).
\paragraph{SimpleInputStack.} We reduce the number of ResNet layers in the feature embedding from 2 to 1 (see \cref{tab:hyperparameters}).
\paragraph{PoolingStabs.} We do not scatter to 2D before pooling in the readout. The result is that we pool across all stabilizers instead of along data qubit rows or columns (corresponding to logical observables).
\paragraph{NoAttBias.} We remove the attention bias, both embedding and event indicator features.
\paragraph{NoNextStabPred.} We remove the next stabilizer prediction loss from the loss.
\paragraph{FewerDims.} We reduce the number of dimensions per stabilizer in the Syndrome Transformer to 30.
\paragraph{FewerLayers.} We reduce the number of layers in the Syndrome Transformer from 3 to 1 for each round.

\subsubsection{Input Ablations}

\paragraph{OnlyEvents} We only give syndrome information as detection events (removing measurements).
\paragraph{OnlyMeasurements} We only give syndrome information as raw qubit measurements (removing events).

\FloatBarrier\subsection{Choice of data and noise model for pre-training}\label{sec:SI-pretraining-ablation}

\begin{figure}[t]
    \centering
    \includegraphics[width=0.8\textwidth]{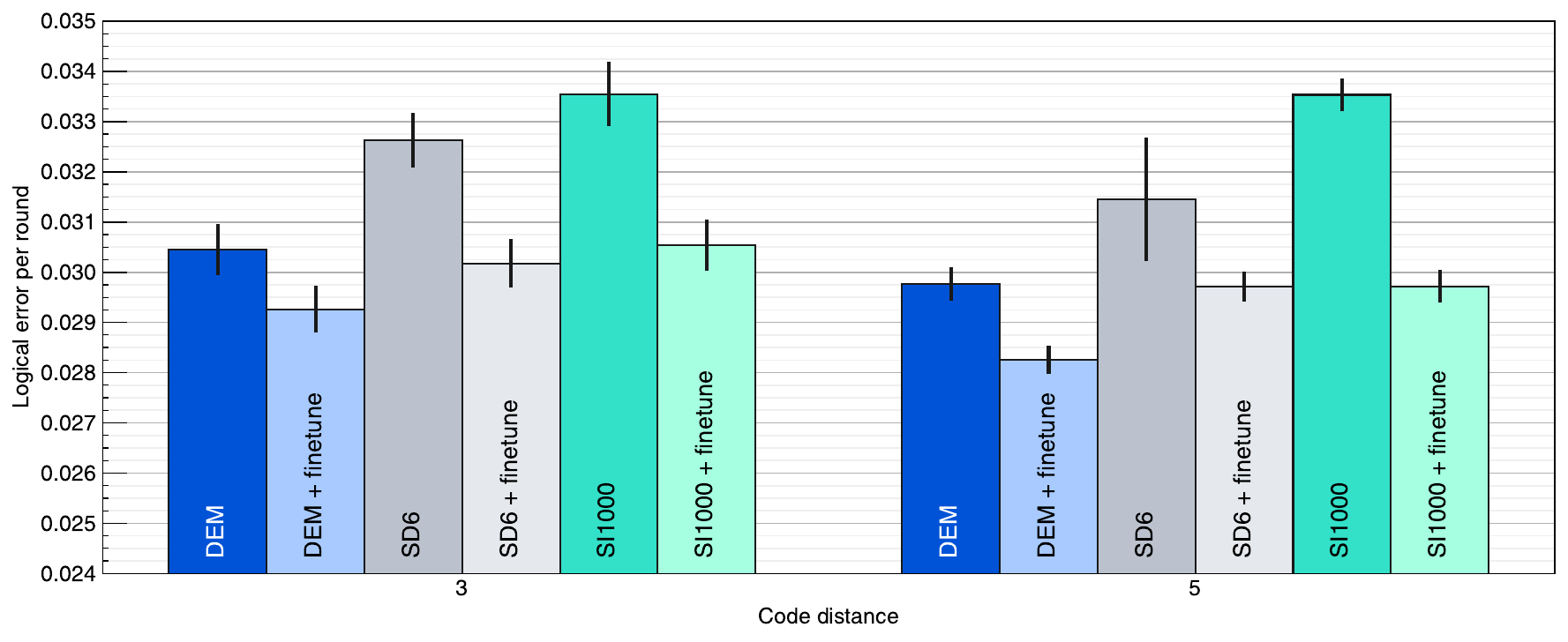}
    \caption{\textbf{Our ML decoder's performance on the Sycamore experiment when pre-training using different noise models.} Each bar represents the logical error per round on experimental data after pre-training in one of experimental-fitted DEM, SD6 depolarizing noise and superconductor-inspired SI1000, and possibly fine-tuning on experimental data.}
    \label{fig:pretrain-ablation}
    
\end{figure}

Obtaining a DEM which matches physical hardware well can be a challenge and often requires significant effort from specialists. On the other hand, weaker noise models such as SI1000 or SD6 provide a simpler alternative but with less expressivity. In order to evaluate the importance of a well-tuned noise model for training our ML decoder we explore the effect of using alternative noise models for pre-training and fine-tuning. To do this, we pre-trained with SI1000 and SD6 noise for code distance 3 and 5 and compare them to the DEM pre-trained and fine-tuned models in Figure \ref{fig:figure-1}. While the DEM models had been pre-trained for a total of 2 billion examples, the SI1000 and SD6 pre-training was early stopped after 500 million examples as the training had completely converged. Furthermore, SD6 and SI1000 pre-training was done for only 5 seeds rather than 20. We then fine-tuned each pre-trained model on a subset of the experimental data as described in Appendix \ref{sec:SI-Training}. We chose the hyperparameters of these noise models such that their event densities roughly match the event density of the fitted DEMs  ($p=0.006$ for SI1000 and $p=0.008$ for SD6).

The results in Figure \ref{fig:pretrain-ablation} show that pre-training with data generated from a matched DEM has a significant advantage over SD6 or SI1000 noise and that this advantage also carries over to the fine-tuned setting. However, fine-tuning models pre-trained using these coarser noise models brings them close to the performance of the tensor network decoder, showing that our neural-network based decoder remains competitive even without access to a carefully tuned noise model. These results further highlight the importance of fine-tuning on real experimental data. It is worth noting that the fine-tuned models only have access to a very limited amount of data and it seems likely that performance would improve further if more data were available. Correspondingly, in the limited data regime, further improving the training noise model may also result in better decoding performance.

\FloatBarrier\subsection{Training Data Efficiency}
\label{sec:SI-training-examples}
\begin{figure}[t!]
    \centering
    \begin{subfigure}[b]{0.70\textwidth}
    \caption{}
    \includegraphics[height=8cm]{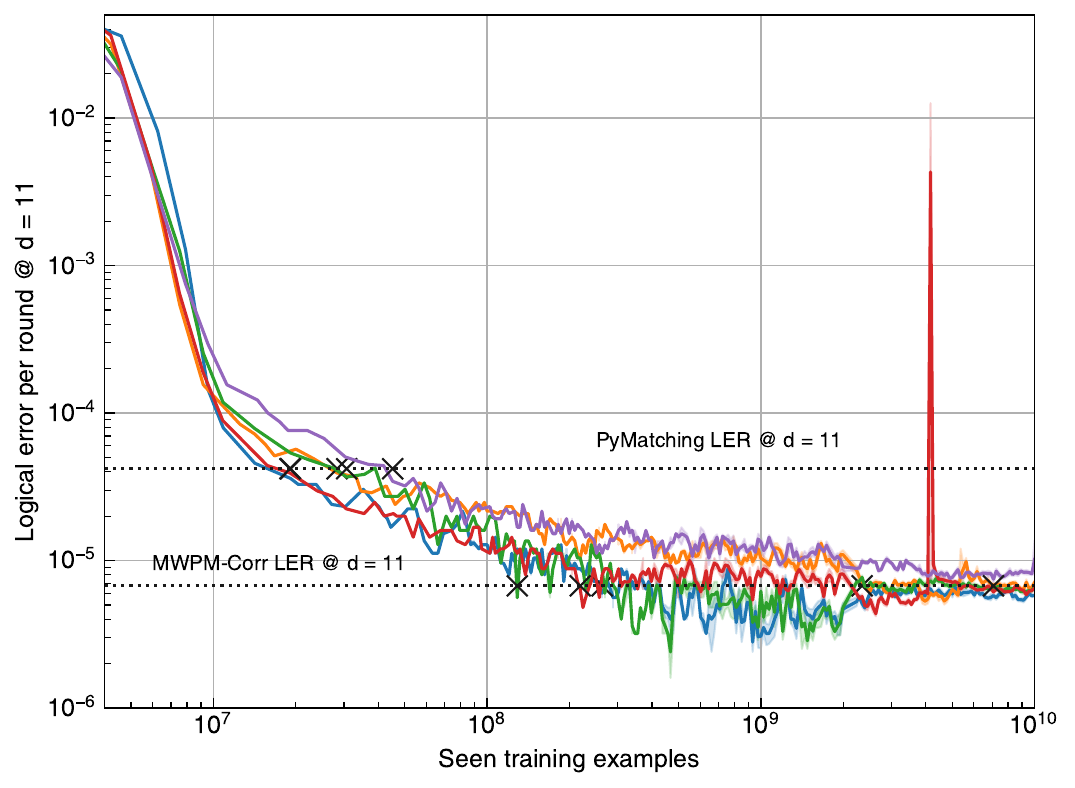}
    \end{subfigure}
    \begin{subfigure}[b]{0.29\textwidth}
    \caption{}
    \includegraphics[height=8cm]{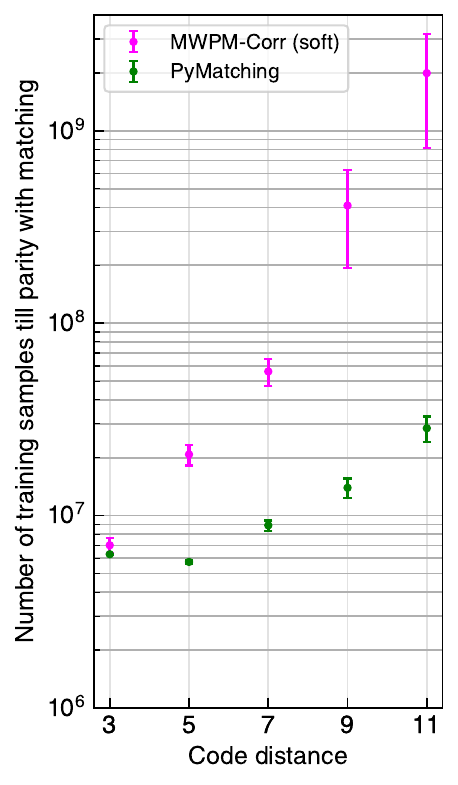}
    \end{subfigure}
    \caption{\textbf{Number of training examples to achieve parity with matching decoders.} 
    (\textbf{A}) Five training curves (with different random seeds) for code distance $11$ for Pauli+ generated data, with SNR $=10$, $t=0.01$ and a $0.1\%$ leakage, showing the validation set LER as training progresses.
    Marked with $\times$ are the crossover points for first reaching parity with either PyMatching or MPWM-Corr with soft inputs.
    (\textbf{B})
    Average first crossover point for different code distances for reaching parity with PyMatching (green) or MPWM-Corr with soft inputs (magenta). Error bars are the standard deviation computed from the five individual seeds.
    }
    \label{fig:training-examples}
    
\end{figure}

A concern raised for ML decoders is whether the training data required scales infeasibly with the code distance~\cite{varsamopoulos2019comparing}. We find that, at certain code distances, when we keep on sampling new data from the simulator, performance continues to increase. Above we arbitrarily limited the training to 2 billion examples regardless of code distance, but Figure \ref{fig:training-examples} shows the number of examples required to achieve parity with PyMatching and MWPM-Corr as a rough guide to the training data required. The sample size needed for our architecture is several magnitudes smaller than prior art~\cite{varsamopoulos2019comparing} at distance equal or larger than 9.  Such scaling advantage can be combined with additional distributed decoding strategy where deep learning is applied to only local error decoding and is then combined with a deterministic decoding algorithm~(such as union find)~\cite{Meinerz_2022,varsamopoulos2020decoding} for global decoding. The parallelism realized through such distributed decoding scheme resembles the type of parameter sharing realized in our decoder through the use of CNN  and RNN. The advantage of our approach is to accommodate more flexible range of correctable errors where no assumptions on the locality of error mechanism is being made.

\end{document}